\documentclass[12pt,showpacs,showkeys,preprintnumbers,amsmath,amssymb,amsfonts,floatfix,aps,reprint,onecolumn]{revtex4-1}

\usepackage{longtable}
\usepackage{graphicx} 
\usepackage{fullpage}
\usepackage{array}
\usepackage{bm}\allowdisplaybreaks[1]
\usepackage{physics}
\usepackage[normalem]{ulem}
\usepackage{adjustbox}
\usepackage{wrapfig}
\usepackage{booktabs}
\usepackage{hyperref}
\hypersetup{colorlinks,allcolors=black}
\usepackage[table]{xcolor}
\usepackage{enumitem}

\renewcommand{\toprule}{\hline\hline}
\renewcommand{\midrule}{\hline}
\renewcommand{\bottomrule}{\hline\hline}

\newcounter{univ_counter}
\addtocounter{univ_counter} {1} 
\edef\YEREVAN{$^{\arabic{univ_counter}}$ }

\addtocounter{univ_counter} {1} 
\edef\ANL{$^{\arabic{univ_counter}}$ }

\addtocounter{univ_counter} {1} 
\edef\BARCELONA{$^{\arabic{univ_counter}}$ }

\addtocounter{univ_counter} {1} 
\edef\BONN{$^{\arabic{univ_counter}}$ }

\addtocounter{univ_counter} {1} 
\edef\CHARLES{$^{\arabic{univ_counter}}$ }

\addtocounter{univ_counter} {1} 
\edef\CAS{$^{\arabic{univ_counter}}$ }

\addtocounter{univ_counter} {1} 
\edef\CASU{$^{\arabic{univ_counter}}$ }

\addtocounter{univ_counter} {1} 
\edef\CNU{$^{\arabic{univ_counter}}$ }

\addtocounter{univ_counter} {1} 
\edef\FIU{$^{\arabic{univ_counter}}$ }

\addtocounter{univ_counter} {1} 
\edef\FSU{$^{\arabic{univ_counter}}$ }

\addtocounter{univ_counter} {1} 
\edef\GWU{$^{\arabic{univ_counter}}$ }

\addtocounter{univ_counter} {1} 
\edef\GLASGOW{$^{\arabic{univ_counter}}$ }

\addtocounter{univ_counter} {1} 
\edef\GSIFFN{$^{\arabic{univ_counter}}$ }

\addtocounter{univ_counter} {1} 
\edef\HUELVA{$^{\arabic{univ_counter}}$ }

\addtocounter{univ_counter} {1} 
\edef\ILL{$^{\arabic{univ_counter}}$ }

\addtocounter{univ_counter} {1} 
\edef\IU{$^{\arabic{univ_counter}}$ }

\addtocounter{univ_counter} {1} 
\edef\INFNBA{$^{\arabic{univ_counter}}$ }

\addtocounter{univ_counter} {1} 
\edef\INFNCAT{$^{\arabic{univ_counter}}$ }

\addtocounter{univ_counter} {1} 
\edef\INFNFE{$^{\arabic{univ_counter}}$ }

\addtocounter{univ_counter} {1} 
\edef\INFNFRA{$^{\arabic{univ_counter}}$ }

\addtocounter{univ_counter} {1} 
\edef\INFNGE{$^{\arabic{univ_counter}}$ }

\addtocounter{univ_counter} {1} 
\edef\INFNMIL{$^{\arabic{univ_counter}}$ }

\addtocounter{univ_counter} {1} 
\edef\INFNPER{$^{\arabic{univ_counter}}$ }

\addtocounter{univ_counter} {1} 
\edef\INFNROMA{$^{\arabic{univ_counter}}$ }

\addtocounter{univ_counter} {1} 
\edef\INFNRO{$^{\arabic{univ_counter}}$ }

\addtocounter{univ_counter} {1} 
\edef\INFNTUR{$^{\arabic{univ_counter}}$ }

\addtocounter{univ_counter} {1} 
\edef\JMU{$^{\arabic{univ_counter}}$ }

\addtocounter{univ_counter} {1} 
\edef\JGU{$^{\arabic{univ_counter}}$ }

\addtocounter{univ_counter} {1} 
\edef\LBL{$^{\arabic{univ_counter}}$ }

\addtocounter{univ_counter} {1} 
\edef\LYON{$^{\arabic{univ_counter}}$ }

\addtocounter{univ_counter} {1} 
\edef\MIT{$^{\arabic{univ_counter}}$ }

\addtocounter{univ_counter} {1} 
\edef\MESSINA{$^{\arabic{univ_counter}}$ }

\addtocounter{univ_counter} {1} 
\edef\MEXICO{$^{\arabic{univ_counter}}$ }

\addtocounter{univ_counter} {1} 
\edef\MILANO{$^{\arabic{univ_counter}}$ }

\addtocounter{univ_counter} {1} 
\edef\MISS{$^{\arabic{univ_counter}}$ }

\addtocounter{univ_counter} {1}
\edef\UNAM{$^{\arabic{univ_counter}}$ }

\addtocounter{univ_counter} {1}
\edef\NANJING{$^{\arabic{univ_counter}}$ }
 
\addtocounter{univ_counter} {1}
\edef\NANJINGINP{$^{\arabic{univ_counter}}$ }
 
\addtocounter{univ_counter} {1} 
\edef\UNH{$^{\arabic{univ_counter}}$ }

\addtocounter{univ_counter} {1} 
\edef\NCW{$^{\arabic{univ_counter}}$ }

\addtocounter{univ_counter} {1} 
\edef\ODESA{$^{\arabic{univ_counter}}$ }

\addtocounter{univ_counter} {1} 
\edef\ORSAY{$^{\arabic{univ_counter}}$ }

\addtocounter{univ_counter} {1} 
\edef\OSAKA{$^{\arabic{univ_counter}}$ }

\addtocounter{univ_counter} {1} 
\edef\PAVIA{$^{\arabic{univ_counter}}$ }

\addtocounter{univ_counter} {1} 
\edef\REGINA{$^{\arabic{univ_counter}}$ }

\addtocounter{univ_counter} {1} 
\edef\ROMAII{$^{\arabic{univ_counter}}$ }

\addtocounter{univ_counter} {1} 
\edef\SACLAY{$^{\arabic{univ_counter}}$ }

\addtocounter{univ_counter} {1}
\edef\ULS{$^{\arabic{univ_counter}}$ }

\addtocounter{univ_counter} {1} 
\edef\SCAROLINA{$^{\arabic{univ_counter}}$ }

\addtocounter{univ_counter} {1}
\edef\SYRACUSE{$^{\arabic{univ_counter}}$ }

\addtocounter{univ_counter} {1} 
\edef\UTFSM{$^{\arabic{univ_counter}}$ }

\addtocounter{univ_counter} {1}
\edef\TENN{$^{\arabic{univ_counter}}$ }

\addtocounter{univ_counter} {1}
\edef\JLAB{$^{\arabic{univ_counter}}$ }

\addtocounter{univ_counter} {1}
\edef\TOKYO{$^{\arabic{univ_counter}}$ }

\addtocounter{univ_counter} {1}
\edef\TORINO{$^{\arabic{univ_counter}}$ }

\addtocounter{univ_counter} {1}
\edef\THU{$^{\arabic{univ_counter}}$ }

\addtocounter{univ_counter} {1} 
\edef\VIRGINIA{$^{\arabic{univ_counter}}$ }

\addtocounter{univ_counter} {1} 
\edef\VUU{$^{\arabic{univ_counter}}$ }

\addtocounter{univ_counter} {1} 
\edef\WM{$^{\arabic{univ_counter}}$ }

\addtocounter{univ_counter} {1} 
\edef\YORK{$^{\arabic{univ_counter}}$ }

\begin{document}

\title{Frascati 22 GeV Workshop Summary}

\author{
A. Accardi,\JLAB$\!\!^,$\CNU\ 
P.~Achenbach,\JLAB\
A.~Afanasev,\GWU\ 
M.~Albrecht,\JLAB\
A.C. Alvaro, \PAVIA\
J.~Arrington,\LBL\ 
H.~Avakian,\JLAB\ 
P.~Barry,\ANL\ 
A.~Bashir,\HUELVA$\!\!^,$\MEXICO\
M.~Bashkanov,\YORK
% ...
\vspace{3.0mm}
}

\affiliation{\YEREVAN A.I. Alikhanyan National Science Laboratory (Yerevan Physics Institute), 0036 Yerevan, Armenia}
\affiliation{\ANL Argonne National Laboratory, Argonne, Illinois 60439}
\affiliation{\BARCELONA Departament de F\'isica Qu\`{a}ntica i Astrof\'isica and Institut de Ci\`{e}ncies del Cosmos, Universitat de Barcelona, c.~Mart\'i i Franqu\`{e}s, 1, 08028 Barcelona, Spain}
\affiliation{\BONN Universit\"{a}t Bonn, D-53115 Bonn, Germany}
\affiliation{\CHARLES Charles University, Prague, Czech Republic}
\affiliation{\CAS Institute of Theoretical Physics, Chinese Academy of Sciences, Beijing 100190, China}
\affiliation{\CASU University of Chinese Academy of Sciences, Beijing 100049, China}
\affiliation{\CNU Christopher Newport University, Newport News, Virginia 23606}
\affiliation{\FIU Florida International University, Miami, Florida 33199}
\affiliation{\FSU Florida State University, Tallahassee, Florida 32306}
\affiliation{\GWU The George Washington University, Washington, D.C. 20052}
\affiliation{\GLASGOW University of Glasgow, Glasgow G12 8QQ, United Kingdom}
\affiliation{\GSIFFN GSI Helmholtzzentrum fur Schwerionenforschung GmbH, D-64291 Darmstadt, Germany}
\affiliation{\HUELVA Universidad de Huelva, Huelva 21071, Spain}
\affiliation{\ILL University of Illinois at Urbana-Champaign, Champaign, IL 61801}
\affiliation{\IU Indiana University, Bloomington, Indiana 47408}
\affiliation{\INFNBA INFN, Sezione di Bari, 70125 Bari, Italy}
\affiliation{\INFNCAT INFN, Sezione di Catania, 95123 Catania, Italy}
\affiliation{\INFNFE INFN, Sezione di Ferrara, 44100 Ferrara, Italy}
\affiliation{\INFNFRA INFN, Laboratori Nazionali di Frascati, 00044 Frascati, Italy}
\affiliation{\INFNGE INFN, Sezione di Genova, 16146 Genova, Italy}
\affiliation{\INFNMIL INFN, Sezione di Milano, 20133 Milano, Italy}
\affiliation{\INFNPER INFN, Sezione di Perugia, 06100 Perugia, Italy}
\affiliation{\INFNROMA INFN, Sezione di Roma, 00185 Roma, Italy}
\affiliation{\INFNRO INFN, Sezione di Roma Tor Vergata, 00133 Rome, Italy}
\affiliation{\INFNTUR INFN, Sezione di Torino, 10125 Torino, Italy}
\affiliation{\JMU James Madison University, Harrisonburg, Virginia 22807}
\affiliation{\JGU Johannes Gutenberg-Universit\"{a}t Mainz, Mainz, Germany}
\affiliation{\LBL Lawrence Berkeley National Laboratory, Berkeley, California 94720}
\affiliation{\LYON Universit\'e Claude Bernard Lyon 1, CNRS/IN2P3, F-69622, Villeurbanne, France}
\affiliation{\MIT Massachusetts Institute of Technology, Cambridge, Massachusetts 02139}
\affiliation{\MESSINA Universit\`a di Messina, 98166 Messina, Italy}
\affiliation{\MEXICO Universidad Michoacana de San Nicol\'{a}s de Hidalgo, Morelia, Michoac\'{a} 58040, Mexico}
\affiliation{\MILANO Universit\`a di Milano, 20133 Milano, Italy}
\affiliation{\MISS Mississippi State University, Mississippi State, Mississippi 39762}
\affiliation{\UNAM Universidad Nacional Aut\'onoma de M\'exico, 04510 Mexico City, Mexico}
\affiliation{\NANJING Nanjing University, Nanjing, Jiangsu 210093, China}
\affiliation{\NANJINGINP Institute for Nonperturbative Physics, Nanjing University, Nanjing, Jiangsu 210093, China}
\affiliation{\UNH University of New Hampshire, Durham, New Hampshire 03824}
\affiliation{\NCW University of North Carolina Wilmington, Wilmington, NC 28403}
\affiliation{\ODESA Odesa Polytechnic National University, 65044 Odesa, Ukraine}
\affiliation{\ORSAY Universit\'{e} Paris-Saclay, CNRS/IN2P3, IJCLab, 91405 Orsay, France}
\affiliation{\OSAKA Research Center for Nuclear Physics, The University of Osaka, Ibaraki, 567-0047, Japan}
\affiliation{\PAVIA Universit\'{a} di Pavia and INFN, 27100 Pavia, Italy}
\affiliation{\REGINA University of Regina, Regina, Saskatchewan, S4S 0A2, Canada}
\affiliation{\ROMAII Universit\`{a} di Roma Tor Vergata, 00133 Rome, Italy}
\affiliation{\SACLAY IRFU, CEA, Universit\'{e} Paris-Saclay, F-91191 Gif-sur-Yvette, France}
\affiliation{\ULS Universidad de La Serena, 1700000 La Serena, Chile}
\affiliation{\SCAROLINA University of South Carolina, Columbia, South Carolina 29208}
\affiliation{\SYRACUSE Syracuse University, Syracuse, New York 13244}
\affiliation{\UTFSM Universidad T\'{e}cnica Federico Santa Mar\'{i}a, Casilla 110-V Valpara\'{i}so, Chile}
\affiliation{\TENN University of Tennessee, Knoxville, Tennessee 37996}
\affiliation{\JLAB Thomas Jefferson National Accelerator Facility, Newport News, Virginia 23606}
\affiliation{\TOKYO University of Tokyo, Hongo, Tokyo 113-0033, Japan}
\affiliation{\TORINO Universit\`a di Torino, 10125 Torino, Italy}
\affiliation{\THU Tsinghua University, Beijing 100084, China}
\affiliation{\VIRGINIA University of Virginia, Charlottesville, Virginia 22901}
\affiliation{\VUU Virginia Union University, Richmond, Virginia 23220}
\affiliation{\WM The College of William and Mary, Williamsburg, Virginia 23187}
\affiliation{\YORK University of York, York YO10 5DD, United Kingdom}

\date{\today}

\maketitle

\thispagestyle{empty}
\tableofcontents

\newpage

\section{Introduction}

This document summarizes the outcomes of the ``Science at the Luminosity Frontier: Jefferson Lab at 22~GeV'' workshop, held at the INFN Laboratori Nazionali di Frascati in December 2024~\cite{workshop-webpage}. The 
primary goal of the workshop was to critically assess and refine the scientific case for a proposed energy upgrade of the Continuous Electron Beam Accelerator Facility (CEBAF) to 22~GeV. This document intends 
to capture the progress on developing the scientific case since the publication of a lengthy “White Paper” in summer 2024 signed by $\sim$450 authors ~\cite{Accardi:2023chb}. Thus, it is organized along the 
following main points:

\begin{enumerate}
\item Capture the changes, refinements, and improvements of motivation, theoretical, and experimental studies since the White Paper;
\item Set milestones for critical-path theoretical or experimental studies in order to be able to assess progress toward these goals at the next 22~GeV workshop; and
\item Summarize, where possible, the anticipated global landscape 10-15 years from now, when the 22~GeV upgrade might begin producing results.
\end{enumerate}


The current 12 GeV program at Jefferson Lab has been highly successful in exploring the structure of hadrons and nuclear matter. However, a 22 GeV CEBAF upgrade would enable a new generation of experiments to address foundational questions in the field that are beyond the reach of the current facility and are an essential complement to the physics programs at future facilities, such as at future Electron Ion Colliders (EIC or EicC)~\cite{AbdulKhalek:2021gbh,Anderle:2021wcy}. Leveraging its unique capabilities in high-luminosity operations and its existing experimental infrastructure, a 22 GeV CEBAF would position the United States to lead in the global quest to understand how Quantum Chromodynamics (QCD) builds the visible matter of our universe.


This summary document reflects the consensus and key findings from the presentations and working group discussions that took place over the course of the week. The workshop focused on four main physics pillars:
\begin{itemize}
   \item Hadron Spectroscopy: The need for complementary data on the production of heavy exotic hadrons, which may illuminate their microscopic structure, was a central theme.
\item Hadron Structure: The study of hadron structure was a primary focus of the program, concentrating on the evolving picture from one-dimensional longitudinal momentum (PDFs) and static spatial distributions (Form Factors) to a complete three-dimensional map of both momentum and spatial information (3D PDFs).
   \item  QCD in Nuclei: Participants explored the exciting new opportunities to study the quark-gluon structure of nuclei and its modifications in the nuclear medium.
   \item Searches for Physics Beyond the Standard Model: Discussions highlighted the potential for a 22 GeV CEBAF to provide unprecedented sensitivity to new particles and forces.
\end{itemize}
Beyond the scientific motivation, the project planning and technical path forward necessary to realize this major upgrade was also presented. Jefferson Lab continues to work with the community to articulate the scientific motivation for positron beams and for an increase in the maximum electron beam energy from 12~GeV to 22~GeV. Such capabilities could be realized in the future through a DOE construction project, which means that advance planning is important now to assess the key elements of a pre-conceptual project plan and the technical feasibility of the design.
An overview of the DOE project life cycle was presented, along with a brief look at early planning activities. The necessary next steps in this preliminary phase focus on two key areas: drafting a preliminary Technical Design Report (pre-TDR), which involves identifying urgent technical challenges for the realization of the accelerator and defining the required R\&D plans to resolve them, and continuing to develop the most compelling science case by addressing critical scientific questions, and ultimately creating the foundation for a 22 GeV Project.

The collaborative effort of the participants led to the development of a robust and compelling scientific program, providing concrete recommendations for future experiments and instrumentation needs. The findings presented here will serve as a foundational resource for the broader community and stakeholders as we move forward in planning for this transformative upgrade.

\clearpage

\section{Meson Spectroscopy}


\noindent  \emph{\underline{Recent developments since the White Paper}}: 
The overarching goals and focus of the meson spectroscopy program remain consistent with those in the White Paper~\cite{Accardi:2023chb}, namely to study the production of potentially exotic charmonium states at JLab 22~GeV. The last two 
decades have produced a significant number of discoveries of new particles in both the charm and bottom sectors by experiments at the $e^+e^-$ facilities BaBar, BESIII, and Belle, as well as LHCb, ATLAS and CMS at
CERN. Available reviews can be found in Refs.~\cite{JPAC:2021rxu,Chen:2022asf,Brambilla:2019esw,Guo:2017jvc,Esposito:2016noz}. Some of these new discoveries, like the observation of excited states of the $\Omega_b$ ($bss$) at LHCb~\cite{LHCb:2021ptx}, have extended our knowledge of conventional hadrons containing heavy quarks, while others, like the charged $Z_c$ tetraquark candidates \cite{BESIII:2013ris,Belle:2013yex,BESIII:2013ouc,LHCb:2014zfx,Belle:2014nuw}, have led to significant discussion of long-standing ideas about the valence quark content of hadrons generated by QCD. These particles, 
often generically referred to as $XYZ$ states, have generated significant discussion in the community about  their underlying structure. At present, the precise microscopic nature of these states is unknown, with multiple possible interpretations in terms of QCD degrees of freedom. The coincidence of nearby multi-particle thresholds may suggest the contribution of important multi-channel dynamics. Furthermore, there are proposed explanations that such configurations are actually shallow bound states with prominent molecular components.

\begin{figure}[h]
\centering
\includegraphics[width=0.8\linewidth]{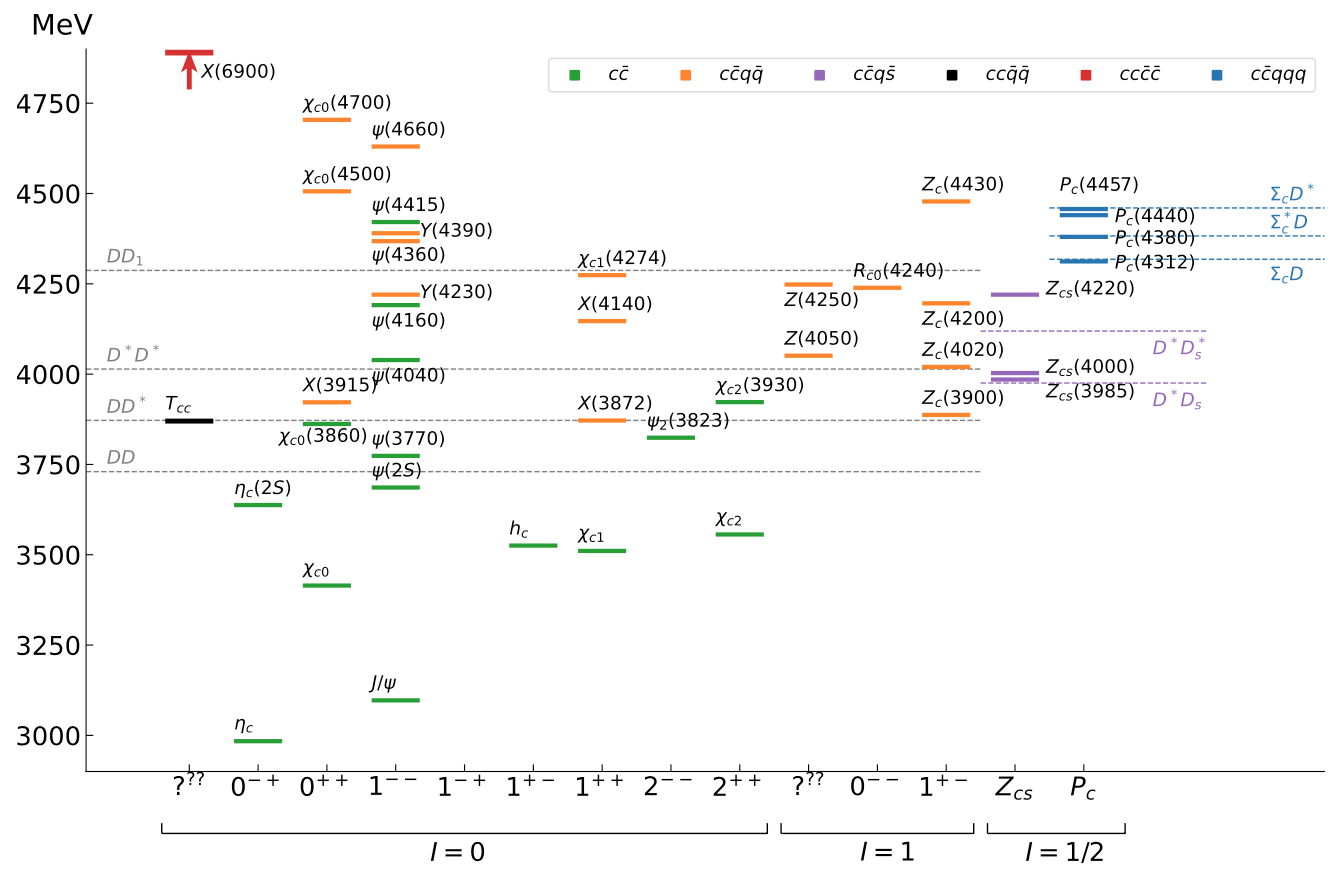}
\caption{Current landscape of conventional $c\bar{c}$ charmonium states and $X,Y,Z$, as well as pentaquark states included in the Particle Data Group listings~\cite{ParticleDataGroup:2024cfk}. Figure from Ref.~\cite{JPAC:2021rxu}.}
\label{xyzlandscape}
\end{figure}

The $XYZ$ states are exotic because their properties are not consistent with those of the well-understood spectrum of charmonium states. Some exotic features are clear: a meson with non-zero electric charge cannot be a $c\bar{c}$ state. Other states are unusual because they have masses or decay properties that do not align with expectations based on our understanding of heavy-quark systems. A common feature to all of the $XYZ$ states is that they have masses where both heavy and light quarks play a key role in their structure and decays, that is, the path to understanding these particles involves QCD in the strongly interacting regime.
Figure~\ref{xyzlandscape} shows the current landscape of the conventionally known $c\bar{c}$ charmonium states, as well as the identified exotic candidates referred to as $X$, $Y$, and $Z$~\cite{JPAC:2021rxu}. 

The main focus on the studies of such heavy exotica at JLab22 focused on the use of a new and complementary reaction mechanism involving real or quasi-real photons. Photoproduction is a well-established
tool for spectroscopic studies and appropriate in the search for states beyond $q\bar{q}$ as:

\begin{itemize}
\item Such a production mechanism can produce final states with any quantum numbers;
\item The $t$-channel production has a kinematically unconstrained initial state that minimizes the role of rescattering (no triangle singularities)~\cite{Szczepaniak:2015eza,Guo:2019twa};
\item It allows access to polarization information to learn about production couplings~\cite{Winney:2019edt};
\item It allows for a probe of the internal dynamics and structure via studies of the $Q^2$ dependence of the reaction mechanism~\cite{Babiarz:2023ebe}.
\end{itemize}

Current photoproduction searches of heavy exotica have taken place at H1~\cite{H1:2002yab} and COMPASS~\cite{COMPASS:2017wql,COMPASS:2014mhq}. At JLab, studies in Hall~B~\cite{hallb-jpsi}, Hall~C~\cite{Hafidi:2017bsg,Meziani:2016lhg}, and Hall~D \cite{GlueX:2023pev} have focused on the production of conventional $c\bar{c}$ states such as the $J/\psi$ and $\psi(2S)$. These studies were limited in statistical precision and beam energy. An upgrade to 22~GeV would enable searches for the $XYZ$ states in photoproduction, which, when combined with global data, may shed light on the nature of these states. \\

\noindent \emph{\underline{Future plans}}:
The simulations of the signal processes are relatively well understood and have already demonstrated the necessary detector resolution to observe structures in the spectra that would be caused by potentially exotic charmonium states. However, background estimations from non-resonant production of $J/\psi \pi$ and $J/\psi \pi\pi$, etc. are limited because there is no previous data in this kinematic regime and theoretical modeling is limited. The modeling of potential non-resonant production is the most import theoretical issue, to both understand the magnitude of this background and its dependence on kinematic variables. There were discussions at the workshop of how limited data from CLAS12 in Hall~B and GlueX in Hall~D at threshold and from COMPASS at higher energy could help set an upper limit on the size of these backgrounds. 

Upcoming key results from the GlueX and CLAS12 Collaborations on improved measurements of the $J/\psi$ cross section and the first observation of the $\chi_c$, are expected to be published in the next 1-2~years.  While open charm final states are a challenge to observe experimentally, an upper limit on $D\Lambda_c$ production is in progress at GlueX and will help to understand if open charm rescattering contributes to charmonium production. GlueX-II will continue through 2026 and increase the precision possible for these measurements. A follow up, third phase of GlueX, GlueX-III, will increase the luminosity in Hall~D to systematically map out the $J/\psi$ and $\chi_c$ cross sections. The paper detailing the first $J/\psi$ quasi-real photoproduction cross sections from CLAS12 on both proton and neutron targets is now in preparation~\cite{hallb-jpsi}.

The current situation opens the door for a new spectroscopy program at JLab22 based on (quasi-real) photoproduction experiments.  With the exception of the $X(3872)$, 
few, if any, $XYZ$ states are observed in multiple production mechanisms.  
This has led to speculation about whether the peaks observed in mass spectra
are due to true resonances or other effects.  Any positive signal in photoproduction, or any other production mechanism, would be a strong indication
that the peak in a particular mass spectrum is not an artifact of the initial state.
If such states can be seen in photoproduction, it is natural to then consider measurements of transition form factors as a function of $Q^2$ to probe the structure of these, possibly exotic, states. Sophisticated theoretical models are not yet available, but a vector meson dominance (VMD) approach has been used to date to make predictions for the cross sections for the production of the $X$, $Y$, $Z$ states~\cite{Albaladejo:2020tzt,JointPhysicsAnalysisCenter:2024pat,Winney:2022tky}. The VMD models have limitations and cross checks to available data for non-exotic $c\bar{c}$ processes will be important moving forward. However, the predicted cross sections are even larger than those that have already been measured for $J/\psi$ in Halls B and D at JLab. A possible experimental program at JLab22 could focus on $X$ and $Z$ production, which is more favorable in JLab22 kinematics. The measurement of $Y$ states is likely better done at the higher energies at the planned Electron-Ion Collider (EIC).

Recent predictions for production of the $X(3872)$ and $Z_c(3900)$ have become available within the coupled-channels framework of Refs.~\cite{Cao:2024nxm,Yang:2021jof,Shi:2022ipx}. These approaches allow explorations of alternative production mechanisms to open charm and show that production of such states is predicted to be sizable in both exclusive and semi-inclusive processes.

The current list of work in progress includes:

\begin{itemize}
\item Simulations are underway to study $X$ and $Z$ production in Hall D with GlueX. Simulations have been done in the GlueX-II configuration based on the 2020 GlueX configuration without the DIRC, TRD, or new high-resolution 
FCAL. Note that GlueX-III is approved to run at two times the luminosity of GlueX-II. Studies have focused on $\gamma p \to J/\psi\pi^+\pi^- p$ running for 1 year, assuming an integrated luminosity of 500 pb$^{-1}$ with a branching 
ratio of 5\%. The expected yields are: N($\psi(2S)$)=900, N($X(3872)$)=2300, N($Y(4260)$)=120. Another study has been done for $\gamma p \to J/\psi\pi p$ to study both charged and neutral $Z_c$ states.
\item Simulations are underway to study $X$ and $Z$ production in Hall B with CLAS12. The baseline version of CLAS12 has been assumed, augmented with a zero-degree photon tagger to enable studies of quasi-real photoproduction (at tagging angles below $1^\circ$). The studies have assumed a luminosity of $10^{35}$~cm$^{-2}$s$^{-1}$ and 50 days of running. The expected yields are: N($X(3872)$)=2-3k, N($Z_c(3900)$)=25k. Work has begun to estimate backgrounds from existing measurements, specifically from data taken at COMPASS on $\mu^{\pm}N$, but more work is needed to make these studies realistic.
\item Photoproduction experiments can have an impact on these studies to confirm states purported to have a non-$q\bar{q}$ nature. The most relevant modes for initial exploration at JLab22 would focus on $X$, $Z$ decays to 
$J/\psi\pi$, $J/\psi \pi\pi$ (these would be the ``discovery” modes). Studies of $\gamma p \to c\bar{c} p$ are optimal using detectors with large acceptance - Hall B and Hall D at JLab. Experimental design and optimization is now in progress, including the development of ``realistic" event generators and Monte Carlo simulations.
\end{itemize}

Questions from discussion:

\begin{itemize}
\item To further the discussion of carrying out such a program at JLab22, the development of realistic event generators is necessary and important. Such studies ongoing for EIC can serve to develop complementary experimental thrusts.
\item It is important to be able to predict the cross sections for both the resonant processes that lead to $X$, $Y$, $Z$ production relative to the non-resonant processes in the same final state to see how well the signal can be separated from the background. A phenomenological model based on a double-Regge approach may provide a reasonable estimate~\cite{Shimada:1978sx}. An inclusive photoproduction model like PYTHIA may provide another avenue to generate non-resonant processes.  Realistic backgrounds must be included in the different simulations to enable more realistic modeling and impact statements to be made, and multiple approaches are welcome in this regard to compare and contrast the findings.
\item Plans for exploiting the measurement of polarization observables need to be advanced in detail.
\item What about establishing strange quark partners of states, e.g. $Z_c \to J/\psi \pi$, $Z_{cs} \to J/\psi K$?
\item What about radiative decays of $X(3872)$? Can these be estimated? Are such studies useful? How do backgrounds affect this channel?
\end{itemize}

\noindent\emph{\underline{The global landscape}}:  
In the next 10 years there are no experiments that will have photo- or electroproduction data to compete with the JLab22 program. Near the end of the 10-15 year time frame there may be competition on that front from AMBER at CERN~\cite{Quintans:2022utc} and the EIC (both in the US and China)~\cite{AbdulKhalek:2021gbh,Anderle:2021wcy}, or from more precise measurements of $XYZ$ branching ratios and lineshapes at the high-luminosity LHCb~\cite{LHCb:2018roe} or the Super Tau-Charm Factory in China~\cite{Achasov:2023gey}. The JLab 22 GeV program will remain unique in the era when these facilities are operational due to its high luminosity and access to near-threshold production where theoretical models predict enhanced cross sections for some states, e.g. $Z_c(3900)$, which was already highlighted in the White Paper.

A 22 GeV electron beam energy provides enhanced capabilities to explore the spectrum of light hadrons, thereby extending the existing 12 GeV program in hadron spectroscopy. Currently, the real photon beam used in the GlueX
experiment that is generated from the 12 GeV electrons has a peak polarization of about 35\% at an energy of about 9 GeV. This configuration is obtained by a choice of orientation of the diamond lattice with respect to the photon beam. Higher polarization can be obtained but it comes with a cost of lowering the energy of the peak intensity. Therefore, an energy upgrade allows not only the obvious increase in photon beam energy but also an option of producing similar energies to the current 12 GeV configuration but with dramatic enhancements in the degree of polarization and the photon flux. The secondary coherent peak in the energy spectrum could also be used. For example, using a 22~GeV electron beam it is possible to configure the beamline such that one has 12 GeV photons with about 70\% linear polarization, while at the same time, having 15 GeV photons with about 50\% linear polarization. This capability allows for exploration of energy-dependent effects.

The increased capabilities of an upgraded machine can be used in a variety of ways. By increasing the polarization of real photons with GlueX, one enhances the dependence of the production amplitude on the orientation of the
decay plane with respect to the production plane. This provides enhanced capability to discriminate between production mechanisms. Conducting meson spectroscopy studies at higher energy also enhances
the kinematic separation between the decay products of produced mesons and excited baryons, i.e., one has much better distinction between the beam fragmentation and target excitation regimes. Finally, one has the ability to study the dependence of the production mechanism on energy. The CLAS12 light hadron spectroscopy program will also greatly benefit from the energy upgrade, providing a high intensity flux of quasi-real photons at high energy and the extra capability of studying the $Q^2$ evolution of any new states produced. 

In summary, an upgraded electron beam at 22 GeV will allow the hadron spectroscopy program at Jefferson Lab to cross the critical threshold into the region where $c\bar{c}$ states can be produced in large quantities, and with additional light quark degrees of freedom. This opens a new opportunity to study production of exotic states and to contribute with potentially decisive information about their internal structure, which would allow us to understand their place in the landscape of hadrons generated by QCD. Moreover, with a 22~GeV electron beam, it will be possible to extend a large class of spectroscopic studies conducted with the 12~GeV program. Many new and unexpected hadrons have been discovered in the last twenty years, and a hadron spectroscopy program at 22 GeV will fully exploit the potential of present and future high-statistics measurements, combining our knowledge of reaction theory, hadron phenomenology, and data analysis to explore the nature of these exotic states.

\clearpage

\section{Hadron Structure}

\subsection{Reconstruction of Pion and Kaon Structure from Global QCD Analysis}

\begin{figure}[h]
\centering
\includegraphics[width=0.8\linewidth]{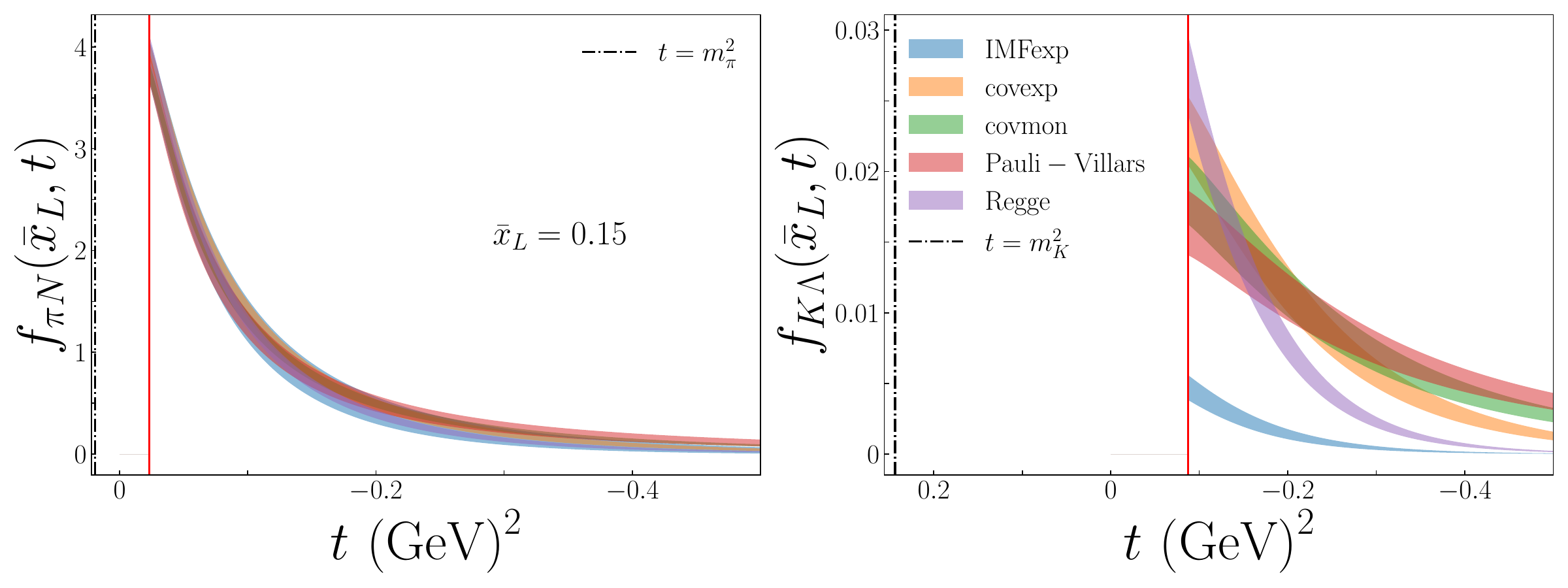}
\caption{Splitting functions for the pion (left) and kaon (right) as a function of momentum transfer squared $t$ over various models in the region of experimental applicability for $\bar{x}_L=0.15$, where $\bar{x}_L$ is the fraction of the proton momentum carried by the virtual meson.}
\label{splfuncs}
\end{figure}

\noindent \emph{\underline{Recent developments since the White Paper}}:  A new analysis to reconstruct simultaneously pion Parton Distribution Functions (PDFs) and $p\to \pi^+ n$ splitting functions has been carried out for the first time using existing Drell-Yan (DY) dilepton production data from hadron-hadron collisions, leading neutron (LN) cross sections in deep-inelastic $ep$ scattering (DIS), and peripheral hadron production data in inclusive $pp$ collisions, $pp \to nX$. The analysis confirmed the universality of the meson splitting functions that appear in both the LN and $pp \to nX$ processes, which are key inputs for the tagged DIS (TDIS) experiments at JLab. The study was further extended to reconstruct kaon splitting functions from existing hyperon production data, $pp \to \Lambda X$, providing new estimates for the expected event rates in the tagged kaon program in the JLab 12-GeV program and beyond. In Fig.~\ref{splfuncs}, the reconstructed splitting functions for pions and kaons are shown, illustrating approximately two orders of magnitude smaller event rates expected for the kaon TDIS program than for pions. Combined with previous findings presented in the White Paper~\cite{Accardi:2023chb}—which showed that JLab 12~GeV kinematics may be rather limited for accessing the partonic structure of the kaon—this new analysis underscores the need for an enhanced phase space and high-luminosity capabilities uniquely available with the JLab 22~GeV energy upgrade.\\

\noindent \emph{\underline{Future plans}}: With the determination of the $p \to K^+ \Lambda$ splitting functions from phenomenology, we plan to carry out impact studies in the near future to assess the potential of kaon TDIS data from JLab at 22~GeV for reconstructing kaon PDFs. Subsequently, we also plan to extend these studies to include TDIS with transverse momentum to provide first access to kaon transverse momentum dependent PDFs (TMDs). This will require further theoretical developments to establish the systematic uncertainties associated with the Sullivan process in the context of TMD observables~\cite{Copeland:2024cgq}.\\

\noindent \emph{\underline{The global landscape}}: Currently, the COMPASS Collaboration has completed data taking for pion-induced DY processes and is working toward preparing these data for public release. These results, together with the pion TDIS program at JLab 12~GeV, will provide a critical test of the theoretical framework used to explore meson structure in TDIS processes. Plans are underway for the AMBER experiment to measure both pion- and kaon-induced DY data with higher precision and extended phase space coverage. The JLab energy upgrade will play a complementary role in this global effort to study meson structure—particularly for kaons at high~$x$—and will enable meaningful comparisons with increasingly reliable lattice QCD results. Ultimately, exploring hadron structure across multiple hadronic systems will deepen our understanding of emergent phenomena in QCD.

\subsection{$d/u$ Ratio at Large $x$}

\begin{figure}[h]
\centering
\includegraphics[width=0.8\linewidth]{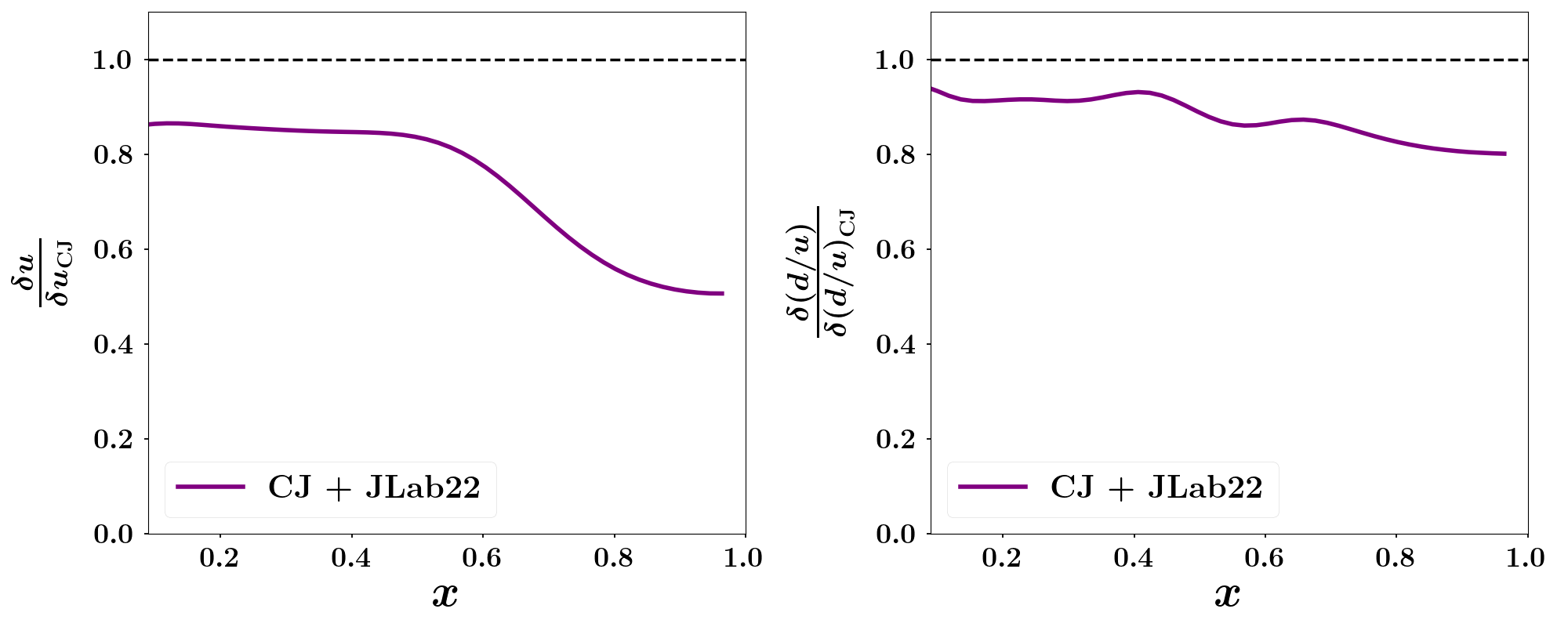}
\caption{Relative PDF uncertainty ratios for $u$ (left) and $d/u$ (right) at $Q^2 = 10$~GeV$^2$, defined as the ratio of uncertainties including JLab 22 GeV pseudo-data to those from the CJ22ht baseline fit (CJ, in the legend): 
$\delta u / \delta u_{\mathrm{CJ}}$ and $\delta (d/u) / \delta (d/u)_{\mathrm{CJ}}$. The impact is especially pronounced at large $x$, with uncertainties reduced by 20--40\% in $u$ and $d/u$.}
\label{jlab22impact}
\end{figure}

\noindent\emph{\underline{Recent developments since the White Paper}}: Recently, the CTEQ-JLab Collaboration performed a new PDF analysis to examine model biases arising from the treatment of higher-twist corrections and light nuclear modifications in DIS at large~$x$. This new work has helped mitigate such biases, enabling a more precise reconstruction of PDFs in the large-$x$ region. With this improved framework in place, a new analysis was carried out to assess the potential impact of the energy upgrade, with particular focus on the $d/u$ ratio, which plays an important role in establishing a robust understanding of proton structure in the valence region. The results, shown in Fig.~\ref{jlab22impact}, demonstrate that the inclusion of JLab 22 GeV pseudo-data leads to a substantial reduction in the relative uncertainties of both the $u$-quark distribution $u(x,Q^2)$ (left panel) and the $d/u$ ratio (right panel) at $Q^2 = 10$~GeV$^2$, especially around $x \simeq 0.6$. This improvement significantly enhances our ability to constrain valence quark distributions and better control higher-twist contributions. These advancements are essential for refining global QCD analyses and for enabling precision tests of the Standard Model, as well as for searches for physics beyond the Standard Model.\\

\noindent\emph{\underline{Future plans}}: While this analysis focused solely on the impact of the energy upgrade, it will be extended by incorporating additional data from JLab 6 and 12 GeV as a baseline for the study.\\

\noindent\emph{\underline{The global landscape}}: While high-energy facilities such as the LHC and the future EIC are optimized for probing low-to-moderate values of~$x$, only JLab provides high-luminosity access to the large-$x$, moderate-$Q^2$ regime. Measurements in this kinematic region, particularly in the context of the energy upgrade, are essential for disentangling the flavor structure of the proton in the valence region, providing key input to global QCD analyses relevant to both precision Standard Model studies and searches for physics beyond the Standard Model.

\subsection{Light Quark Sea in the Intermediate-$x$ Region}

\begin{figure}[h]
\centering
\includegraphics[width=0.45\linewidth]{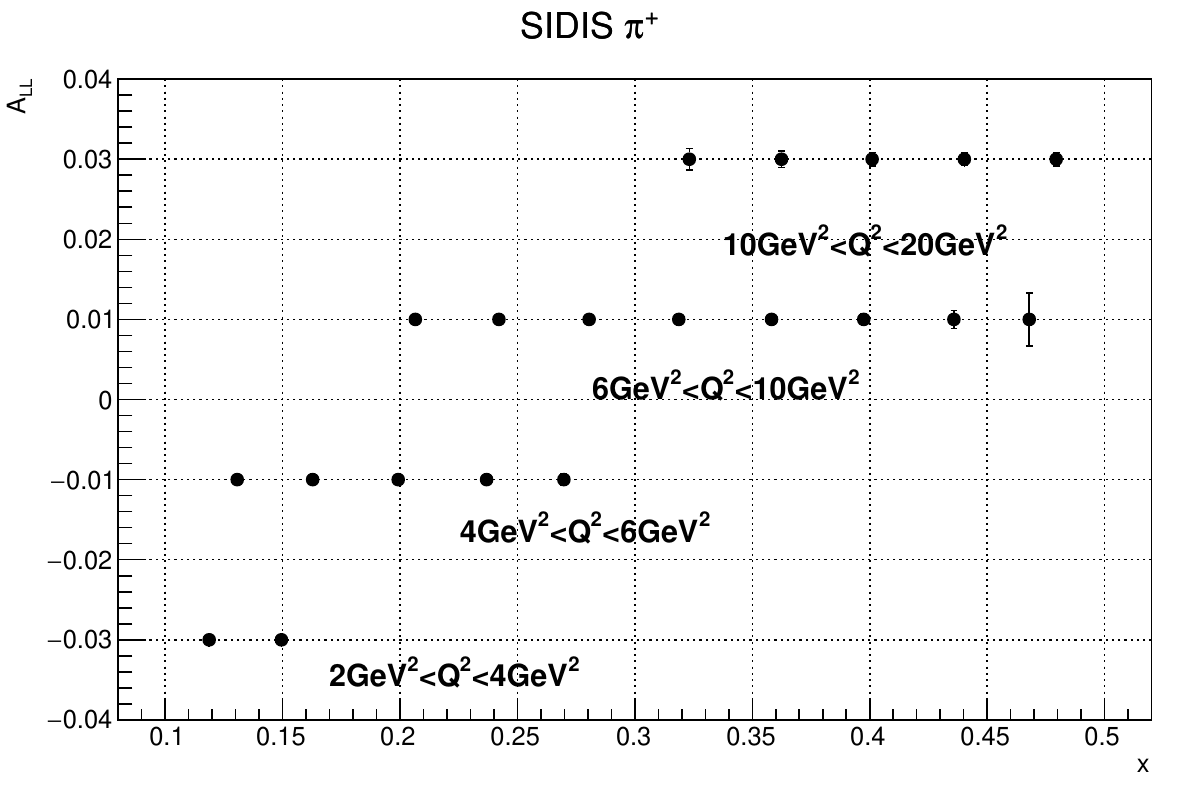} 
\includegraphics[width=0.5\linewidth]{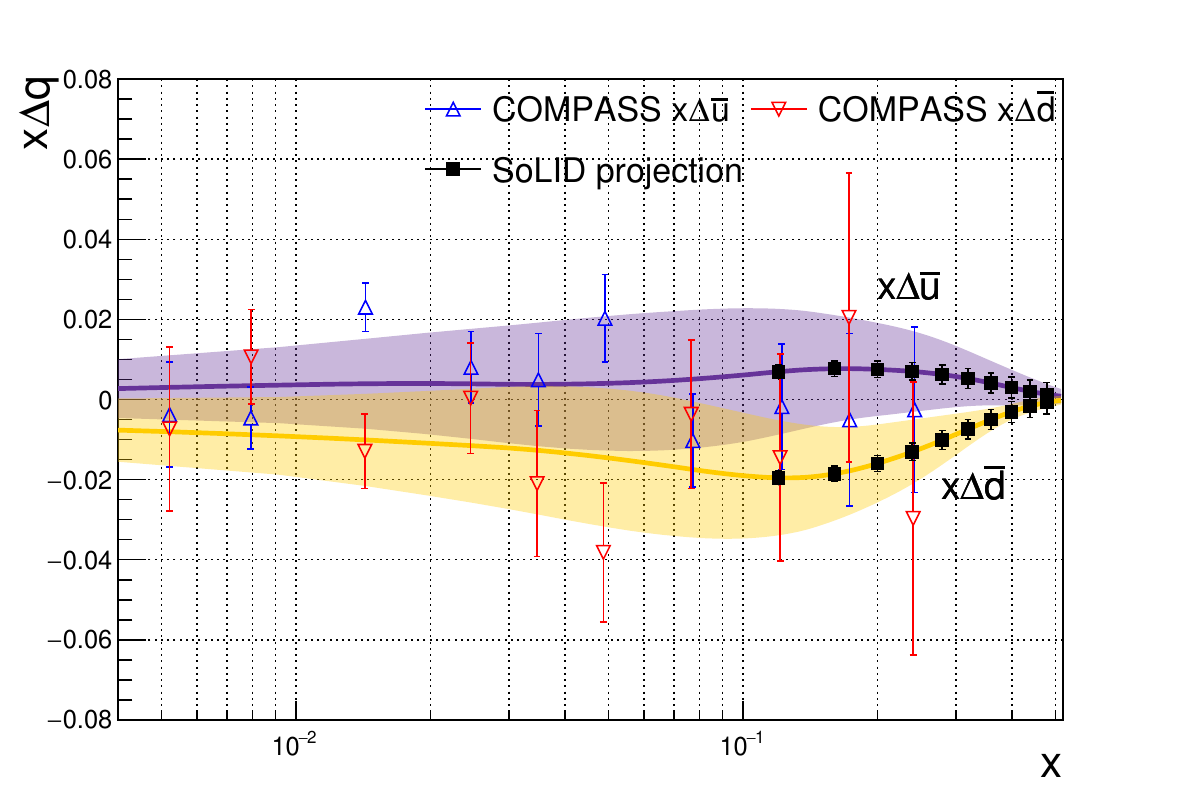}
\caption{(Left) Projected statistical uncertainties for pion $A_{LL}$ measurements with a polarized $^3\text{He}$ target with 100 PAC days using the SoLID spectrometer. (Right) Projected impact of the SoLID measurements on the sea quark helicity distribution using leading-order formalism and comparisons with extractions from COMPASS~\cite{COMPASS:2010hwr}. The bands represent the 68\% confidence level of NNPDFpol1.1~\cite{nocera2014}.}
\label{fig:enter-label}
\end{figure}

\noindent\emph{\underline{Recent developments since the White Paper}}: Probing light sea quark distributions with SIDIS at JLab 22~GeV was briefly discussed in the White Paper~\cite{Accardi:2023chb}. Since then, realistic simulations have been performed for measurements with both unpolarized and polarized SIDIS cases to study light sea quark distributions. The unpolarized case uses the HMS-SHMS in Hall C, while the polarized case uses the proposed SoLID spectrometer in Hall~A~\cite{JeffersonLabSoLID:2022iod}. Recently published SeaQuest results~\cite{cite-key} show that the excess $\overline{d}(x)$ sea (over the $\overline{u}(x)$ sea) continues from medium $x$ into high $x$, in contrast to earlier published results~\cite{PhysRevD.64.052002}. Proposed possible explanations~\cite{soffer2019, kumano2002} have different predictions for the helicity distributions that have been observed in $W^\pm$ double spin asymmetries at RHIC \cite{Cocuzza:2022jye, STAR:2010xwx, STAR:2014afm, STAR:2018fty, PHENIX:2015ade, PHENIX:2018wuz}. By using data acquired from charged pion SIDIS reactions $(e,e’\pi^{\pm})$ on both polarized $^{3}$He and polarized proton targets, one gains sensitivity to light quark helicity PDFs. Figure~\ref{fig:enter-label} (left) shows the projected statistical uncertainties for the $\pi^{+}$ $A_{LL}$ measurements with a polarized $^3\text{He}$ target using the SoLID spectrometer with 100 PAC days. The energy upgrade will improve the precision of polarized SIDIS measurements in the intermediate to high $x$ region by one order of magnitude compared to existing world data~\cite{COMPASS:2010hwr}, as indicated in Fig.~\ref{fig:enter-label} (right). The impact of the upgrade on the sea quark helicity distributions is evaluated using a leading-order formalism with pseudo-data for $\pi^{\pm}$ on both polarized $^{3}$He and polarized proton targets. While initial studies can begin with the current JLab energies, the 22~GeV electron beam will significantly expand the clean region of SIDIS for extracting PDFs and reduce the theoretical uncertainties from effects such as contamination from non-current fragmentation and higher-twist effects.\\

\noindent\emph{\underline{Future plans}}: 1) Work with JLab Theory Group to perform the impact study in the context of global data and the inclusion of higher-order QCD effects. 2) Continue the studies for the uncertainty quantification of experimental systematic uncertainties. 3) Extend the study to the strange sector with kaon SIDIS simulations.\\

\noindent\emph{\underline{The global landscape}}: Precision experimental determinations of the partonic structure of the nucleon are actively ongoing at world-class facilities, including the LHC, RHIC, and FNAL, and will continue with planned future facilities. These experiments have provided, and will continue to provide, precision data on helicity distributions at small-$x$ and large $Q^2$ \cite{aschenauer2023, COMPASS:2010hwr, abdulkhalek2022}, but data at intermediate to high $x$ will have limited precision. With the large acceptance of the SoLID spectrometer and the energy upgrade, JLab will provide the most precise extractions of the light sea quark helicity distributions at intermediate and high $x$.

\subsection{Precision Measurement of QCD Strong Coupling}

\begin{figure}[h]
\includegraphics[width=0.6\textwidth]{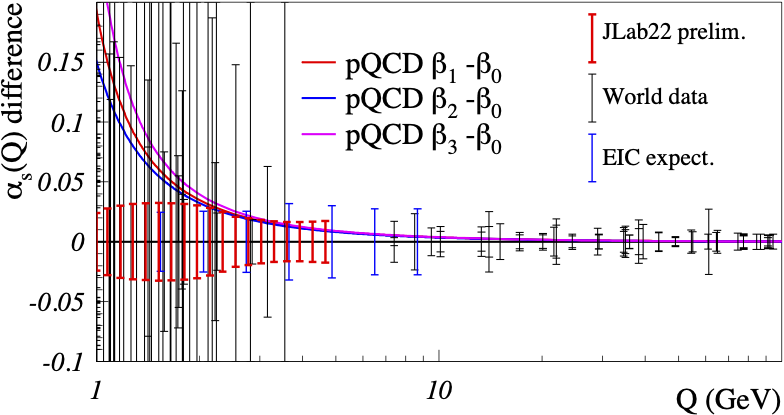}
\caption{Higher-order multiple-loop corrections to the running of the strong coupling are shown relative to the leading-order baseline. Uncertainties on $\alpha_s$ from the world data are shown by the black error bars, while the expected uncertainties from EIC and JLab at 22 GeV + EIC are shown by the blue and red error bars, respectively.}
\label{fig:alpha_s_loops}
\end{figure}

\noindent \emph{\underline{Recent developments since the White Paper}}: In the White Paper~\cite{Accardi:2023chb}, a preliminary study of the QCD strong coupling constant $\alpha_s$ at JLab22 was reported. Combined with low-$x$ data from the EIC, an accuracy of $\Delta\alpha_s(M_Z)/\alpha_s(M_Z) \sim 0.6\%$ can be reached, improving upon the current world knowledge of about $\sim 0.8\%$ that combines all global data~\cite{ParticleDataGroup:2024cfk, Deur:2023dzc}.
Since then, this study has been extended to investigate the $Q^2$-evolution of $\alpha_s$ with JLab22 kinematics, focusing on a kinematic range of around a few GeV$^2$, where higher-order quantum corrections to the strong coupling are sizable, as shown in Fig.~\ref{fig:alpha_s_loops}. We conclude that JLab22 is currently the only place in the world capable of providing high-precision polarized double-spin asymmetry data, and for the first time, directly accessing kinematic regions sensitive to higher-order quantum effects, thus allowing tests of the Standard Model (SM) and opening new opportunities to explore physics beyond the SM (BSM). This is because deviations from the SM in the strong coupling are expected to occur starting at two-loop QCD corrections.\\

\noindent \emph{\underline{Future plans}}: The extraction of $\alpha_s(Q)$ at 22 GeV with the SoLID detector~\cite{JeffersonLabSoLID:2022iod} will be fully simulated, similar to what was done for the EIC studies \cite{Kutz:2024eaq}. The sensitivity of the determination of $\alpha_s(Q)$ at JLab+EIC to BSM physics will be computed.\\

\noindent \emph{\underline{The global landscape}}: Precision experimental determinations of the QCD strong coupling constant are an active research area at CERN and will also be a focus at the EIC (U.S.)~\cite{AbdulKhalek:2021gbh} and EicC (China)~\cite{Anderle:2021wcy}. The JLab 22-GeV upgrade will have a unique capability to reconstruct this fundamental quantity, which is connected to a broad range of physics applications beyond hadronic physics, including cosmology and astrophysics. The community aims to reach a goal of 0.1\% accuracy in the next decades~\cite{dEnterria:2022hzv}, which is only possible through a combination of independent experimental extractions. Therefore, the envisioned energy upgrade can contribute crucially to this goal, thanks to its high precision and distinct experimental methods compared to LHC data.

\subsection{Meson Structure with TDIS}

\begin{figure}[h]
\centering
\includegraphics[width=0.35\linewidth]{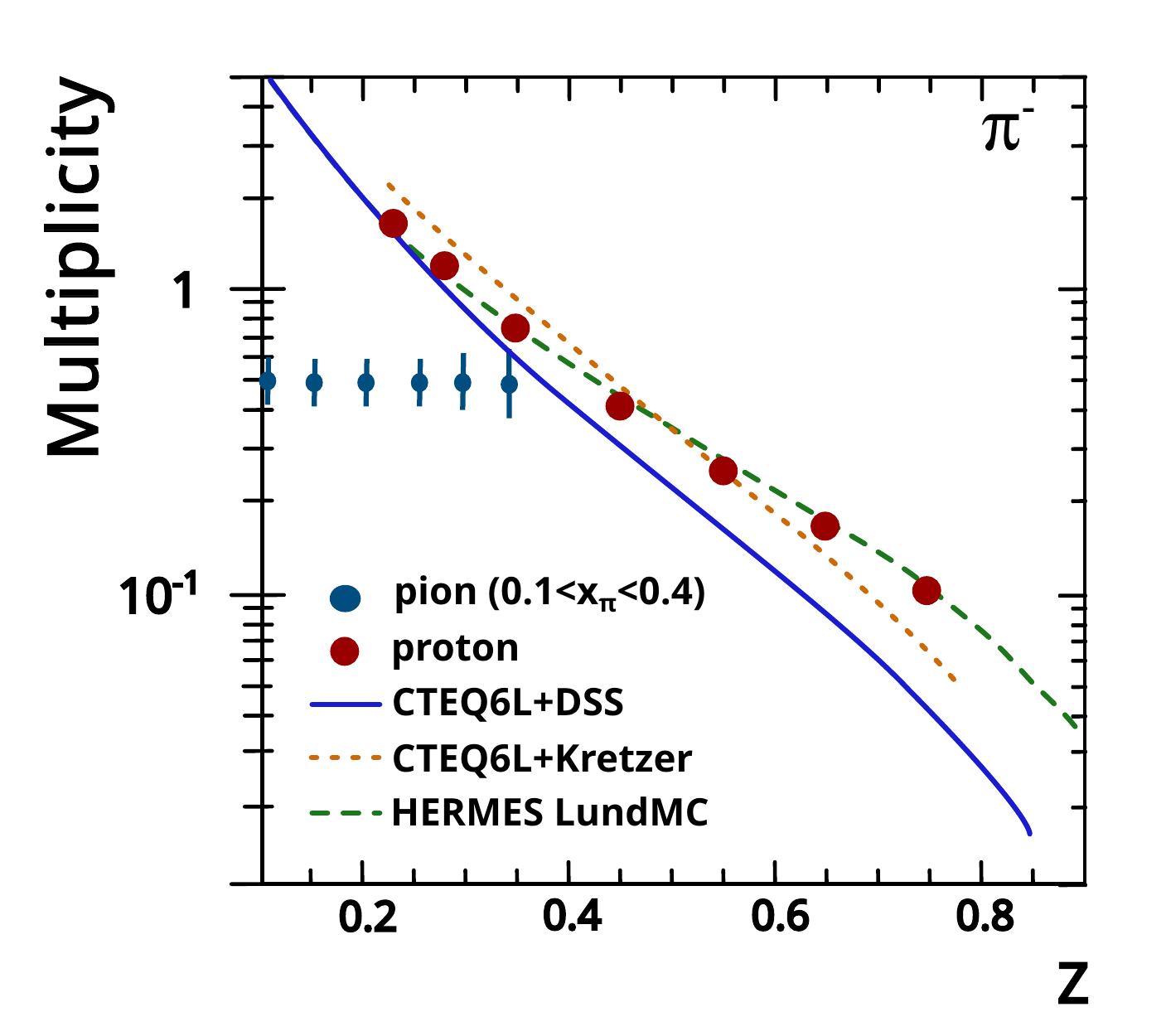}
\includegraphics[width=0.5\linewidth]{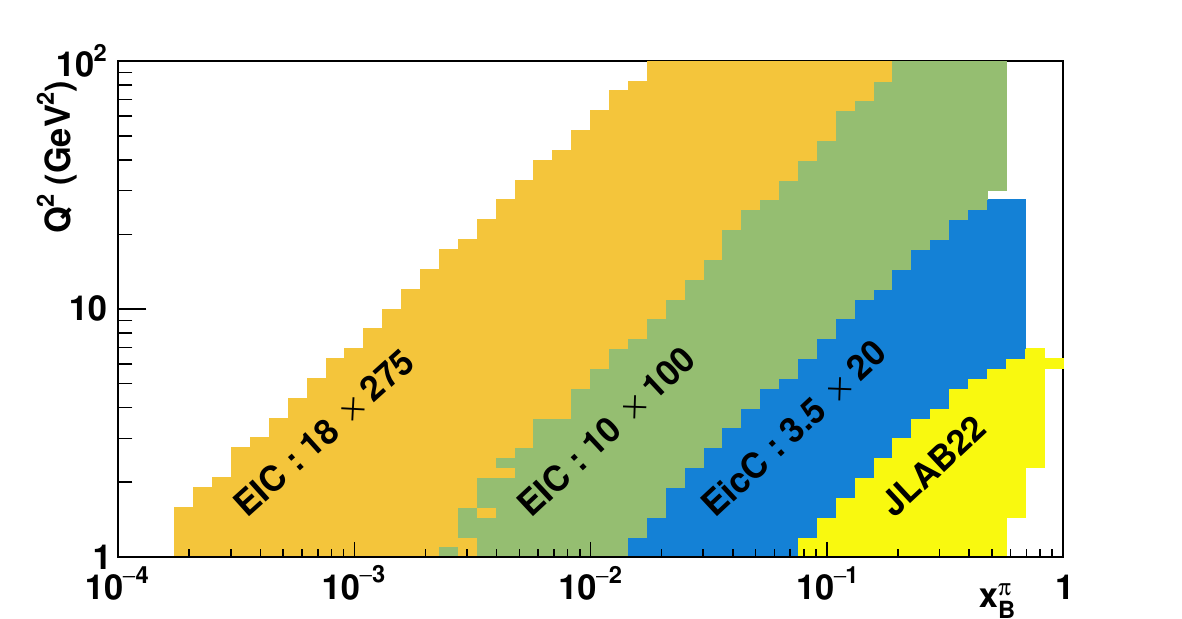}
\caption{New TDIS opportunities at 22~GeV. (Left) Projected pion SIDIS multiplicities based on 50 days of beam time. (HERMES results~\cite{PhysRevD.87.074029} to demonstrate existing data from SIDIS by the proton only, not for comparison). (Right) Pion DVCS phase space, complementary to future collider data.}
\label{fig:solid-tdis}
\end{figure}

\noindent\emph{\underline{Recent developments since the White Paper}}: 
Tagged Deep Inelastic Scattering (TDIS) will be pivotal for shedding light on the internal structure and quark-gluon dynamics of light mesons via the Sullivan process, deepening our understanding of the emergent phenomena of QCD. This is being studied by the TDIS Collaboration and the Meson Structure Function Working Group. The science case in the White Paper~\cite{Accardi:2023chb} remains the same. A 22-GeV program expands the available phase space for meson PDFs, building on the 11~GeV proof-of-principle experiment. It also unlocks a novel opportunity to access light meson TMDs via SIDIS on virtual mesons. Another opportunity, not included in Ref.~\cite{Accardi:2023chb} but reported at the Frascati 2024 workshop, is pion DVCS via the Sullivan Process~\cite{PhysRevLett.128.202501}, where 22~GeV extends the $\pi N$ invariant mass range to yield more statistics above the nucleon resonance region. Both SIDIS (see Fig.~\ref{fig:solid-tdis} (left)) and DVCS (see Fig.~\ref{fig:solid-tdis} (right)) are only available with TDIS at 22~GeV, not 11~GeV, revealing an exciting opportunity for a multi-dimensional meson imaging program, covering PDF, TMD, and GPD studies.\\

\noindent\emph{\underline{Future plans}}:
Theoretical developments concerning light meson structure, including PDF, TMD, and GPD extractions and modeling, are very active. The TDIS Collaboration and Meson Structure Function Working Group include experimentalists and theorists. We expect modeling of 22\,GeV TDIS topics is possible. The next steps, therefore, will focus primarily on simulations that further demonstrate the experimental feasibility for extraction of the observables. The existing set-up could be installed in either Halls A or C, and is sufficient for meson PDF studies. However, for SIDIS and DVCS, additional final-state particles must be measured, necessitating detector additions. Our proposal for final-state pions is to surround the existing TDIS recoil detector with a barrel of High Voltage Monolithic Active Pixel Sensors (HVMAPS)~\cite{Rudzki:2021smh, NILOY2024169874}, but the design is in progress and has to be studied. The detection of final state DVCS photons needs consideration.\\

\noindent\emph{\underline{The global landscape}}:
JLab 22 GeV TDIS data, combined with existing Drell-Yan data from CERN~\cite{CERNDY} and FNAL~\cite{FermilabDY}, and future 11~GeV JLab results and AMBER measurements at CERN, will add to an extremely sparse world dataset in the valence region for PDF studies, especially for the kaon. TDIS is unique, as the only direct probe of the meson cloud of the nucleon in this region, offering tests of universality. Complementary low-$x$ data exists from HERA~\cite{H1tdis, Zeustdis}, and will be collected from the EIC~\cite{EICtdis} and potentially a future Chinese Electron-Ion Collider. JLab is unique in its luminosity capabilities in the valence region, providing improved uncertainties. SIDIS/DVCS opportunities with TDIS at JLab 22 GeV are novel and further complementary to collider opportunities (see Fig.~\ref{fig:solid-tdis} (right)). 

\subsection{Pion Electroproduction at Very High Transverse Momenta}

\begin{figure*}[htb]
\begin{center}
\includegraphics[width=0.85\textwidth]{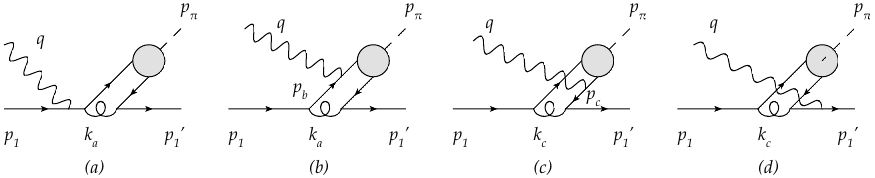}
\caption{Quark level diagrams for isolated pion production.} 
\label{fig:isodiagrams}
\end{center}
\end{figure*}

\noindent \emph{\underline{Recent developments since the White Paper}}: The White Paper~\cite{Accardi:2023chb}, in its coverage of SIDIS, concentrates on semi-inclusive meson electroproduction produced by hadronization, which is the production of observed hadrons proceeding from a struck quark fragmenting into a spray or jet of hadrons. In addition, we argue, when observing the highest transverse momentum mesons, they are produced by the perturbatively calculable higher-twist process of direct or isolated meson production. The idea was put forth a long time ago by Baier and Grozin~\cite{Baier:1980yk}, and is relevant today for its potential dominance of meson production in suitable kinematic regions, and for its potential to elucidate the distribution amplitude of the meson and the PDFs of the target at high momentum fraction $x$. Specializing, direct pion production~\cite{Carlson:1993ys,Afanasev:1999xk,Afanasev:2003ne,myfix,AC2025} is shown in Fig.~\ref{fig:isodiagrams}. In these diagrams, the pion is produced from an existing quark in the target plus an antiquark produced from an emitted gluon. The pion is produced in kinematic isolation, with no accompanying jet of other hadrons. If the transverse momentum of the pion is large, production via fragmentation is suppressed by the fall-off of the fragmentation function, and direct production becomes the main way a pion can be produced. The cross section for the overall electroproduction process can be given as a product of a virtual photon flux factor times a cross section for $\gamma^* p \to \pi + X$, and cross sections for transversely polarized virtual photons striking an unpolarized proton are shown, with cross sections from fragmentation and from a vector meson dominance (VMD) process  included for comparison.\\

\noindent \emph{\underline{Future plans}}: Data on unpolarized nucleons can verify the reaction mechanism, bridge the gap between deep inelastic and deep exclusive reactions, and study target and meson structure. In addition, the White Paper provides details about measurements with polarized targets, and we can extend the direct pion calculations to this case also. Still further, calculations for direct $\rho$ meson production can be done, and will be important for di-hadron SIDIS in certain kinematic regions.

\subsection{Kaon Semi-Inclusive Deep Inelastic Scattering}

\begin{figure}[h]
\begin{center}
\includegraphics[width=\textwidth]{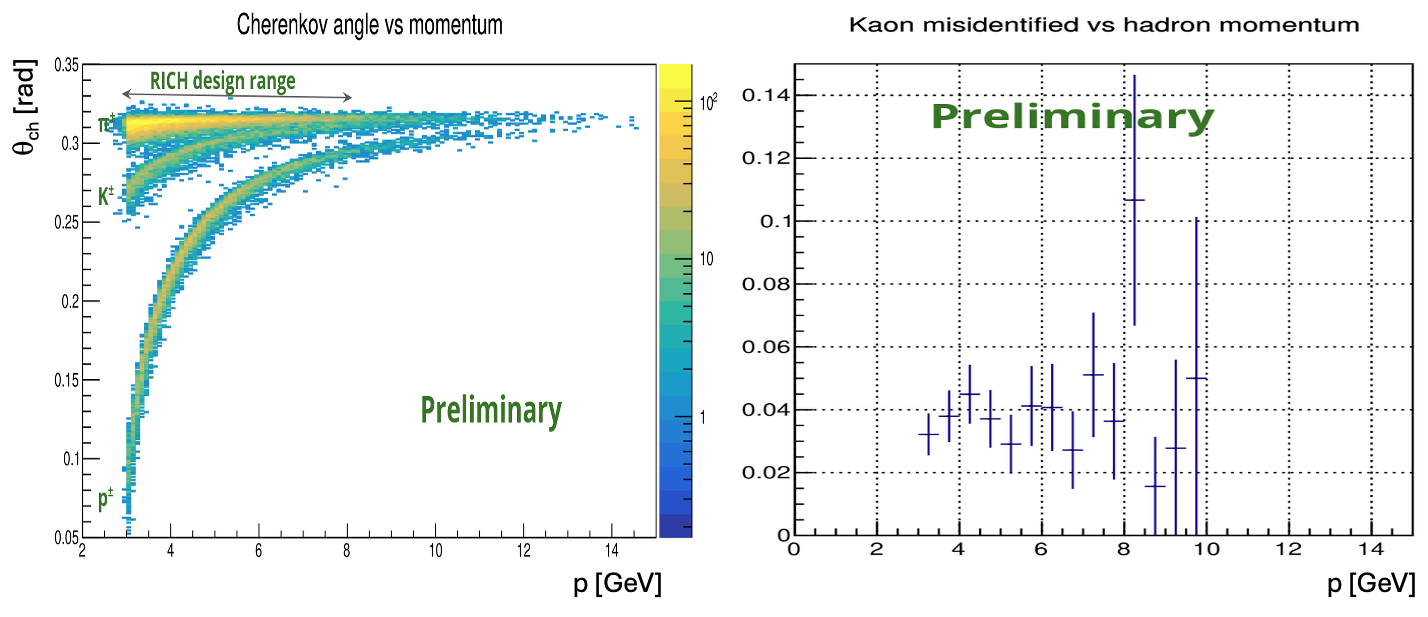}
\caption{(Left) Average Cherenkov angle vs.\ hadron momentum. (Right) Pion contamination in the kaon sample as a function of hadron momentum. No fiducial or PID cuts applied.}
\label{fig_vallarino}
\end{center}
\end{figure}

\noindent \emph{\underline{Recent developments since the White Paper}}: A significant portion of the White Paper~\cite{Accardi:2023chb} emphasized the importance of precise, multi-dimensional SIDIS measurements to access TMD PDFs and Fragmentation Functions (FFs). Since its publication, new simulation studies have been carried out to assess the performance of the CLAS12 detector in the extended phase space that will become accessible with the proposed 22 GeV energy upgrade. In particular, the focus of this study has been on evaluating the performance of the Ring Imaging Cherenkov (RICH) detector for high-momentum $\pi^\pm/K^\pm$ separation. Two RICH modules were integrated into CLAS12 in 2018 and 2022. The system is designed to achieve a pion rejection factor of 1:500 in the kaon sample within the 3-8 GeV momentum range, corresponding to a $4\sigma$ separation between $\pi^\pm$ and $K^\pm$ signals~\cite{CLAS12RingImagingCherenkov}. Figure~\ref{fig_vallarino} (left) shows the simulated distribution of the Cherenkov angle as a function of the hadron momentum. The pion misidentification rate in the kaon sample is shown in Fig.~\ref{fig_vallarino} (right). While these results are obtained without applying any fiducial cuts, they already indicate that pion contamination remains under control up to 10 GeV. The development and optimization of fiducial cuts, particularly including constraints on the hadron track that has a significant effect on the RICH ray-tracing algorithm, will be crucial to maintain identification performance while maximizing statistical yield.\\

\noindent \emph{\underline{Future plans}}:  The next steps involve consolidating simulation studies with larger statistics and more realistic detector conditions. Ongoing work aims to refine the evaluation of kaon identification efficiency, pion misidentification rates, and resolution effects across the full momentum range expected at 22 GeV. In particular, it will be important to assess the impact of the energy upgrade on key observables such as the beam Single Spin Asymmetry (BSA) in SIDIS for charged kaon electroproduction. This kind of measurement at 12~GeV is already available~\cite{CLAS:2025asy}, and will serve as a valuable benchmark when data at 22 GeV become available. While the RICH detector will be significantly affected by the increased hadron momentum range, similar dedicated studies should also be performed for other CLAS12 subsystems to ensure optimal performance and robust event reconstruction in the upgraded kinematic regime.\\

\noindent \emph{\underline{The global landscape}}: 
The performance of CLAS12 with the 22 GeV beam, especially if combined with a future luminosity upgrade, will need to be carefully validated to ensure reliable measurements across the full accessible phase space. The extended kinematics will enable high-precision studies in regions that are only marginally populated at 12 GeV, particularly in the high-$x_B$ and high-$Q^2$ domains. With its unique combination of high luminosity, wide acceptance, and powerful particle identification capabilities, CLAS12 at 22 GeV will continue to play a leading role in global efforts to map the three-dimensional structure of the nucleon using kaon SIDIS. This program will provide crucial input complementary to what is expected from COMPASS, Belle-II, and the EIC. In conclusion, the 22 GeV upgrade represents a major opportunity to expand the physics reach of kaon SIDIS studies at JLab. The ongoing simulation work is essential to fully exploit the detector capabilities in the new energy regime, and will be key in demonstrating the feasibility and scientific impact of precision TMD measurements with strange hadrons.

\subsection{The Sivers Effect}


\noindent \emph{\underline{Recent developments since the White Paper}}: SIDIS is a powerful tool for accessing TMDs. Precise extraction of flavor-dependent TMDs requires large luminosities, broad $Q^2$ and $x$ coverage, polarization of both protons and neutrons in the transverse and longitudinal directions, and various hadron production processes sensitive to different quark flavors. However, existing SIDIS data, from experiments like HERMES, COMPASS, and JLab, have large uncertainties. Global analyses, combining ($e^{+}$,$e^{-}$) and DY data, have provided limited information on Collins and Sivers TMDs. Current and upcoming JLab experiments (e.g., Hall~A SBS, Hall~B CLAS12, and Hall~C) are expected to provide high-quality SIDIS data but face limitations in kinematic coverage, luminosity, and the ability to acquire data from both polarized proton and neutron targets.

The SoLID experiment that has been proposed to deliver both large luminosities ($10^{37} - 10^{39}$~cm$^{-2}$s$^{-1}$) and 4$\pi$ acceptance~\cite{JeffersonLabSoLID:2022iod}, will measure SIDIS with pion and kaon production using both transversely and longitudinally polarized proton and neutron targets. Recent studies combining theoretical models, world data, and future SoLID SIDIS pion-production pseudo-data suggest that SoLID data can improve the extraction of the $u$ and $d$ quark tensor charges, Transversity, and Sivers TMDs in the valence-quark region ($x>0.3$) by at least a factor of 10~\cite{Ye:2016prn, Liu:2017olr}.

\begin{figure}[h]
\centering
\includegraphics[width=1.0\textwidth]{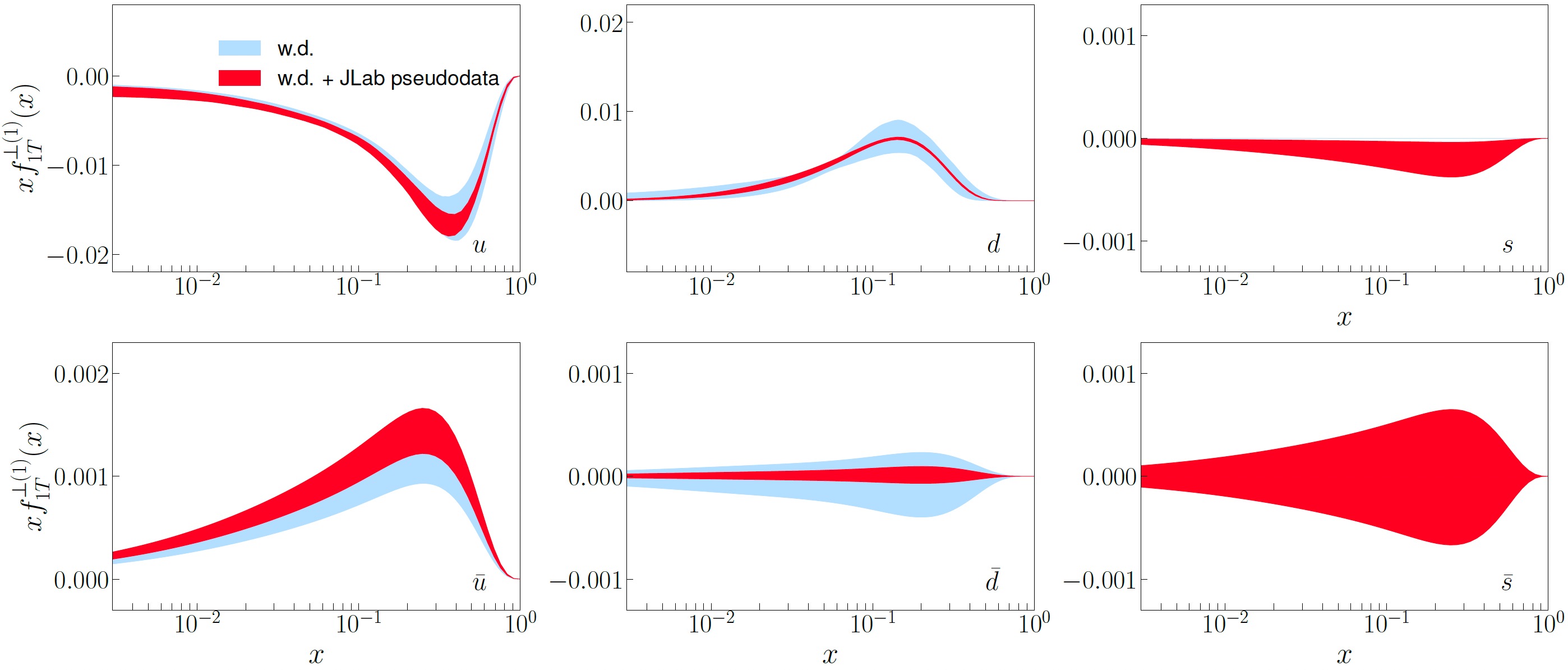}
\caption{Extraction of Transversity TMDs for light quarks using the world data and the JLab SoLID pseudo-data, including both the pion and kaon SIDIS data.}
\label{fig:solid-sidis}
\end{figure}

Since the White Paper~\cite{Accardi:2023chb}, similar studies using the 22 GeV electron beam show that kinematic coverage can extend to $Q^2<30~{\rm GeV}^2$ and $x>0.05$. Additionally, kaon-SIDIS production, previously limited by statistics with the 11~GeV electron beam, can now be more extensively measured with similar kinematic coverage as pion-production data, while maintaining low statistical uncertainties. A new impact study using the 22 GeV SoLID SIDIS simulated data was performed. The results shown in Fig.~\ref{fig:solid-sidis} suggest that the energy upgrade will significantly improve the extraction of both valence and sea-quark Sivers functions relative to the current global knowledge.\\

\noindent \emph{\underline{Future plans}}: 1) Perform simulation of the SIDIS reaction with a 22 GeV beam with more final states including $p$, $\bar{p}$, $\Lambda^{\pm}$, and other vector mesons. 2) Study the impact of the energy upgrade on the flavor-separation of different types of TMDs beyond the Sivers function.\\

\noindent\emph{\underline{The global landscape}}:
Upcoming SIDIS data from CLAS12, Hall~A SBS, and Hall~C with 11 GeV electron beams, along with the future SoLID SIDIS program, will provide the most precise and comprehensive data on hadron structure using proton and neutron targets with transverse and longitudinal polarization, constituting a new era of exploration in hadronic physics. These efforts will explore TMDs in the valence quark region and at relatively low $Q^2$. The energy upgrade will not only be more effective in studying TMDs and FFs in the current fragmentation region, but it will also allow for the constraint of the sea sector of TMDs.

\subsection{Back-to-Back Hadron Production and ``$\rho$-Free'' SIDIS }


\noindent  \emph{\underline{Recent developments since the White Paper}}: Measurements of the SIDIS cross sections for various hadron production processes and different configurations for initial lepton and target nucleon polarizations provide essential information about the underlying quark distributions and their interactions within the nucleon. Measurements of multiplicities and asymmetries of multi-particle final states in wide kinematics, including di-hadrons and vector mesons, will be crucial for a separation of different structure functions, and different dynamical contributions to specific structure functions,  used in phenomenological studies of SIDIS.

\begin{figure}[h]
\begin{center}
\includegraphics[width=0.4\textwidth]{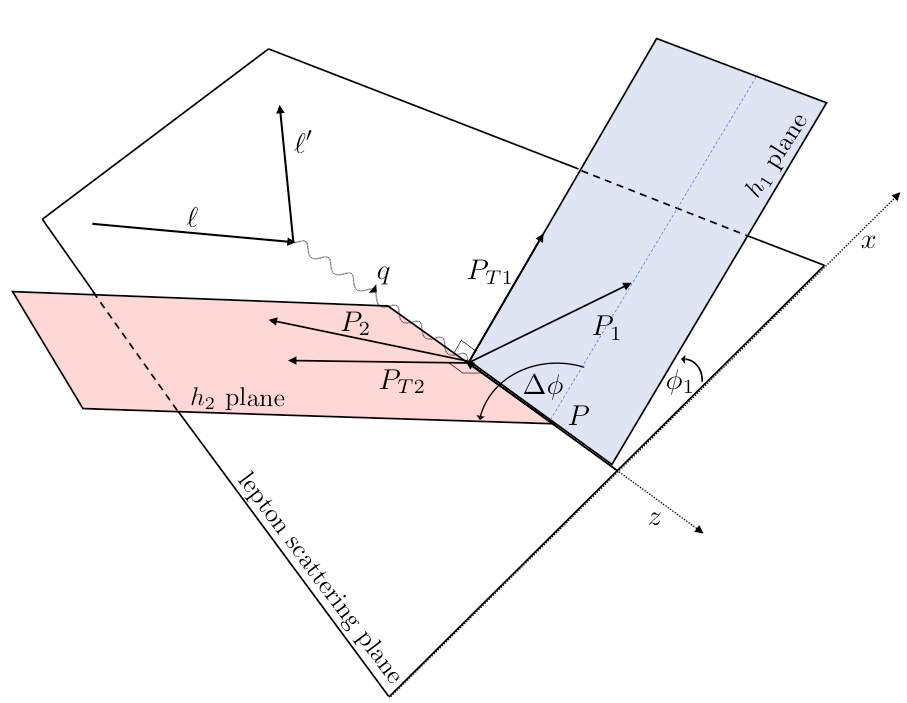}
\includegraphics[height=0.36\textwidth,width=0.48\textwidth]{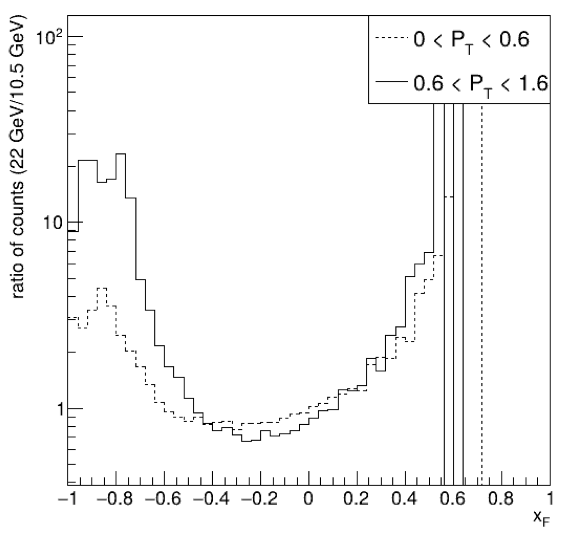}
\caption{The kinematics of back-to-back production of hadrons (left) and the ratio of distributions of TFR protons collected for the same time interval by CLAS12 for 22 GeV and 10.6 GeV electron beam energies.}
\label{b2b-kin}
\end{center}
\end{figure}

Recently a new class of measurements has been introduced, involving detection of two hadrons,  produced in opposite hemispheres along the $z$-axis in the center-of-mass frame, with the first hadron produced in the current-fragmentation region (CFR) and the second in the target-fragmentation region (TFR). Possible interpretations of back-to-back hadron production involve fracture function formalism~\cite{Anselmino:2011ss}, and most recently the ``semi-exclusive'' formalism, assuming the TFR baryon is exclusive~\cite{Guo:2023uis}. Since most of the TFR protons are indeed exclusive in the TFR, this formalism looks very attractive for accessing GPDs/GTMDs on the target side and TMD FFs on the CFR side. Measurements of correlations between hadrons in the TFR and CFR~\cite{CLAS:2022sqt}, thus, open new possibilities to quantify the relationship between the spin and transverse momenta of quarks in the nucleon and provide a new avenue for studies of the complex nucleonic structure in terms of quark and gluon degrees of freedom. Since both interpretations require separation of TFR and CFR processes, increasing the beam energy will significantly improve the separation. In addition, higher energies open the phase space for higher transverse momenta of hadrons. Typically the SSAs observed in electroproduction are proportional to the product of the transverse momenta of the final state hadrons. That makes the improvement with an upgrade of JLab to 22 GeV in back-to-back hadron production with the first hadron produced in the CFR and the second in the TFR very significant. In particular, Fig.~\ref{b2b-kin}  shows that enhancement of events in the most critical part of the kinematics, namely large transverse momenta and large $x_F$, can be a factor of 10 for the same run time. 

Studies of partonic and hadronic correlations  are becoming increasingly important in the interpretation of electroproduction data, in general, and the hadronization process of quarks, in particular. Currently there are major challenges in the theoretical description of large transverse momenta of hadrons in polarized SIDIS~\cite{Gonzalez-Hernandez:2018ipj} and it is very important to understand the impact of possible assumptions used in phenomenology. More significantly than originally anticipated, the fraction of pions coming from correlated di-hadrons indicated by recent measurements at JLab and supported by various realistic models describing the hadronization process, may have a significant impact on various aspects of data analysis, including the modeling, composition, and interpretation of SIDIS data, as well as calculations of radiative corrections (RC). Recent studies of exclusive vector mesons performed at JLab and COMPASS indicate understanding of the impact of exclusive vector mesons is absolutely critical for interpretation of SIDIS, and will likely be even more relevant for measurements performed or planned at higher energies. 

\begin{figure}[h]
\begin{center}
\includegraphics[width=0.95\textwidth]{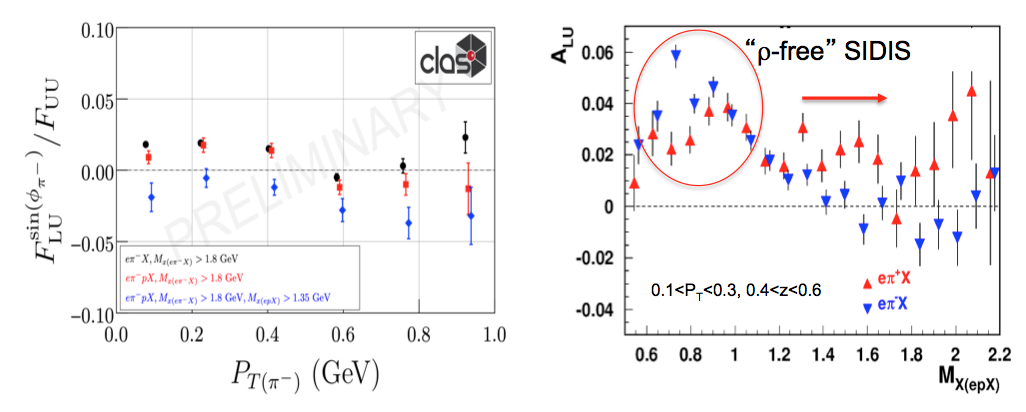}
\caption{(Left) Illustrates the transverse momentum dependence of the beam SSA for inclusive $\pi^-$ for 3 cases: 1) canonic SIDIS sample with detection of $e'\pi^- X$ (black), 2) the same sample but with proton detected in the target fragment, $e'p\pi^- X$ (red), 3) the same as 2), but with additional cut on the missing mass of the proton to eliminate exclusive $\rho$s from the inclusive $\pi^-$ sample (blue). (Right) Actual beam SSAs of the samples of semi-inclusive $\pi^+$ (red) and $\pi^-$ (blue) with detection of an additional proton in the TFR, versus the missing mass of the $e'pX$, clearly showing the region of low masses, where the $\rho^0$ dominates, showing the same, large positive asymmetry for both particles. The kinematic bin in this plot is defined by the subsample of pions in a bin $0.1<P_T<0.3$~GeV, $0.4<z<0.6$.}
\label{fig:impact}
\end{center}
\end{figure}

Exclusive $\rho$ contributions, which are not included in the current formalism of 3D studies~\cite{Boussarie:2023izj}, can indeed dramatically change single- and, most importantly, even double-spin asymmetries for charged pions measured by the HERMES~\cite{HERMES:2003jyc} and COMPASS Collaborations~\cite{COMPASS:2022xig,COMPASS:2019lcm}. For proper comparison with the theory, the data should be cleaned up from exclusive vector mesons, in particular, from exclusive $\rho^0$ contributions. An example of the impact of exclusive $\rho^0$ contributions on semi-inclusive $\pi^-$ production is shown in Fig.~\ref{fig:impact}. The effect is significant and can even reverse the sign of the single-spin asymmetries (SSAs) at relatively low transverse hadron momenta, where pions from $\rho$ decay dominate. Similar effects are expected for other polarized target observables, especially for Sivers-type asymmetries.

\begin{figure}[h]
\begin{center}
\includegraphics[width=0.9\textwidth]{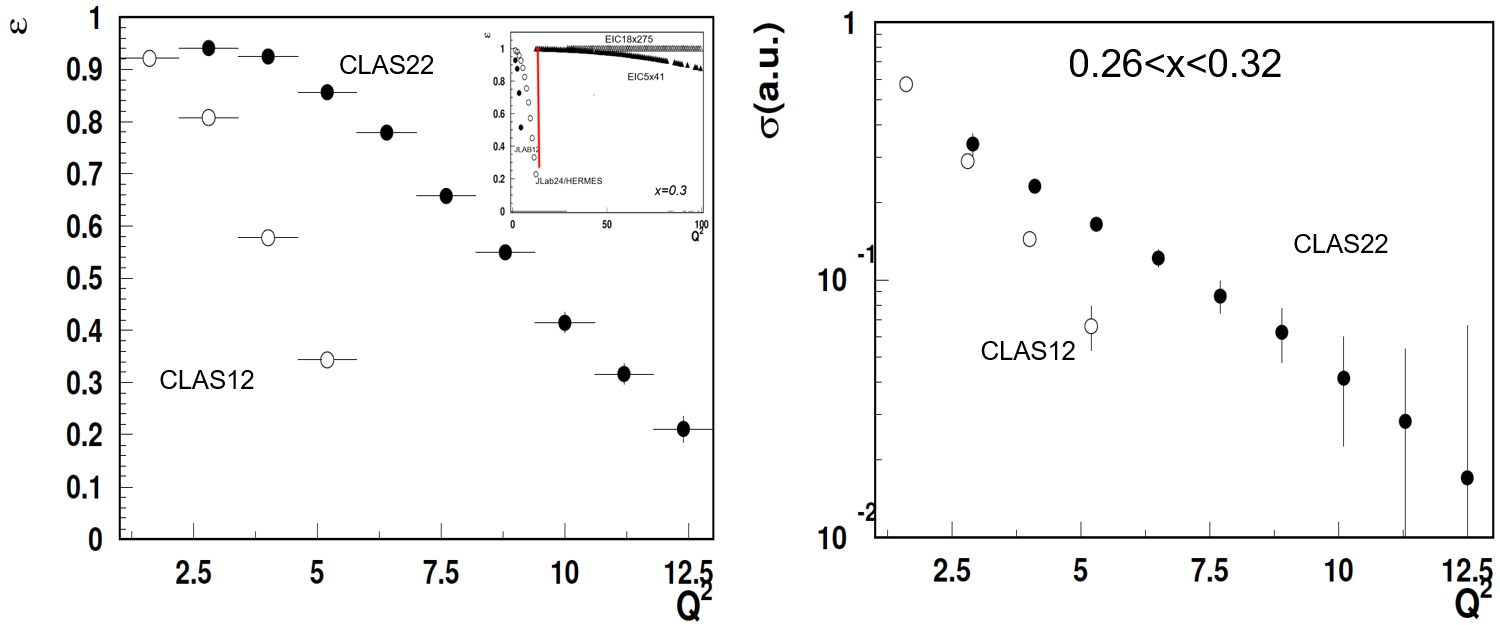}
\caption{Comparison of JLab at 10.6 GeV and 22 GeV for the $Q^2$-dependence of the virtual photon polarization $\epsilon$ (left) and the total counts (right).}
\label{rho-q2}
\end{center}
\end{figure}

A JLab energy upgrade to 22 GeV will significantly increase the phase space, in particular, providing access to a much wider range in $Q^2$. Figure~\ref{rho-q2} shows the major improvement in the kinematic range and the overall counts of exclusive $\rho$s with an upgraded JLab. The extended range would allow also the gap between JLab and the lowest energy of EIC to be spanned, providing an important bridge to link the measurements. That will provide important information, critical for interpretation of EIC data, as at higher energies the extraction of exclusive $\rho$ and the separation from semi-inclusive background, as well as separation of contributions from longitudinal and transverse photons, will be extremely challenging. The detection of the target fragment proton allows, in addition, the introduction of a new analysis framework, called ``$\rho$-free'' SIDIS. It has been demonstrated, that detection of the TFR proton in coincidence with the CFR hadron indeed could be ``$\rho$-free'', as a simple cut on the missing mass of the proton can easily eliminate events with exclusive vector meson production from inclusive pion SIDIS (see Fig.~\ref{fig:impact}). Detailed cross-checks with more traditional ``$\rho$-subtracted'' SIDIS measurements will also provide a mechanism for the validation of extraction frameworks, which is important for proper evaluation of systematic uncertainties and critical for extension of phenomenology to higher transverse momenta of hadrons, providing direct access to transverse momenta of partons. \\
 
\noindent \emph{\underline{Future plans}}: SIDIS measurements can be divided into 3 classes:

\begin{enumerate}
    \item Standard SIDIS ($eN \to e'hX$, $h=\pi, K, ...$) within the full accessible kinematics, corrected for acceptance and RC, measured in the multidimensional space.
    \item Standard SIDIS ($eN \to e'\pi X$) within the full accessible kinematics, corrected for acceptance and RC, measured in the multi-dimensional space, with subtracted in multi-dimensional bins for exclusive $\rho^0$ contributions (``$\rho$-subtracted'' SIDIS).
    \item SIDIS subsamples ($eN \to e'p \pi X$, $eN \to e'\pi \pi X$) within the  fully accessible kinematics, allowing clear elimination of the exclusive  $\rho^0$ contributions using cuts on the missing masses of $e'pX$ or $e'\pi\pi X$. 
\end{enumerate}
Measurements indicate that the bias introduced in the $e'\pi X$ observables by the ``$\rho$-free'' SIDIS is smaller and makes a big difference with the original observables with no cuts on the $\rho$ contributions. Proper accounting of the contributions of the $\rho$ can dramatically change the phenomenology of the $e'\pi X$ studies. In fact, the beam SSA for the ``$\rho$-free'' SIDIS sample can have the opposite sign to the standard ``contaminated'' SIDIS (see Fig.~\ref{fig:impact}). All of these different sub-samples will have different biases with respect to the existing theory, and detailed understanding of their differences will help in planning for future measurements. In particular, the kinematics where they agree with each other can be considered properly cleaned up from diffractive $\rho$ contributions and adequate for application of the standard TMD theory.

To summarize, JLab is the only facility capable of separating different structure functions involved in polarized SIDIS in the valence region, including longitudinal photon contributions in general and diffractive vector mesons, in particular. The upgrade of JLab to 22 GeV would allow extension of the phase space accessible for TMD phenomenology to higher transverse momenta of hadrons, which is critical for interpretation of the polarized SIDIS data collected by HERMES, COMPASS, and JLab, as well as future experiments at JLab22~\cite{Accardi:2023chb} and EIC~\cite{AbdulKhalek:2021gbh}. Additionally, the detection of multi-particle final states and the study of multiplicities and asymmetries of di-hadrons and vector mesons will offer crucial insights into the source of single-spin asymmetries and the dynamics of the polarized quark hadronization process, as well as the elusive ``diffractive'' DIS content.\\

\noindent\emph{\underline{The global landscape}}: The White Paper emphasizes performing experiments over a wide range of kinematics. For the direct pion process, data at fixed $Q^2$ and varying $p_{\pi T}$ will come from a range of Bjorken $x$ in the valence region and will measure the shape of the target quark PDFs in that region. Also, the distribution amplitude enters within an integral that is $Q^2$ dependent, so that data selection at a variety of $Q^2$ at fixed $t$ (the invariant momentum transfer squared from the photon to the pion) can pin down the distribution amplitude of the pion or that of other mesons.

\subsection{Extracting Flavor Dependence of TMDs}


\noindent \emph{\underline{Recent developments since the White Paper}}: The White Paper~\cite{Accardi:2023chb} emphasized the importance of multi-dimensional SIDIS measurements for accessing the unpolarized quark TMD PDFs, $f_1(x, \boldsymbol{k}_{\perp})$. A relevant impact study using projections from the 22 GeV JLab program was presented in that document, based on the MAPTMD22 extraction \cite{Bacchetta:2022awv}. The study showed a clear reduction in uncertainty bands; however, although already significant at the time, it was performed without flavor separation and did not include PDF uncertainties, as these had not yet been incorporated into TMD fits. In a subsequent and more comprehensive analysis by the MAP Collaboration, named MAPTMD24 \cite{Bacchetta:2024qre}, we introduced both flavor dependence and the propagation of uncertainties from collinear PDFs. This extraction was carried out at the highest available perturbative accuracy, namely $\text{N}^3\text{LL}$, and included all relevant experimental data. For SIDIS observables in particular, we analyzed measurements from the HERMES and COMPASS Collaborations. The error analysis was performed using the same Monte Carlo method and $\chi^2$ definition as in MAPTMD22. Relying on the NNPDF3.1 set (NNPDF31\_nnlo\_pch\_as\_0118) \cite{NNPDF:2017mvq} and the MAPFF1.0 NNLO FFs \cite{AbdulKhalek:2022laj}, we generated 100 replicas of the experimental data and associated each replica of the TMDs accordingly. The fit quality is excellent, with $\chi^2/N_{\text{data}} = 1.08$ for the central (unfluctuated) replica. A key novelty of this analysis is the clear distinction observed in the $\boldsymbol{k}_\perp$-dependence of the TMD PDFs across different quark flavors, as well as in the TMD FFs across different fragmentation channels. Another important finding is that SIDIS data with identified hadrons in the final state exhibit the highest sensitivity for disentangling the flavor-dependent behavior of TMD distributions.\\

\begin{figure}[h]
\centering
\includegraphics[width=0.8\linewidth]{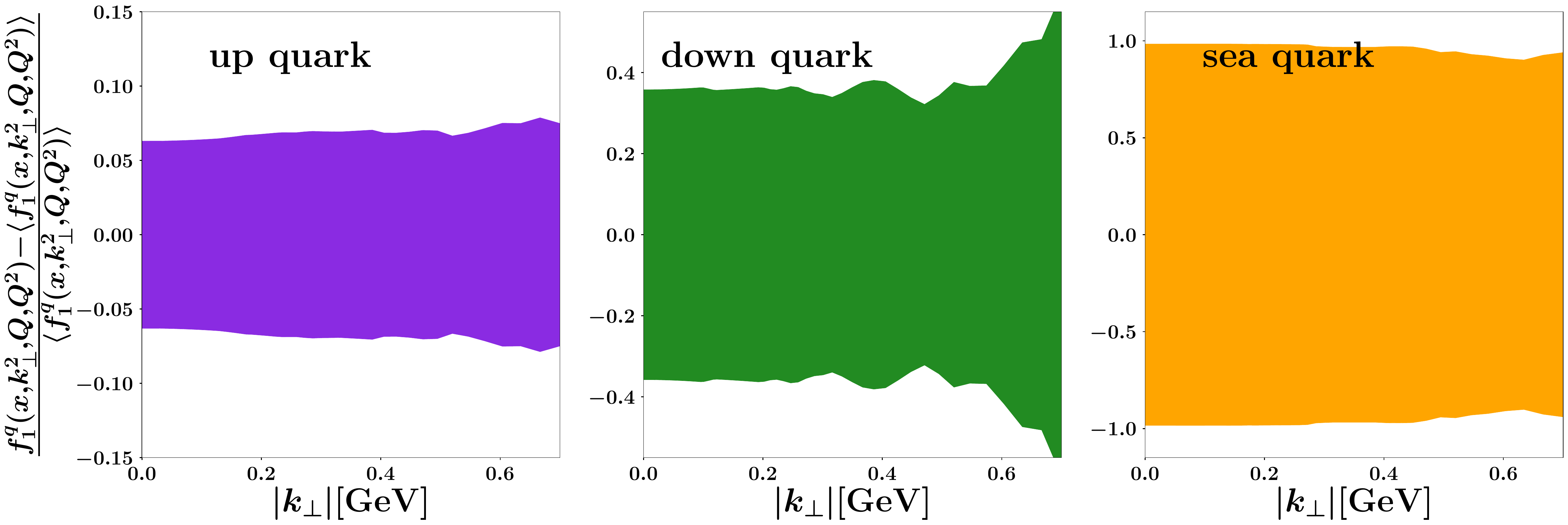}
\caption{Relative error bands for unpolarized TMD PDFs of up (left), down (middle), and sea (right) quarks at $Q = 5$~GeV and $x= 0.5$. The bands show the current 68\% C.L. uncertainties from MAPTMD24.}
\label{f:MAP24_unc}
\end{figure}

\noindent \emph{\underline{Future plans}}: As already mentioned, future measurements of the SIDIS multiplicities, $\mathcal{M}(x, z, |\bm{P}_{hT}|, Q)$, or more directly of structure functions $F_{UU,T}(x, z, |\boldsymbol{k}_\perp|, Q)$, with identified hadrons in the final state, are expected to have a major impact on the flavor separation of both TMD PDFs and TMD FFs. At present, this separation is limited by the scarcity of sufficiently sensitive data. In addition, the upcoming measurements at CLAS12 will serve as a fundamental probe of the region of large values of $x$ in TMDs, which is currently unexplored. This is illustrated in Fig.~\ref{f:MAP24_unc}, where the 68\% relative uncertainties of the TMD PDFs for up, down, and sea quarks are shown at $x = 0.5$ and $Q = 5$ GeV. Apart from the up quark, the other two flavors are still poorly constrained in this part of the kinematic domain. This region overlaps the one that will be accessed by the 22 GeV JLab program. Furthermore, building on the work presented in Ref.~\cite{Bacchetta:2025ara}, where we fitted DY data using neural networks, we have planned to extend this innovative use of artificial intelligence techniques to SIDIS data as well. One of the goals is to investigate the behavior in the intermediate region of the transverse momentum spectrum, where theoretical considerations suggest the presence of a transition between the $W$ and $Y$ terms.\\

\noindent \emph{\underline{The global landscape}}: Although the unpolarized TMD PDFs have now reached a high level of perturbative accuracy, a detailed understanding of $f_1(x, \boldsymbol{k}_\perp)$ across different quark flavors remains essential for constructing a complete picture of the three-dimensional momentum structure of the nucleon. The flavor dependence of the intrinsic transverse-momentum distribution provides unique insight into non-perturbative QCD dynamics, potentially revealing differences in the confinement mechanisms acting on various quark flavors. Disentangling such effects requires high-precision and flavor-sensitive data, which are still scarce, particularly in the regions of large $x$ and low $Q^2$. In this context, the future experimental program at JLab is expected to play a central role. Thanks to its ability to perform multi-differential SIDIS measurements with hadron identification in the final state, the proposed 22-GeV upgrade will access the relevant kinematic regions with the luminosity needed to significantly constrain the flavor structure of $f_1(x, \boldsymbol{k}_\perp)$ with unprecedented precision. These data will make it possible to study in detail the transverse-momentum distributions of up, down, and sea quarks individually, and to determine whether different flavors exhibit systematically broader or narrower $\boldsymbol{k}_\perp$-profiles.

\subsection{Probing the Transverse Momentum of Longitudinally Polarized Quarks}


\noindent \emph{\underline{Recent developments since the White Paper}}: The White Paper~\cite{Accardi:2023chb} emphasized the importance of multi-dimensional SIDIS measurements for accessing helicity-dependent TMD PDFs $g_1(x,\boldsymbol{k}_\perp)$. Projections in the kinematics of the 22 GeV JLab program using the CLAS12 detector were presented to show the accessible range in transverse momentum, which is expected to reduce vector meson background contributions and provide high-precision measurements. Since then, the MAP Collaboration has performed the first extraction of the helicity TMDPDF from a fit of SIDIS longitudinal-spin asymmetry $A_1$~\cite{Bacchetta:2024yzl}. This analysis, called MAP22pol, was performed at the highest current perturbative accuracy, i.e. next-to-next-to-leading logarithmic (NNLL) accuracy. Importantly, this can be seen as the most consistent fit of the helicity TMD because it is obtained using the same computational framework as for the extraction of the unpolarized TMDs. In fact, the TMD-factorized expression of the SIDIS asymmetries requires the knowledge of unpolarized TMD PDFs in the denominator and TMD FFs in both numerator and denominator. To maximize the consistency of the extraction, we used the MAPTMD22~\cite{Bacchetta:2022awv} unpolarized TMDs at the same accuracy. In this analysis, we included measurements of the $A_1$ asymmetry from HERMES~\cite{HERMES:2018awh} and we imposed the same kinematic cuts as the MAPTMD22 fit. We excluded COMPASS and CLAS6 data because they are not compatible with such cuts. The main feature of this work is the model adopted for the nonperturbative part $g_{1 {\rm NP}}$ of the helicity TMD. Specifically, our parameterization consists in the product of the nonperturbative model for the unpolarized TMD PDF $f_{\rm NP}$ and a Gaussian with an $x$-dependent width. We fixed the MAPTMD22 parameters for the $f_{\rm NP}$ and we fitted 3 parameters for the Gaussian function. Importantly, we imposed the positivity constraint $g_1(x,\boldsymbol{k}_\perp) \leq |f_1(x,\boldsymbol{k}_\perp)|$ at the level of the nonperturbative model through a suitable analytical method. 

\begin{figure}[h]
\centering
\includegraphics[width=0.55\linewidth]{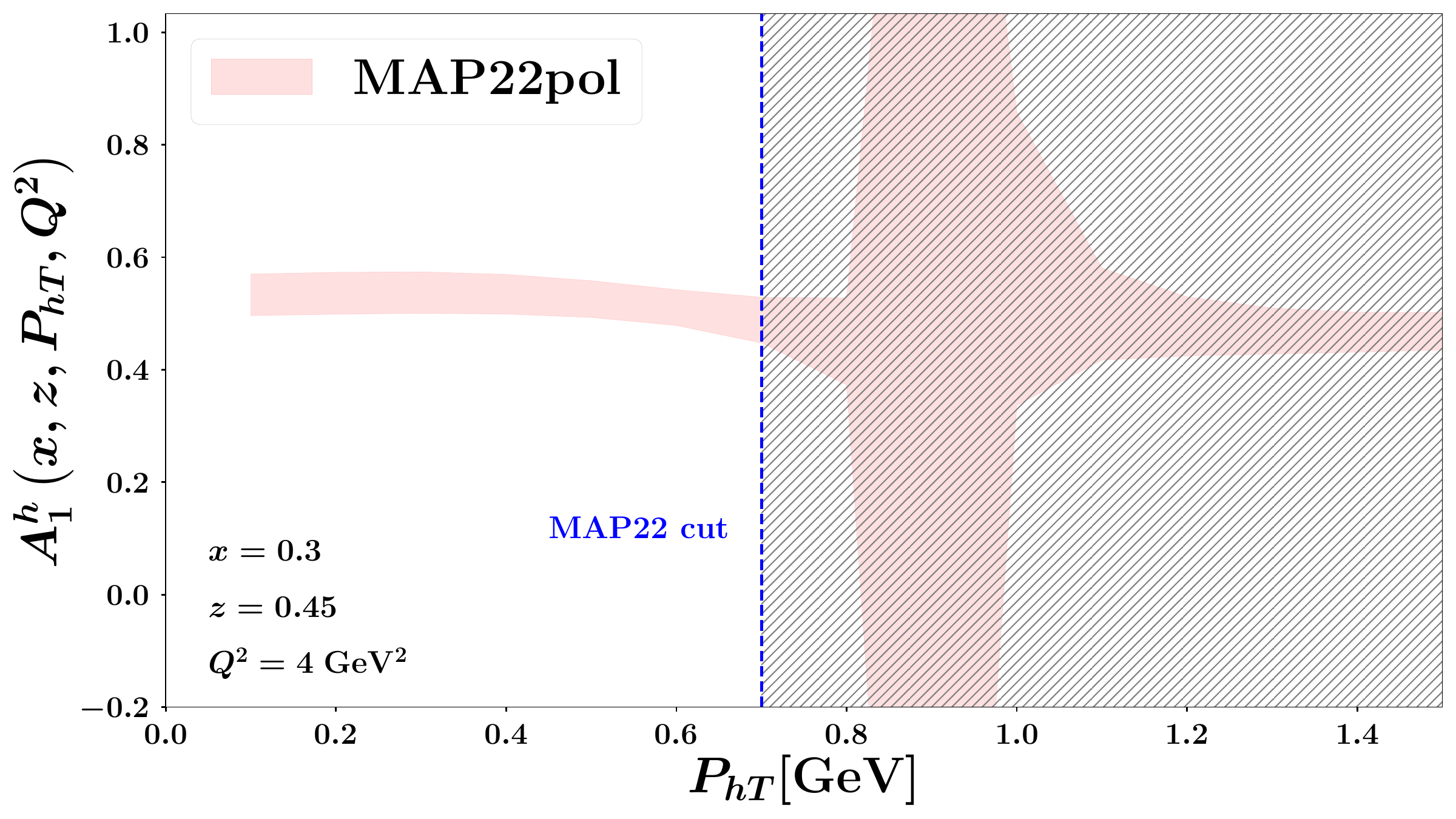}
\caption{Predictions of the $A_1$ asymmetry for charged pion production in the kinematic region corresponding to Fig.~16 of the White Paper~\cite{Accardi:2023chb}, based on the MAP22pol extraction. The dashed vertical line indicates the cut on transverse momentum applied in the fit. The pink band represents the 68\% confidence interval.}
\label{f:JL22_MAP22pol-pred}
\end{figure}

At the end, we fitted 3 parameters to 291 experimental data of the $A_1$ asymmetry. The error analysis is performed with the same Monte Carlo method and $\chi^2$ definition of the MAPTMD22 analysis. According to the NNPDFpol1.1~\cite{Nocera:2014gqa} polarized collinear PDFs set as input, we generated 100 replicas of the experimental data and we associated the $i^{th}$ replica of the helicity PDF and the corresponding extracted helicity TMD PDF to the same replica of the unpolarized TMDs in the MAPTMD22 extraction. The quality of the fit is very good, with a $\chi^2$ per data of 1.09 for the central replica (unfluctuated data). Given the small number of included data and their relatively large uncertainties, the parameters extracted in our analysis are affected by large uncertainties. The extracted $|\boldsymbol{k}_\perp|$-distribution of the ratio between helicity and unpolarized TMDs $g_1(x,\boldsymbol{k}_\perp) / f_1(x,\boldsymbol{k}_\perp)$ manifests a behavior that depends on $x$: flat at small $x$; sharper as $x$ increases. This appears to be in fair agreement with calculations of this quantity in lattice QCD. The agreement improves when comparing to new preliminary lattice results. We used the extracted helicity TMD to calculate the size and shape of the $A_1$ asymmetry in the kinematic region of the reported in Fig.~16 of the White Paper~\cite{Accardi:2023chb}. We extrapolated our results at large transverse momentum, even far beyond the kinematic cut imposed in our fit, to study the behavior of our formalism in a region where TMD factorization is expected to fail. In Fig.~\ref{f:JL22_MAP22pol-pred}, we report the result of our $A_1$ asymmetry calculation for measured pions. The pink band represents the 68\% uncertainty and the dashed vertical line the maximum value of measured transverse momenta that fulfills the kinematic cut imposed in our fit. The dashed vertical line indicates the $P_{hT}$ cutoff used in our MAP22pol fit; beyond this value, our predictions are unreliable, as a different formalism than TMD factorization is required to describe the observable. We note that the predictions manifest an almost flat behavior. This is consistent with the extracted $g_1/f_1$ TMD PDF ratio and is a consequence of the small number of datasets included in our analysis, along with the relatively large experimental uncertainties. \\

\noindent \emph{\underline{Future plans}}: Future efforts to extract the helicity TMD PDF depend strongly on the availability of new, high-precision experimental data. Once measurements from CLAS12, and potentially the SoLID detector, become available, we plan to test new models for the nonperturbative structure of the helicity TMD. In particular, we aim to implement a more flexible parameterization that allows $g_1(x, \boldsymbol{k}_\perp)$ to change sign across different regions of $\boldsymbol{k}_\perp$. Preliminary investigations of this feature have been conducted, though current HERMES data appears to disfavor such behavior. Moreover, we aim to quantify the impact of JLab22 simulated data on the extracted TMD helicity ratio in the valence region and, when perturbative calculations will be available, improving the accuracy of the extraction to NNNLL. \\

\noindent \emph{\underline{The global landscape}}: The knowledge of the helicity TMD PDF allows us to understand whether quarks with spin parallel to the proton spin tend to have smaller or larger transverse momentum than quarks with antiparallel spin. This is particularly relevant for JLab, given that one of its main goal is to provide data to improve the determinations of the spin dependence of multi-dimensional quark distributions. The availability of new precise experimental data of the $A_1$ asymmetry differential in transverse momentum is of utmost importance to study in more detail the transverse-momentum dependence of the helicity ratio. The experimental measurements at the possible 22 GeV JLab upgrade will cover the kinematic region and have the luminosity required to enormously reduce the uncertainties in the extraction of this quantity. Understanding the helicity TMD PDF sheds light on whether quarks with spin aligned to the spin of the proton tend to carry more or less transverse momentum than those with opposite spin orientation. This question is particularly relevant for JLab, where a central objective is to provide high-precision data to improve our knowledge of the spin dependence in multi-dimensional quark distributions. The availability of accurate measurements of the $A_1$ asymmetry as a function of transverse momentum is essential for investigating the detailed $|\boldsymbol{k}_\perp|$-dependence of the helicity ratio. The potential 22 GeV upgrade at JLab is expected to access the relevant kinematic region with sufficient luminosity to significantly reduce uncertainties in the extraction of this quantity. Moreover, such precision studies of spin-momentum correlations could provide sensitivity to subtle effects beyond the Standard Model~\cite{Bacchetta:2023hlw}, particularly if deviations from the expected helicity behavior emerge in future global analyses.

\subsection{Spatial Structure and Mechanical Properties}


\noindent  \emph{\underline{Recent developments since the White Paper}}: The main areas of active inquiry of greatest relevance to the 22 GeV upgrade have concentrated on the following: 1) Tomographic imaging of the nucleon with DVCS, Double DVCS, and TCS; 2) Gluonic mass and momentum distributions of the nucleon from threshold charmonium production; 3) $\pi^+$ and $K^+$ structure studies. All of these areas of focus have high scientific potential, and were discussed in the White Paper~\cite{Accardi:2023chb}. The basic scientific thrust has not changed significantly since 2022.

In some cases, there have been improved theoretical developments enabling improved QCD conclusions to be drawn from the data, but in other areas the theoretical tools are still insufficient. Examples of areas with improved theoretical support include: 1) new tools to fit GPD information from Compton scattering observables, leading to an improved understanding of the gravitational structure of the proton and 2) new GPD analyses of threshold $J/\psi$ electroproduction cross sections in terms of gluonic GPDs.

Older $J/\psi$ analyses relied on holographic analysis tools \cite{Mamo:2022eui,Hatta:2018ina}, which are appropriate in the limit of large $N_c$ and strong $\alpha_s$. New GPD analyses \cite{Guo:2023qgu} are more flexible, particularly if the terms beyond the leading moments are retained. Gluon gravitational form factors were extracted from the existing GlueX $J/\psi$ data in both approaches. The resulting form factors from both approaches were in broad agreement. Additional $J/\psi$ data with improved precision and over a broader $t$ range would clarify the limitations in either approach.

\begin{figure*}[htbp]
\begin{center}
\raisebox{12.0mm}{\includegraphics[width=0.28\columnwidth]{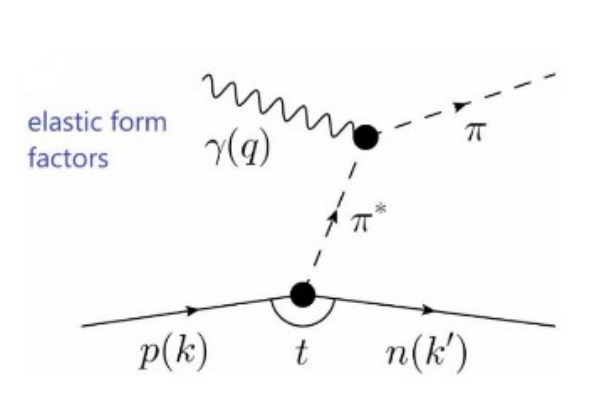}}
\raisebox{12.0mm}{\includegraphics[width=0.28\columnwidth]{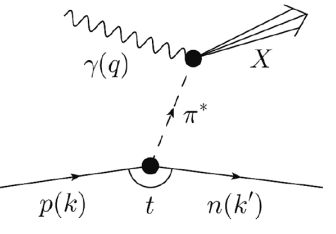}}
\includegraphics[width=0.4\columnwidth]{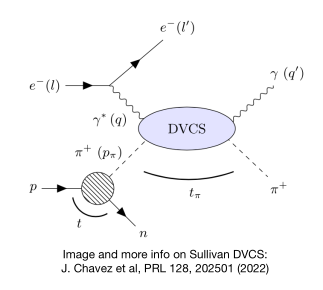}
\vspace{-3mm}
\caption{The Sullivan processes needed to acquire meson structure information  above $Q^2>0.5$~GeV$^2$ for (left) exclusive $\pi^+$ production yielding form factor information, (center) leading neutron process yielding $\pi^+$ PDF information, and (right) the $\pi^+$ DVCS process.}
\label{fig:sullivan}
\end{center}
\end{figure*}

Another area of intense theoretical activity is the modeling of meson form factors and PDFs. There has been recent recognition that as the pseudo-Goldstone bosons of QCD, the internal structure of the $\pi^+$ and $K^+$ provide a wealth of information on emergent hadron mass from QCD dynamics and the role of color confinement. While there has been much work on the modeling of meson form factors and PDFs from basic QCD principles, the less glamorous tools needed to extract such information from experiment suffer neglect. The issue is that meson structure information can only be indirectly accessed by experiment -- due to their short lifetimes, $\pi^+$ and $K^+$ targets do not exist and one must utilize the process where the virtual photon probe interacts with the virtual mesons keeping the nucleus intact (Sullivan process). In extractions of meson form factors from exclusive $p(e,e'\pi^+)n$ and $p(e,e'K^+)\Lambda$ data, it is greatly preferred to conduct a longitudinal/transverse (Rosenbluth) separation of the experimental cross section according to the polarization components of the virtual photon (see Fig.~\ref{fig:sullivan} (left)). In extractions of meson PDFs from inclusive data, the leading neutron (or $\Lambda$) electroproduction cross sections must be determined (see Fig.~\ref{fig:sullivan} (center)). Two theoretical issues that require significantly more theoretical investment include: the splitting functions that determine the effective ``virtual meson flux'' in the initial state, and the modeling of the off-shell nature of the struck meson and how this affects the cross section. The current models of the meson splitting functions neglect the $t$-dependence introduced by the virtual nature of the struck meson, and in the cases of both the meson form factors and PDFs, the theoretical uncertainties in the linkage to the experimental observables are inadequately understood.

There has been significant progress on all topics listed above.

\begin{enumerate}
\item Regarding GPD information from Double DVCS (DDVCS), simulations have been performed on the $Q^2$ vs. $Q'^2$ coverage for two scenarios: CLAS12+$\mu$EMCAL and SoLID+$\mu$Det (see Fig.~\ref{fig:DDVCS}), where $Q^2$ and $Q'^2$ refer to the incoming and outgoing virtual photons. The specific advantage of 22 GeV is that it gains much more access to the resonance-free (2-3 GeV) region compared to 11 GeV, which is crucial to the GPD interpretation of the DDVCS data. The much greater $Q^2-Q'^2$ coverage allows for tests of scaling and GPD evolution with greater authority, and allows higher-twist effects to be studied for the first time.

\begin{figure*}[htbp]
\begin{center}
\includegraphics[width=0.65\columnwidth]{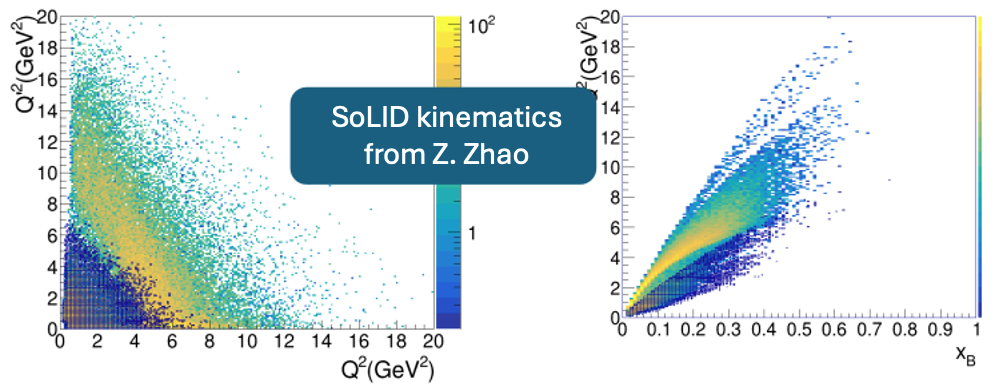}
\vspace{-3mm}
\caption{Kinematic coverage for the DDVCS process with SoLID+$\mu$Det. (Left) $Q^2$ vs. $Q'^2$. (Right) $Q^2$ vs. $x_B$. The dark region is the coverage with an 11 GeV beam, while the colored region is the expanded coverage with a 22 GeV beam.}
\label{fig:DDVCS}
\end{center}
\end{figure*}

\item Regarding threshold charmonium production, simulations for Hall D using both 17 and 22 GeV electron beams, show a significantly increased figure of merit for linearly polarized photons above the $J/\psi$ threshold (for a 12 GeV beam, much of the polarized photon flux is below threshold (see Fig.~\ref{fig:GlueX}). A 22~GeV beam will also allow polarization measurements to be obtained for threshold $\chi_c$ and $\psi'$ production for the first time. Simulations for $J/\psi$ electroproduction with SoLID have also been carried out. While an 11~GeV beam would enable SoLID to measure threshold $J/\psi$ production from $Q^2$=0.3-2~GeV$^2$, a 17 or 20 GeV beam would enable unique measurements up to $Q^2=8$ GeV$^2$, and $\psi$(2S) electroproduction up to $Q^2=1.5$ GeV$^2$. SoLID also opens up the possibility of threshold Charmonium measurements with a polarized NH$_3$ target.

\begin{figure*}[htbp]
\begin{center}
\includegraphics[width=0.65\columnwidth]{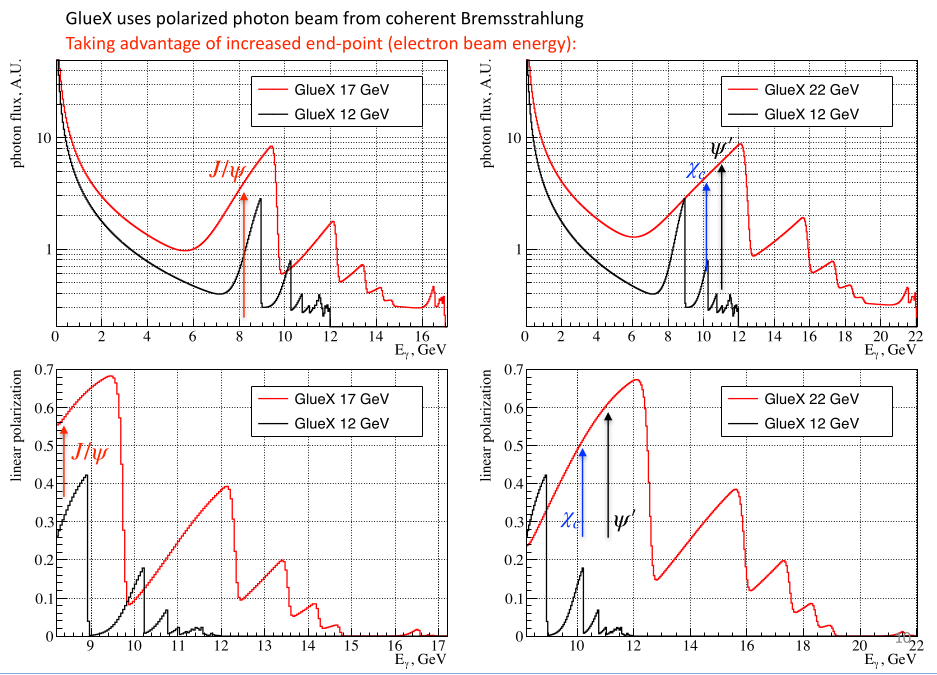}
\vspace{-3mm}
\caption{Linearly polarized photon flux at GlueX with 12, 17, and 22 GeV electron beams. (Left) Above the $J/\psi$ threshold, (right) above the $\chi_c$ and $\psi'$ threshold.}
\label{fig:GlueX}
\end{center}
\end{figure*}

\item Projections for $\pi^+$ and $K^+$ form factors utilizing $L/T$ separations in Hall C have been completed. As in 2022, a two-phase upgrade scenario was presented, where the developments since then are mainly in better understanding the tradeoffs between expanded kinematic coverage with the existing HMS+SHMS spectrometers and possible scenarios should the HMS be replaced with an upgraded spectrometer. Projections for a 22~GeV $\pi^+$ and $K^+$ TDIS experiment show that a drastically expanded $x_B$ range of the data would significantly add to the sparse $K^+$ dataset compared to 11 GeV (see Fig.~\ref{fig:piKTDIS}). A 22 GeV beam also would allow SIDIS studies on the virtual meson to become possible for the first time.

\begin{figure*}[htbp]
\begin{center}
\includegraphics[width=0.8\columnwidth]{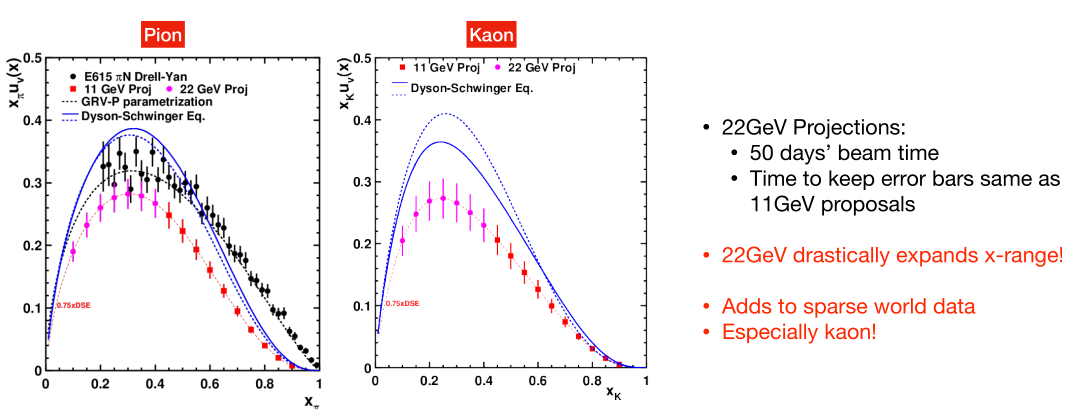}
\vspace{-3mm}
\caption{PDF projections from the TDIS experiment with 11 GeV (red squares) and 22~GeV (magenta circles) electron beams for (left) $\pi^+$ PDF $x_{\pi}u_{v}(x)$, (right) $K^+$ PDF $x_{K}u_{v}(x)$. The 22 GeV projections assume 50 days of beam, needed to keep the statistical uncertainties the same as for the 11 GeV proposals.}
\label{fig:piKTDIS}
\end{center}
\end{figure*}

A very exciting study enabled by 22 GeV would be $\pi^+$ DVCS, which would open up the possibility for multidimensional imaging of the pion (see Fig.~\ref{fig:sullivan} (right)). The issue, similar to that described above for DDVCS, is the ability to take significant data above the $\pi N$ resonance region. The projections used a $m_{\pi N} > 1.8$~GeV cut, which is sparsely populated with an 11 GeV beam. Projected uncertainties were shown for the beam spin asymmetry from the pion DVCS process with 22 GeV for $1<Q^2<2$~GeV$^2$. However, much work is still needed here, both experimentally and theoretically, including how to measure the final state $\pi^+$ and $\gamma$ and experimentally isolate the Sullivan process in $p(e,e'\pi^+\gamma)n$, and the reliable modeling of the pion splitting functions for the virtual pion cloud in the DVCS process.\\
\end{enumerate}

\noindent \emph{\underline{Future plans}}: Further work needs to be done on the DDVCS and TCS simulations. The DDVCS simulations using CLAS12 and SoLID are disjoint and better coordination between the two working groups would give a better understanding of the relative merits of the two measurements. No new work was presented at the workshop on Timelike Compton Scattering (TCS), however, if its feasibility is established, it could become one of the flagships of the 22 GeV upgrade.

Additional work on the feasibility of the pion DVCS process should also be supported. There are many open questions, both theoretically and experimentally, on whether it is possible to measure the $p(e,e'\pi^+)\gamma$ reaction with the required cleanliness and statistics, and whether the desired pion GPD information can be reliably extracted from such data. If feasible, such measurements would provide incredible detail on the internal structure of the $\pi^+$, but the challenges are also significant.

GlueX-III will provide a wealth of threshold $J/\psi$ photoproduction data, while SoLID would provide a huge increase in electroproduction data beyond what has been obtained from Hall~C. Initial results from the CLAS12 and SoLID DDVCS experiments will also be important for establishing the reliability of such measurements before embarking on a more ambitious 22-GeV program.\\

\noindent\emph{\underline{The global landscape}}: The EIC is expected to mount a substantial exclusive reaction program covering the same topics discussed here, including for $x_B>0.05$~\cite{AbdulKhalek:2021gbh}. The EIC timeline has various uncertainties, but only the earliest initial results might be expected 15 years from now. The complementarity and overlap of future 22~GeV JLab and EIC measurements should be investigated in greater detail. As an example, the comparison of threshold $J/\psi$ production from 22 GeV to threshold $\Upsilon$ production from EIC versus $Q^2$ will likely provide significant advances to our understanding of the QCD trace anomaly contribution to the proton mass. A second example would be TCS, which has not yet been investigated in any detail at EIC, and also needs significantly more investigation for 22 GeV.

The EicC also plans an ambitious exclusive reaction program covering these topics, but its timeline and possible capabilities are even more uncertain~\cite{Anderle:2021wcy}. The merits of EicC measurements relative to a 22 GeV JLab program are very difficult to predict, but it seems likely that 22 GeV would have a significant luminosity advantage over EicC, while EicC would have an advantage over polarized beam flexibility.

The AMBER program at CERN intends to measure $\pi^+$ and $K^+$ structure PDFs via pion and kaon-induced DY. This complementary process to the TDIS experiment is a very important cross-check on the theoretical uncertainties in the Sullivan process, and will allow the sea quark content of the pion to be accurately measured for the first time. AMBER also plans to measure pion-induced $J/\psi$, which accesses the gluon and quark PDFs. Both sets of data are currently scheduled to be acquired in the 2030-2031 physics run, so it would be reasonable to expect first results in the 15 year timeframe. The 22 GeV meson structure program would benefit greatly from the information coming from AMBER.

It is absolutely essential that the 22 GeV upgrade preserve the high current capabilities of CEBAF. Maintaining the 100 $\mu$A current capability of Halls A and C is essential to the feasibility of the 22 GeV program, because cross sections drop rapidly with momentum transfer. This requires a doubling of the available beam power. EIC will of course have some advantage in terms of the flexibility of arbitrarily polarized nucleon and $^3$He beams, but it appears that the EIC will not have the ability to take measurements with polarized deuterium beams, at least not without a further upgrade. If the technical challenges can be achieved, then the 22 GeV capabilities will enable precision measurements of small cross sections over a very wide kinematic range, which are absolutely essential for GPD studies from the various Compton scattering processes and for the measurements of meson structure.

\subsection{Nucleon Resonance Structure and Emergence of Hadron Mass}
\label{sec:nstar}

\begin{figure*}[ht]
\begin{center}
\includegraphics[width=0.55\columnwidth]{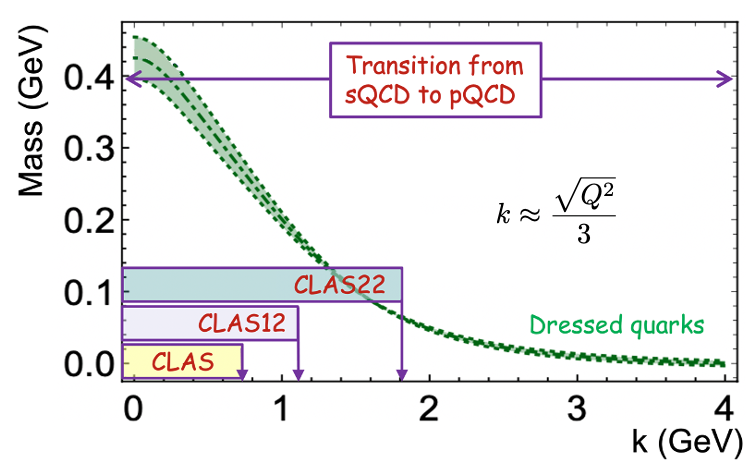}
\caption{Experimental reach for mapping the momentum dependence of the dressed quark mass in studies of the $\gamma_v pN^*$ electrocouplings using existing data from CLAS, the expected results from CLAS12, and projections for a potential JLab energy upgrade to 22~GeV, presented in terms of the accessible ranges of the absolute values of the quark propagator four-momenta $k$, where the mass $m(k)$ is the dressed quark running mass.} 
\label{kranges}
\end{center}
\end{figure*}

\noindent  \emph{\underline{Recent developments since the White Paper}}: 
While the primary goals of the program to explore the structure of nucleon excited states from different exclusive reaction channels to better understand the strong interaction dynamics underlying their generation remain unchanged since the White Paper~\cite{Accardi:2023chb} was released, new experimental results and recent developments with the Continuum Schwinger methods (CSMs) open up additional opportunities to achieve these objectives. Studies of $N^*$ structure from the evolution of their $\gamma_vpN^*$ electrocouplings with $Q^2$, provide a unique opportunity to explore many facets of the strong interaction seen in generating the spectrum of $N^*$ states with different structural features \cite{Mokeev:2022xfo}. These studies are crucial for understanding the emergence of hadron mass (EHM)~\cite{Carman:2023zke,Ding:2022ows}. According to the EHM concept developed within CSMs, dressed quarks with running masses are generated in quark-dressed gluon interactions when the transition from small (i.e. the pQCD regime) to large (i.e. the strongly coupled regime) QCD running coupling takes place as shown in Fig.~\Ref{kranges}. 

More than 98\% of the mass of the ground and excited $N^*$ states is created by the emergent components of the three dressed quarks. The extension of the results on the $Q^2$-evolution of the electrocouplings, as well as on the combined analysis of pseudoscalar meson elastic form factors and PDFs, will allow us to establish either universality or sensitivity of the dressed quark mass function to the structural environment and to explore di-quark correlations of different spin-parities. The experimental studies of both meson and baryon structure coupled with advances in CSMs offer a promising tool 
to understand the emergence of hadron mass and structure \cite{Ding:2022ows,Raya:2024ejx}.

A successful description of the electrocouplings of the $\Delta(1232)3/2^+$ and $N(1440)1/2^+$ has been achieved for $Q^2 < 5$~GeV$^2$ by employing the same dressed quark mass function shown in Fig.~\ref{kranges} that was also used to successfully describe the experimental results on the pion and nucleon elastic electromagnetic form factors~\cite{Carman:2023zke,Mokeev:2023zhq}. The CSM predictions for the electrocouplings of the $\Delta(1600)3/2^+$ \cite{Lu:2019bjs} were confirmed by the later experimental results~\cite{Mokeev:2023zhq}. These successes highlight the relevance of dressed quarks with running masses predicted as active components of baryon structure and demonstrate the capability of gaining insight into EHM from the experimental results on the $Q^2$-evolution of the electrocouplings. 

\begin{figure*}[htbp]
\begin{center}
\includegraphics[width=0.70\textwidth]{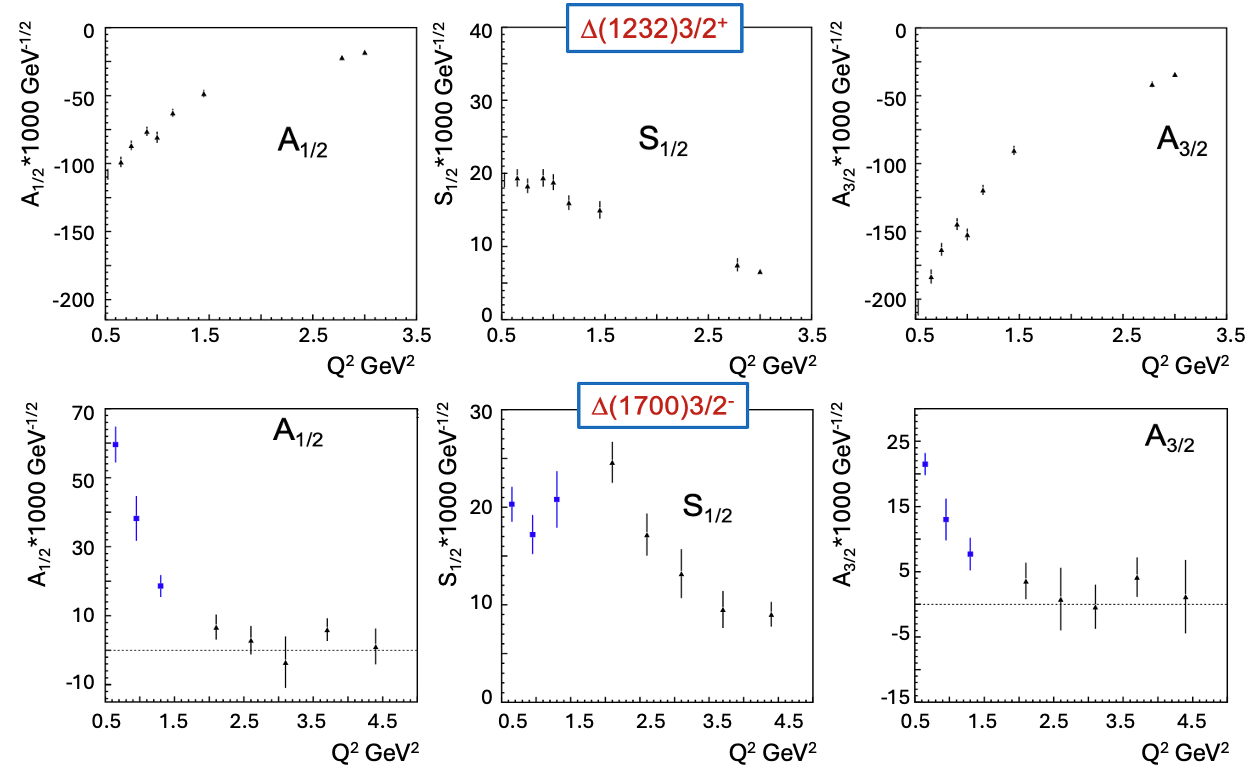}
\caption{(Top): Electrocouplings of the $\Delta(1232)3/2^+$ obtained from the analysis of $\pi N$ electroproduction data~\cite{CLAS:2009ces}. (Bottom): Electrocouplings of the $\Delta(1700)3/2^-$ obtained from the $\pi^+\pi^-p$ channel (blue) from combined studies of photo- and electroproduction~\cite{Mokeev:2020hhu} and (black) the new preliminary results for $Q^2$ from 2 to 5~GeV$^2$~\cite{mokeev-nstar24}.} 
\label{d33_p33ground_el}
\end{center}
\end{figure*}

The results on the electrocouplings of the $\Delta(1232)3/2^+$~\cite{CLAS:2009ces} and the new preliminary results for its chiral partner $\Delta(1700)3/2^-$~\cite{mokeev-nstar24} are shown in Fig.~\ref{d33_p33ground_el}. In the chiral limit, the electrocouplings of these resonances should be related by a chiral rotation. However, the experimental results reveal significant differences. Analyzing these differences presents a promising avenue for understanding the emergence of baryons from QCD and the connection between EHM and dynamical chiral symmetry breaking.

Advances in CSMs make possible predictions for the nucleon gravitational form factors with basic CSM ingredients checked from the studies of nucleon and pion elastic form factors and electrocouplings for the range of $Q^2 < 30$~GeV$^2$ accessible only after the 22-GeV upgrade. Studies of baryon structure within this broad range of $Q^2$ will benefit understanding of the mechanical properties of the nucleon and pion \cite{Yao:2024ixu}. In fact, the importance of pursuing a program of extracting the nucleon resonance electrocouplings from different exclusive reaction channels over the broadest range of $Q^2$ has only increased. Such data spanning $Q^2$ up to 30~GeV$^2$ will offer the only foreseeable opportunity to explore the full range of distances where the dominant part of hadron mass and $N^*$ structure emerge from QCD. This makes CEBAF at 22~GeV unique and potentially the ultimate QCD facility at the luminosity frontier.\\

\noindent \emph{\underline{Future plans}}:
The simulations carried out to date for the 22~GeV program for exclusive $\pi N$, $K^+Y$, and $\pi^+ \pi^- p$ processes have employed the configuration of the existing CLAS12 detector. Recently improved simulations of the $\pi^+ \pi^- p$ electroproduction channel based on the existing CLAS12 detector at the current luminosity have confirmed that the acceptance and resolution are sufficient to reliably measure the exclusive cross sections needed to extract the $\gamma_v p N^*$ electrocouplings. However, a number of hardware and software upgrades are currently in progress that will improve the charged particle tracking resolution and efficiency for increased luminosity operations. As these designs advance, further improved simulations of what we call ``CLAS22" will need to be carried out. These simulations should be possible in the next 3-4 years as more details on the full set of improvements become available.

Forging synergistic efforts between experiment, phenomenology, and QCD-based hadron structure theory is essential for addressing the key open problems in the Standard Model regarding hadron structure and the emergence of hadron mass. The success of the CSM framework in advancing our understanding of hadron mass and structure generation paves the way for broader theoretical efforts. These efforts will be critical in supporting the interpretation of experimental results, helping to resolve long-standing challenges within the Standard Model, and potentially offering insights into the deeper complexity of Nature that lies beyond its current boundaries. In this regard this topic of study is being developed in partnership with an international collaboration of theorists from the EU, Mexico, and China working on the description of hadron spectra/structure. A new review paper has recently been prepared to provide an overview of our current results and future plans from the 6-GeV and 12-GeV programs with CLAS and CLAS12, to motivate an extended program at 22~GeV \cite{Achenbach:2025kfx}.

The kinematic reach of the 6-GeV era experiments of $Q^2$ up to 5~GeV$^2$ allows us to map the dressed quark mass within the range of distances where approximately 30\% of the emergent hadron mass is generated. The first measurements of inclusive $(e,e'X)$ cross sections with CLAS12 at $W < 2.5$~GeV for $Q^2$ from 2.5--10.5~GeV$^2$ have revealed the presence of resonance-like structures at $W \approx 1.5$~GeV and 1.7~GeV across the entire $Q^2$ range \cite{CLAS:2025zup}. CLAS12 is the only facility in the world capable of extending information on the electrocouplings of prominent $N^*$ states up to a mass of 2.5~GeV for $Q^2$ up to 10~GeV$^2$. These electrocouplings will ultimately be obtained from $\pi N$, $\eta p$, $KY$, and $\pi^+\pi^-p$ electroproduction data and first preliminary electroproduction cross sections have become available in the $\pi^+\pi^-p$ channel. The results on the $\gamma_vpN^*$ electrocouplings for $Q^2 < 10$~GeV$^2$ will allow for exploration of EHM within the range of distances where around 50\% of the emergent hadron mass is generated. The anticipated results and outcome of their analyses within CSM, lattice QCD, and Hamiltonian Effective Field Theory (HEFT) will help to solidify the scientific case for the experiments in the 22-GeV era, conclusively demonstrating the need to provide information on the evolution of the strong interaction dynamics within the full range of distances where the transition from the pQCD to strongly coupled regimes is expected and the dominant portion of the $N/N^*$ mass and structure is generated.\\

\noindent\emph{\underline{The global landscape}}:  
The analysis of the CLAS12 data already collected is well underway for exclusive processes in the $\pi N$, $\eta N$, $KY$, and $\pi^+\pi^-p$ channels. These data will ultimately extend the range of the extracted electrocouplings up to $Q^2 \approx 10\,$GeV$^2$. The full set of proposed measurements from the 12-GeV era will be available within the next 10 years. Completion of the 12-GeV program is essential in the lead up to the 22-GeV program.

Increasing the JLab energy up to 22~GeV will enable extension of the accessible $Q^2$ range in the resonance region up to 30~GeV$^2$. Monte Carlo simulations based on the existing CLAS12 detector have so far confirmed that the acceptance and resolution are adequate to reliably measure the exclusive cross sections needed to extract the electrocouplings. This will make CLAS22 the only foreseen facility capable of carrying out such a program. The results on the evolution of the $\gamma_v p N^*$ electrocouplings for $Q^2$ up to 30~GeV$^2$ will allow for exploration of the full range of distances where the dominant portion of emergent hadron mass and hadron structure is generated. This will establish JLab@22~GeV as the ultimate QCD facility at the luminosity frontier.

\clearpage

\section{QCD in Nuclei and Associated Nuclear Modifications and Dynamics}


\subsection{Prospects of Medium Modification Studies at JLab22}

\noindent \emph{\underline{Recent developments since the White Paper}}: 
A recent focus has been on spin-1 targets, which exhibit several novel properties not found in the spin-1/2 nucleon. The best known spin-1 nucleus is the deuteron, however, of the more than 250 stable 
nuclei only two others have spin-1, $^6$Li and $^{14}$N. The novel properties of spin-1 nuclei include: a quadrupole electromagnetic form factor and associated quadrupole moment that measures deformation, 
a new unpolarized quark distribution called $b_1^q(x)$ that is associated with the tensor polarization of a spin-1 nucleus, a leading-twist gluon transversity distribution that cannot appear in the nucleon because of 
helicity conservation, and several new GPDs and TMDs associated with tensor polarization.

From a nuclear structure perspective these spin-1 nuclei are associated with two-body and four-body clustering, where the deuteron is the simplest two-body cluster, $^6$Li is to a good approximation a bound 
$^4$He--deuteron system, and $^{14}$N can be considered a bound system of three $^4$He nuclei with a deuteron. Missing in this picture is the bound system of two $^4$He nuclei with a deuteron, but this can be identified 
with $^{10}$B, which is spin-3 and here the deuteron is in a $L=2$ state. Therefore, the set of four nuclei defined by the deuteron, $^6$Li, $^{10}$B, and $^{14}$N possess interesting spin and clustering structures.
Understanding the quark and gluon structure of these nuclei at JLab 22~GeV would provide a crucial bridge between traditional low-energy nuclear structure and QCD dynamics. Plans for studies of these systems in the 22~GeV kinematic regime have been further developed.\\

\begin{figure}[h]
\centering
\includegraphics[width=0.55\textwidth]{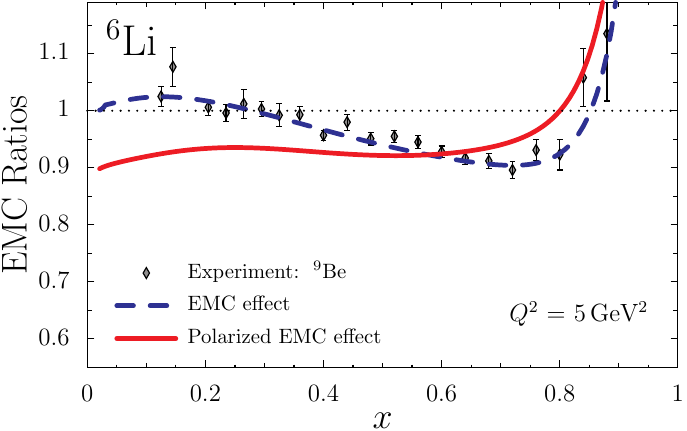}
\caption{Results for the EMC and polarized EMC effects in $^6$Li~\cite{Cloet:2006bq}. These results are compared to unpolarized experimental data for $^9$Be because $^6$Li data is not available.}
\label{fig:emcLi6} 
\end{figure}

\noindent \emph{\underline{Future plans}}:
As a first step in gaining insight into this ensemble of nuclei we have calculated for the first time the EMC and polarized EMC effects in $^6$Li. From Fig.~\ref{fig:emcLi6} we see that the EMC and polarized EMC effects
in $^6$Li are comparable in size in the valence region~\cite{Cloet:2006bq}. This is already surprising from a nuclear structure and short-range correlation (SRC) perspective, because the spin is almost completely carried by the valence nucleons, which to a good approximation, is described by a deuteron loosely bound to a spin-0 $^4$He core. From this perspective one would expect the polarized EMC effect to be much smaller than the familiar EMC effect. In addition, for SRCs there is no leading mechanism for a SRC pair to align its spin with the spin of the nucleus and therefore SRCs are depolarizing, which again should give a small polarized EMC effect. Further studies in this direction are in progress and will continue.\\

\noindent\emph{\underline{The global landscape}}:
JLab at 22~GeV can also explore the sea-quark region, where we see that the EMC effect has the familiar anti-shadowing, whereas for the polarized EMC effect this does not appear. Measuring both EMC effects in the sea-quark region would reveal interesting aspects on how sea-quarks are impacted by the nuclear medium and if these effects impact polarized sea-quarks differently from an unpolarized sea.

\subsection{The EMC Effect of Light Nuclei within the Light-Front Hamiltonian Dynamics}

\noindent \emph{\underline{Recent developments since the White Paper}}: 
In 1983, the European Muon Collaboration (EMC) discovered a significant deviation from unity in the ratio between the structure functions of nuclei and those of deuterium in the valence region, i.e., $0.3 \leq x \leq 0.7$, 
a phenomenon now known as the EMC effect~\cite{Aubert:1983}. In fact, at high energy, given the binding energy per nucleon, it was expected that the partonic structure of bound nucleons would be identical to that of free 
nucleons. However, the observed deviation indicates that valence quarks in free nucleons move, on average, faster than those in nuclei. Despite four decades of investigation, a definitive explanation for this phenomenon 
remains elusive. 

\begin{figure}[htb]
\includegraphics[width=7.8 cm]{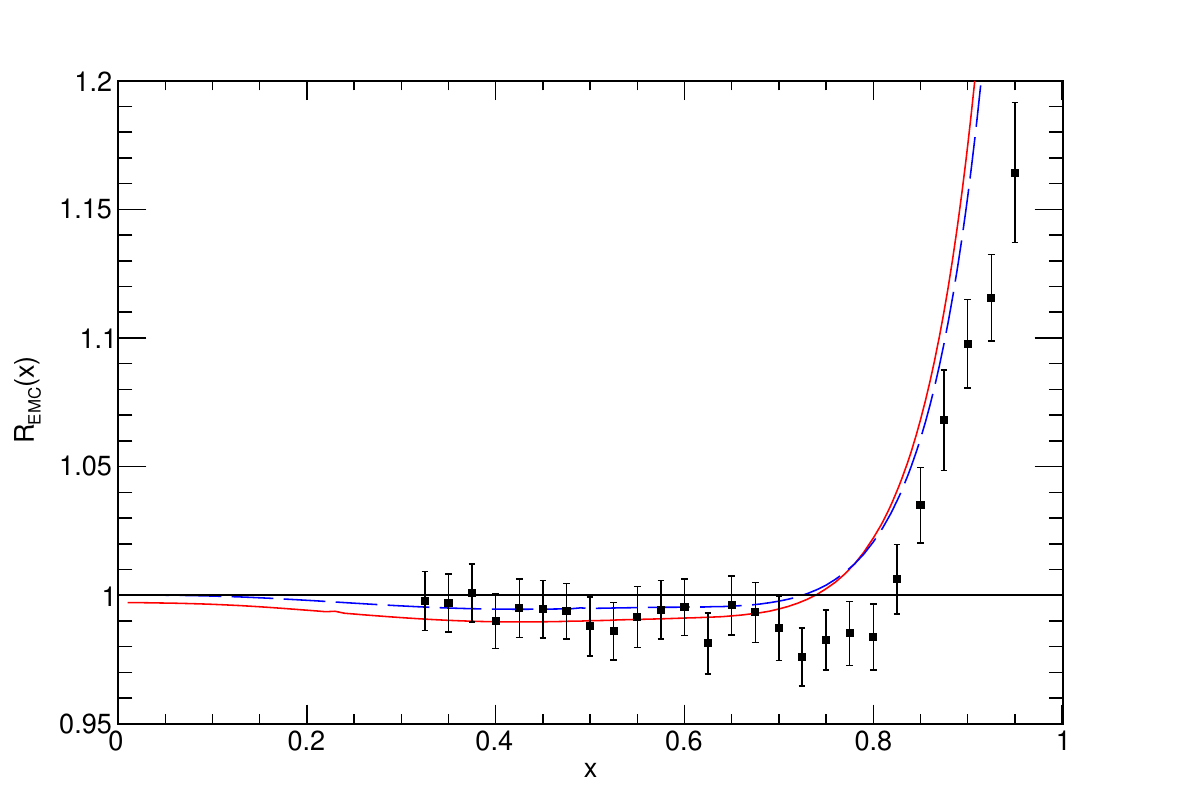}
\includegraphics[width=7.8 cm]{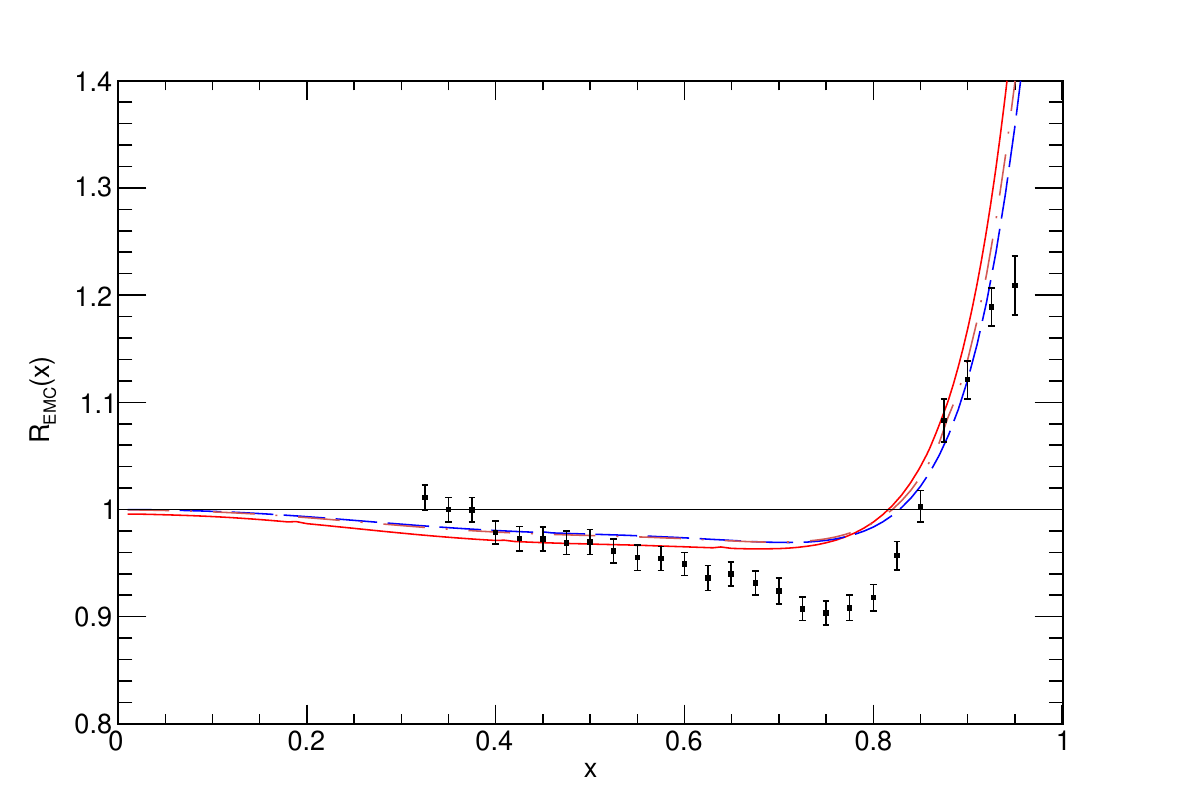}
\caption{The EMC ratio calculation within the light-front Poincar{\'e} approach \cite{Pace:2022qoj,Fornetti:2023gvf}. The different lines correspond to the use of a different nuclear wavefunction obtained from a different 
nuclear potential. Details can be found in Ref. \cite{Fornetti:2023gvf}. Left panel for $^3$He. Right panel for $^4$He.}
\label{fig:EMCratio}
\end{figure}

Notably, conventional non-relativistic nuclear physics models fail to adequately describe the existing experimental data. Conversely, several alternative models have been proposed that successfully 
reproduce these observations. For example, the off-shellness, which leads to a modification of the inner structure of bound nucleons with respect to the free ones. A comprehensive understanding of the 
dynamical origin of this effect is essential for several significant reasons. Notably, since neutron structure functions can only be extracted from measurements on nuclear targets, a precise characterization of nuclear 
effects is imperative for accurate interpretation of the data. Within this context, a rigorous calculation of the EMC effect using an approach that formally satisfies both baryon number and momentum sum rules is 
imperative. Such calculations, implemented with state-of-the-art nuclear wavefunctions, are essential to precisely delineate the potential contribution of exotic mechanisms. To this aim, in 
Refs.~\cite{Pace:2022qoj,Fornetti:2023gvf}, the $^3$He, $^3$H and $^4$He SFs have been evaluated within a Poincar{\'e} covariant Light-Front approach \cite{Dirac:1949cp}. Due to this procedure, macroscopic locality, baryon
number and momentum sum rules are fulfilled.  Moreover, due to an appropriate construction of the Poincar{\'e} generators \cite{Bakamjian:1953kh}, one can use them in the calculation of the nuclear wavefunctions obtained 
from realistic nuclear potentials. Therefore, these calculations represent a baseline for any further inclusions of exotic effects beyond the conventional ones.   The analysis indicates that for $^3$He, where recent
experimental investigations are found in Refs.~\cite{JeffersonLabHallATritium:2021usd,JeffersonLabHallATritium:2024las}, conventional nuclear physics frameworks can largely account for the corresponding EMC effect (see left panel of 
Fig.~\ref{fig:EMCratio}).

\begin{figure}[htb]
\includegraphics[width=6.8 cm]{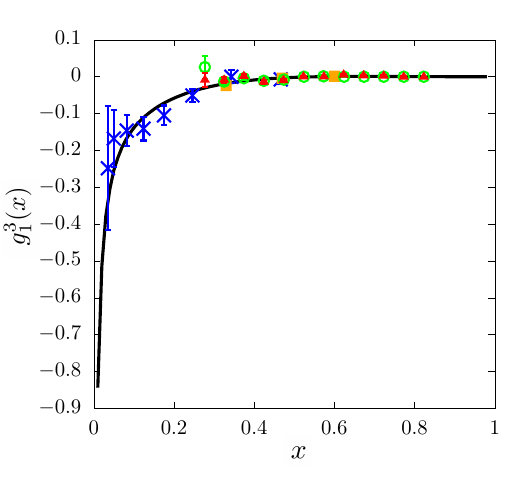}
\includegraphics[width=6.8 cm]{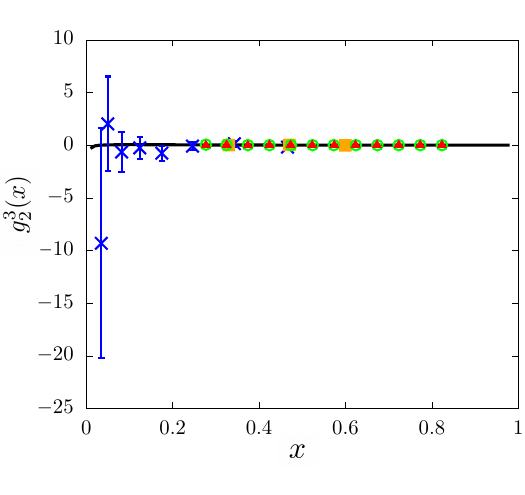}
\caption{The $^3$He spin-dependent structure function. The full lines are the calculations of Ref. \cite{Proietti:2024bhb}. Details on the experimental data are given in Ref.~\cite{Proietti:2024bhb}. Left panel: $g^3_1(x)$. Right panel: $g^3_2(x)$.}
\label{fig:EMCratio2}
\end{figure}

However, in the case of $^4$He, despite the fact that conventional effects predict a structure-function ratio less than unity, the experimental data cannot be adequately reproduced. This discrepancy strongly suggests 
that additional non-conventional mechanisms may need to be incorporated into theoretical frameworks (see right panel of Fig. \ref{fig:EMCratio}) Therefore given the still open questions, the increased statistics 
offered by the JLab 22 GeV upgrade will be crucial to highlight the possible role of effects beyond the conventional ones. Moreover, increasing the data in the high $x$ region would also help in extracting the 
neutron structure functions in this poorly known region. These data could also help in properly characterizing the description of higher-twist effects \cite{Cerutti:2025yji}. Furthermore, this approach 
has been successfully extended in the polarized case for the $^3$He obtaining a good agreement with present data (see  Fig. \ref{fig:EMCratio2}) \cite{Proietti:2024bhb}. Such a procedure is crucial to extrapolate the 
spin-dependent neutron structure function. Moreover, a detailed knowledge of these quantity is fundamental also for the studies of the partonic spin structure of hadrons, e.g. the extraction of the orbital angular momentum of partons. 
For this purpose precise extraction of polarized PDFs and GPDs functions is mandatory. \\

\noindent \emph{\underline{Future plans}}:
The results discussed herein underscore the necessity for a JLab 22 GeV upgrade to enhance statistical precision in future measurements. Current efforts are underway to integrate Light-Front methodology into 
phenomenological parameterizations of nuclear parton distribution functions and structure functions.\\

\noindent\emph{\underline{The global landscape}}:
It is noteworthy that forthcoming data in the high-$x$ region, where the EMC effect deviates from linearity, will contribute significantly to our fundamental understanding of Fermi motion phenomena and to extract the 
neutron structure functions at high $x$ with precision. Furthermore, the extraction of the spin-dependent structure functions of $^3$He is of paramount importance for both theoretical and experimental investigations of proton spin structure,
particularly regarding the determination of parton orbital angular momentum within hadrons. Consequently, the proposed JLab 22 GeV upgrade would prove invaluable for elucidating novel approaches to investigating the 
transition between nucleonic and partonic degrees of freedom.

\subsection{Continuing the Search for 3N~SRCs}

\noindent \emph{\underline{Recent developments since the White Paper}}: 
Precision observations of two-nucleon (2N) SRCs, followed by studies of isospin and nuclear dependence, as well as probe independence, made up a very successful program of study at JLab at 6~GeV and continue to do so at 12~GeV. 
Multiple attempts have been made to look for a signature of three-nucleon (3N) SRCs in inclusive cross section ratios of $A/^3$He at $x>2$.  However, no second scaling plateau has so far been observed. With the exception of the 6~GeV 
Hall C data~\cite{Fomin_2012}, all other measurements were at $Q^2 < 2$~GeV$^2$, which we now know does not reach far enough in $\alpha_{3N}$ (below 1.6) to probe 3N SRCs.  The Hall C data were on the edge of 
threshold 3N SRC kinematics, but suffer from insufficient precision.\\

\begin{figure}[htb]
\centering
\includegraphics[width=0.45\textwidth]{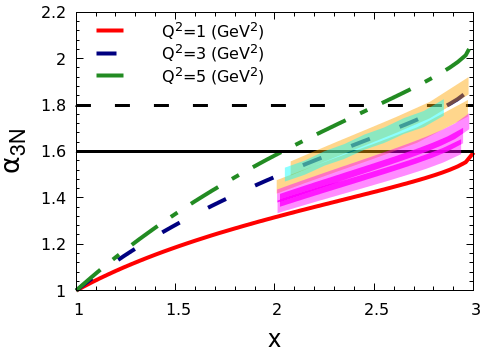}
\caption{Kinematic coverage of previous inclusive 3N SRC searches at JLab. The magenta region is for Hall A data \cite{HallA:2017ivm}, the light blue region is the statistics-limited Hall~C 6 GeV data \cite{Fomin_2012}, and the 
yellow region represents the yet to be published data from a 12~GeV experiment in Hall C.}
\label{fig:alpha-kine}
\end{figure}

\noindent \emph{\underline{Future plans}}:
Recently, E12-06-105 (part of the XEM2 Run Group in Hall C) collected high-precision data on several nuclear targets (A=3, 4, 9, 12, 40) at the highest $Q^2$ values so far, reaching above 3~GeV$^2$. The coverage in
$\alpha_{3N}$ can be seen in Fig.~\ref{fig:alpha-kine}. If the scaling expected sets in at the lowest possible $\alpha_{3N}$, the experiment may be able to resolve a second plateau. \\ 

\noindent\emph{\underline{The global landscape}}:
The 12 GeV Hall~C data is unlikely to provide a definitive result as it will not tell us enough about the nuclear or $Q^2$ dependence of 3N SRCs. Dedicated experimental searches are needed - the highest $Q^2$ data we have right now comes from 
experiments whose main goals were to measure other kinematics, and 3N SRC runtime was, therefore, minimal. Higher $\alpha_{3N}$ are accessible with 12~GeV at JLab, but the run times rise exponentially fast. A 22~GeV 
energy upgrade, if coupled with upgrades to the Super High Momentum Spectrometer in Hall C could make the difference. High-resolution spectrometers are absolutely vital for these searches, as the community learned from
Refs.~\cite{Egiyan_2006,Higinbotham:2014xna}.

\subsection{Studying the Tensor-Polarized Deuteron System in the 22~GeV Era}

\noindent \emph{\underline{Recent developments since the White Paper}}: 
The deuteron is the simplest bound nuclear system and serves as a distinctive probe for examining the strong interaction at different resolutions and for testing the limits of 
nuclear models based solely on nucleons. Despite its importance, the tensor component of this spin-1 nuclear system has hardly been explored. A comprehensive understanding of the deuteron's tensor structure promises to 
reveal novel insights into the partonic landscape of these nuclei.

The tensor structure of the deuteron, a fundamental spin-1 nucleus, has remained largely unexplored, primarily due to the experimental challenges associated with achieving suitable tensor polarization for electron 
scattering studies. Recent advancements, however, have demonstrated the feasibility of achieving significant tensor polarization (up to 30\%)~\cite{Keller:2020wan,Clement:2023eun}, leading to a resurgence of interest 
in the physics potential of this system. Fostering these efforts has continued since the release of the 22-GeV White Paper~\cite{Accardi:2023chb}.\\

\noindent \emph{\underline{Future plans}}:
The JLab 12-GeV program has approved two dedicated experiments focusing on the $b_1$ structure function in DIS~\cite{E12-13-011} and the $A_{zz}$ 
quasi-elastic asymmetry~\cite{E12-15-005}. Nevertheless, growing interest in other reaction channels necessitates not only a suitable tensor-polarized deuteron target but also experimental conditions conducive to high data rates.

Two programs that would be well-suited for the 22-GeV energy upgrade include first a focus on the exploration of spin-1 TMDs and, second, an examination of the high-energy component of nucleons in deuteron
electrodisintegration.

Existing TMD experiments have provided significant insights into the spin-$1/2$ properties of nucleons. In contrast, the tensor spin 
aspects of spin-1 systems, exemplified by light nuclei, have received considerably less attention. This gap in our understanding of the tensor structure at the parton level motivates the exploration of tensor-polarized
structure functions as a novel direction in spin physics. These functions hold the potential to reveal unique features of parton dynamics that are not accessible through studies of spin-$1/2$ nucleons alone 
\cite{PhysRevD.82.2010_Kumano,proceeding_Kumano_2022}. Although the spin-1 deuteron is often treated as a basic proton-neutron bound system, the fundamentally tensor nature of a spin-1 particle, which is absent in 
spin-$1/2$ systems, indicates that tensor-polarized structure functions can open a new avenue in hadronic physics, offering valuable information on QCD and the structure of nuclei.
JLab's 12-GeV program will initially focus on longitudinally polarized deuterium targets through a three-step approach to measure structure functions. First, an approved CLAS12 analysis will investigate the unmeasured 
tensor component in vector-enhanced deuterons (up to 20\% tensor polarization) to estimate its size~\cite{Poudel:2025tac}. This will be followed by a dedicated experiment in Hall C and a comprehensive program utilizing 
SoLID for detailed measurements~\cite{Poudel:2025nof}. These studies aim to provide the first experimental determination of the deuteron's tensor structure.
 
Recent studies have indicated that the use of a tensor-polarized target will enable unique isolation of the node in the $S$-wave distribution of the deuteron in a high-$Q^2$ 
electrodisintegration processes~\cite{Sargsian:2024hyx}. The observable is the $A_{node}$ asymmetry, which is related to the tensor asymmetry $A_d^T$ by:

\begin{equation}
    A_{node} = 1 +  \frac{2A_{d}^{T}(p_m, \theta _{N})}{3\text{cos}^{2}(\theta _{N} ) - 1} = \frac{u^2 (p_m) + 2\sqrt{2}u(p_m)w(p_m)} {u(p_m)^2 + w(p_m)^ 2},
\end{equation}
\noindent 
where $p_m$ is the momentum of missing momentum and $\theta_N$ is the direction of the internal momenta with respect to the polarization axis of the deuteron. $A_{node}$ is also related to the $S$- and $D$-waves, $u$ and 
$w$, respectively. This asymmetry goes to zero at two points, first where $u(p_1) = -2\sqrt{2}w(p_1)$, and second where $u(p_2) = 0$. In the first case, $p_1 \sim 180$~MeV, which corresponds to relatively small internal
moments for which the deuteron wavefunction is well known, such that different models agree. The second one occurs at $p_2 > 350$~MeV where the Paris~\cite{Lacombe:1981eg}, AV-18~\cite{Wiringa:1994wb} (hard), and 
CD-Bonn~\cite{Machleidt:2000ge} (soft) $NN$ potentials disagree.

\begin{figure}[htbp]
\centering
\includegraphics[scale=0.3]{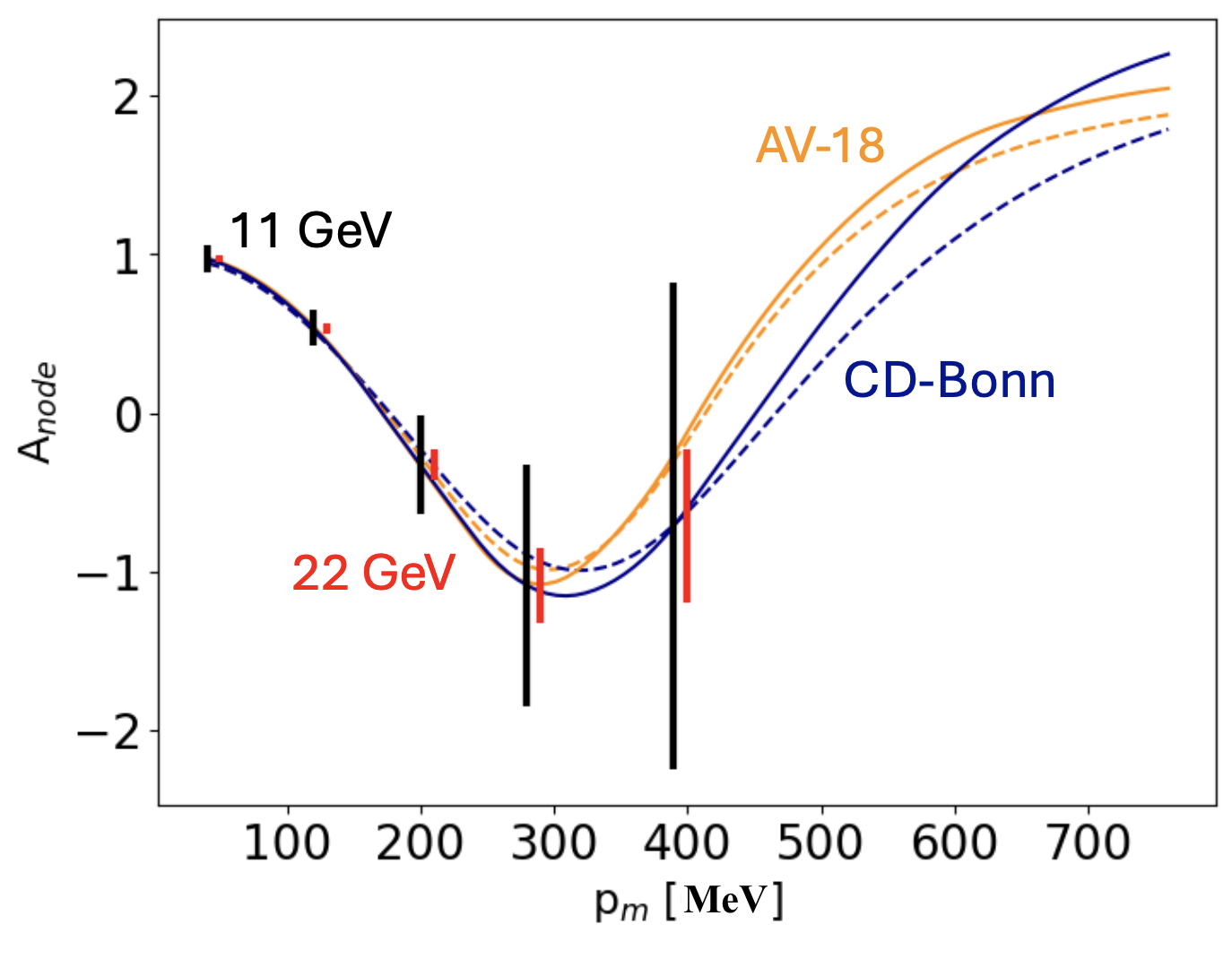}
\caption{$A_{node}$ as a function of $p_m$ for $Q^2 = 2.5$~GeV$^2$. The dashed lines show the plane wave impulse approximation (PWIA) calculations, and the solid lines include final state interactions (FSI)~\cite{Sargsian:2024hyx} for the CD-Bonn (blue) and AV-18 (yellow) potentials. 
The black points correspond to 11 GeV projections in Hall C, and the projections for 22 GeV, with similar considerations to the spectrometers, are shown in red for two weeks of data taking.}
\label{fig:Anode}
\end{figure}

This measurement is particularly challenging due to the low rates encountered at momentum $p_2$, coupled with the requirements for a tensor-polarized target. Figure~\ref{fig:Anode} shows projections of statistical 
uncertainties for $A_{node}$ under the same target conditions as the approved Hall C experiments~\cite{E12-13-011,E12-15-005} for two weeks, compared with the same conditions at 22~GeV. To minimize final state interactions, 
the recoil angle is selected such that $0^{\circ} < \theta _{rq} < 35^{\circ}$. We observe a significant improvement in the error bars simply by increasing the beam energy. However, in this case, we could not utilize the same 
Hall~C detectors because they would be outside their coverage. Therefore, we need to consider using the SoLID detector for the electrons and an additional detector to measure the protons. Further studies of the detector 
requirements need to be studied.
 
\begin{figure}[htb]
\centering
\includegraphics[width=0.45\textwidth]{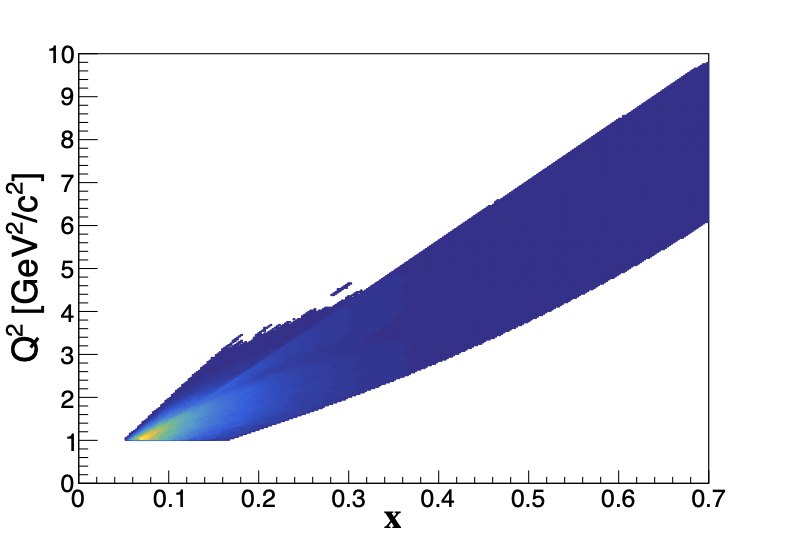}
\includegraphics[width=0.45\textwidth]{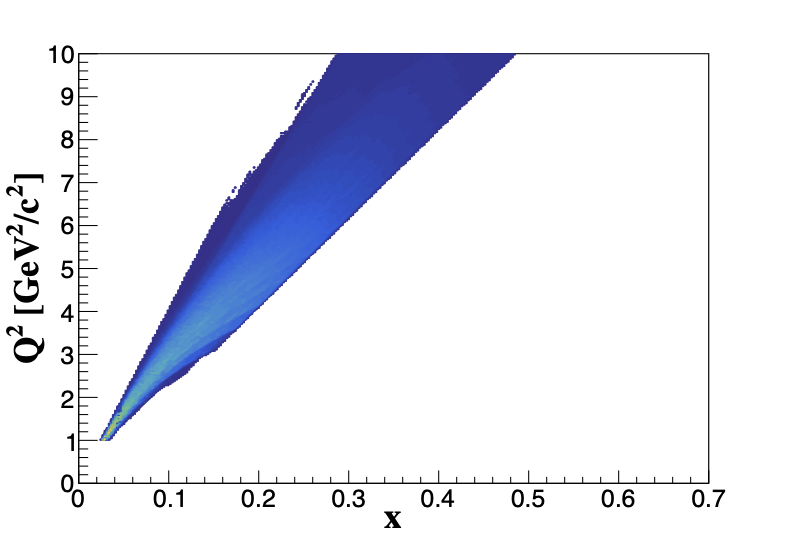}
\caption{$Q^2$ vs. $x$ kinematic coverage for the SoLID detector using the 11 GeV beam (left) and 22 GeV beam (right) in the reaction $eD \to e’ \pi^-X$.}
\label{fig:Spin1-SIDIS}
\end{figure}

\noindent\emph{\underline{The global landscape}}:
Figure~\ref{fig:Spin1-SIDIS} shows the kinematic coverage for the SoLID detector using the 11~GeV beam (left) and 22~GeV beam (right). The future 22-GeV experimental program holds significant promise for advancing our
understanding of TMD physics. By providing measurements across key energy and transverse momentum scales, it will be instrumental in refining QCD factorization theorems and elucidating the intricacies of hadronization.
Moreover, the anticipated data will allow for a clearer separation of the TMD and collinear QCD regimes. A particularly unique aspect of 22 GeV is its capacity for detailed exploration of the intermediate kinematic 
region, where the transition between the well-understood perturbative and non-perturbative regimes of QCD occurs.

\subsection{Probing the EMC Effect Through Measurement of Super-Fast Quarks in Nuclei}

\noindent \emph{\underline{Recent developments since the White Paper}}: 
The general goal of the super-fast quark (SFQ) studies is largely unchanged from what was reported in the White Paper~\cite{Accardi:2023chb}. Nuclear PDFs in the super-fast quark region, $x \gtrsim 1$, are predicted to have extremely large 
nuclear effects beyond those due to the motion of the nucleons, and different models predict significant suppression or enhancement of the PDFs in this regime. If the PDFs can be extracted in this region, comparison to
calculations for the deuteron can be used to look for potentially large nuclear effects, while target ratios and the $Q^2$ dependence can be used to provide further discrimination between different models.

\begin{figure}[htb!]
\centering
\includegraphics[width=0.49\textwidth]{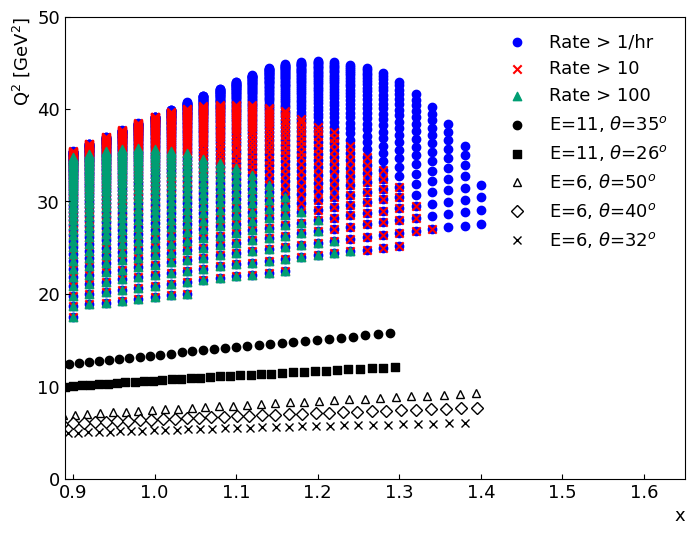}
\includegraphics[width=0.49\textwidth]{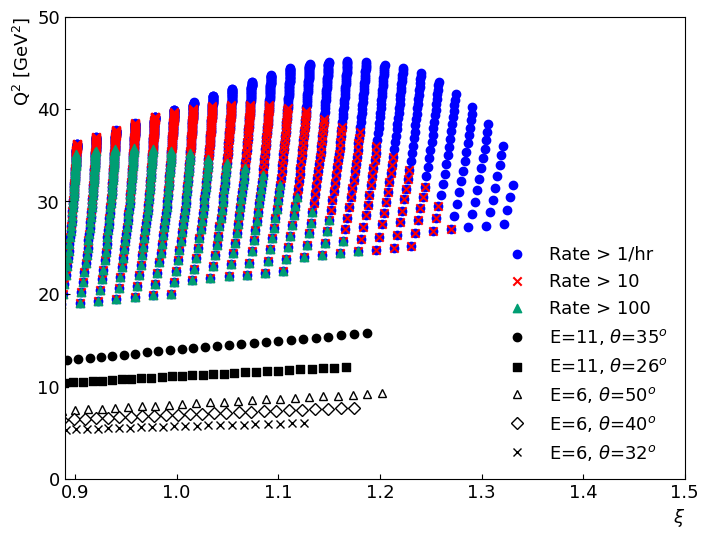}
\caption{Kinematic coverage for 22 GeV SFQ measurements compared to existing JLab kinematics. The left panel shows the coverage as a function of $x$ and $Q^2$, while the right panel shows the 
kinematics vs. $\xi$ and $Q^2$, which provides a more meaningful comparison of $Q^2$ range and statistics for fixed $\xi$ values over a range of $Q^2$.}
\label{fig:SQF_kine}
\end{figure}

There has been some progress relative to the White Paper. Figure~\ref{fig:SQF_kine} shows updated comparisons of the 22~GeV kinematic coverage as a function of $x$ ($x$ is the quark longitudinal momentum fraction and called the Bjorken scaling variable) (left) and $\xi$ ($\xi$ is equivalent to $x$ in the Bjorken limit and called the Nachtmann scaling variable) (right). For extraction of the nuclear 
PDFs, examining the $Q^2$ range and statistics as a function of $Q^2$ is more appropriate for fixed $\xi$ values. The updated study demonstrates that 22~GeV provides both a factor of three increase in $Q^2$ (and more 
than a factor of two in $Q^2$ with very good statistics) and a significant increase in the $\xi$ range, with the predicted nuclear modification in some models growing significantly between $\xi=1.2$ and 1.3. Detailed 
simulations to examine the impact of spectrometer resolution are still required, but these effects are not expected to be significant based on previous measurements at lower energies~\cite{Fomin:2010, E12-06-105}.\\

\noindent \emph{\underline{Future plans}}:
Additional work on the theory side will be important in interpreting these results. As a first step, data on the deuteron structure function will be compared to calculations of the expected deuteron structure function 
in the absence of effects beyond smearing. Reliable calculations with reasonable estimates of the uncertainties in these predictions are needed for these studies. More can be learned by examining target ratios in the SFQ
regime, and by examining the $Q^2$ dependence of these ratios, but evaluation of the models of nuclear effects for a range of nuclei and $Q^2$ is required. Work towards baseline predictions for the deuteron is in 
progress. Once these calculations exist, with well-understood conventional input, evaluations of nuclear models should be relatively 
straightforward. Additional input and a more formalized plan and timeline will help advance these studies.

The other requirement is that we can quantify how reliably these data can be used to extract the PDFs. At the large $\xi$ values of interest, data at lower $Q^2$ values have large contributions from the resonance region 
and quasielastic scattering. Data at 6~GeV indicated that approximate scaling was observed even when the data is dominated by scattering with $W^2<4$~GeV$^2$, as a consequence of quark-hadron duality. Recent 11~GeV data 
taken by E12-06-105~\cite{E12-06-105} will provide a quantitative test of scaling, and evaluate sensitivity to the PDFs, for kinematics that are dominated by DIS but which still have significant contributions from 
scattering at lower $W^2$. This should allow us to demonstrate that the data at 22~GeV will not be constrained by questions of finite-$Q^2$ corrections to the structure function. The limited $\xi$ range of E12-06-105, 
combined with significantly larger contributions from non-DIS scattering, makes it likely that the measurement will not be able to make a strong statement on nuclear effects unless the nuclear effects show a significant
enhancement before $\xi=~1.2$. While the EIC will be able to extract the structure function at large $\xi$ and very high $Q^2$ values~\cite{AbdulKhalek:2021gbh}, it is not expected to make significant measurements above $\xi = 1-1.1$. So while 
the larger $Q^2$ values make the interpretation cleaner, it does not extend into the region where the nuclear effects are expected to strongly discriminate between different models.\\ 

\noindent\emph{\underline{The global landscape}}:
The SFQ studies are conceptually very closely connected to the idea of hidden color, and measurements looking at scattering from pairs of high-momentum baryons have been proposed as a way to search for hidden 
color and related effects. Such studies were considered for lower energies. However, the lack of statistics and issues of interpretation at lower-$Q^2$ values were considered to be severely limiting. Such approaches 
should be significantly better for the 22~GeV beam, but evaluation of these ideas would need significant work in terms of modeling and simulation.

\subsection{Searching for Color Transparency Effects at 22~GeV}

\begin{figure}[htb!]
\centering
\includegraphics[width=0.5\textwidth]{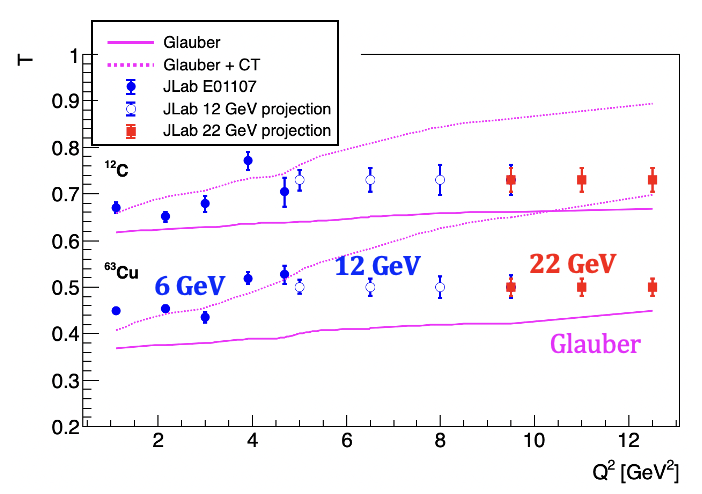}
\caption{A 17.6~GeV electron beam on Hall C production targets of $^{12}$C and $^{63}$Cu for 200~hours of beam time is shown. The maximum $Q^2$ is 12.5~GeV$^2$.}
\label{fig:ct-pion} 
\end{figure}

\noindent \emph{\underline{Recent developments since the White Paper}}: 
The overarching goals in studies of the onset of color transparency (CT) phenomena remain largely the same as those mentioned in the White Paper~\cite{Accardi:2023chb}, but there are new ideas for exploring different kinematics 
and reaction types at 22~GeV. To explore the onset of CT effects, an experimental focus has developed to explore different reaction mechanisms and different kinematics, which may increase the sensitivity for the 
observation in protons. In assuming that there is no upgrade or change to the existing spectrometers, these studies will not necessarily benefit from the full 22~GeV upgrade, but they will significantly benefit from the 
increased beam energies that will be available beyond the current 12~GeV electron beam. The currently approved 12-GeV era CT experiments will dictate the motivation and need for experiments under a 22-GeV upgrade. No 
further theoretical input is currently required, but continuous theoretical interpretation of the current and upcoming experiments will be needed to interpret the results and determine what remains to be understood. 
In the mesonic sector, confirmation of the continued increase in the $\rho^0$ and $\pi^+$ transparencies will indicate the need for further study at 22~GeV. It is very likely that any increase in the transparencies 
in the current 12-GeV era experiments will benefit from studies at higher beam energies to further evaluate and confirm the offset. While the projections have not changed since the White Paper, the theoretical projections 
for the $\pi^+$ have been updated in Fig.~\ref{fig:ct-pion}, which shows the accuracy achievable with an upgrade for a 200~beam-hour experiment with a 17.6~GeV electron beam and 3\% uncertainties 
on $^{12}$C and $^{63}$Cu targets. To remain in kinematics with reduced FSIs, the squared four-momentum transfer $t$ must be less than 1~GeV$^2$. The maximum useful $Q^2$ for this study is 12.5~GeV$^2$.\\

\begin{figure}[htb!]
\centering
\includegraphics[width=0.5\textwidth]{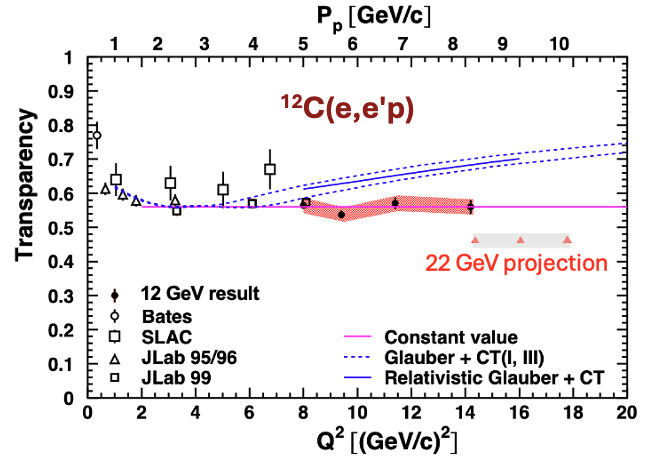}
\caption{Proton knockout from carbon in parallel kinematics at high $Q^2$ in Hall C. The red points can be obtained in a 180~hour beam on target experiment at 13~GeV electron beam energy.}
\label{fig:ct-proton} 
\end{figure} 

\noindent \emph{\underline{Future plans}}:
In the baryonic sector, holographic light-front QCD makes predictions that we should observe proton CT at $Q^2>14$~GeV$^2$~\cite{Brodsky:2022bum}. This is also consistent with the lack of observed CT in the recent 
Hall~C experiment, which only measured up to $14$~GeV$^2$~\cite{Dutta:2021}. Therefore, by measuring quasielastic proton knockout reactions from carbon with a beam energy of 13~GeV in Hall C, one can attain high 
statistical precision to extend the previous studies up to a $Q^2$= 17.4~GeV$^2$ in about 180~hours of the experimental beam on target. The increase in $Q^2$ can probe more rapid expansions of the point-like configuration
proton and is shown in Fig.~\ref{fig:ct-proton}.

A more recently approved experiment since the 22~GeV White Paper proposes to explore proton CT in maximal rescattering kinematics using a deuterium target. The benefit of such a reaction is that the production of the 
point-like configuration can be studied somewhat separately from its subsequent expansion by controlling the inter-nucleon distances after production~\cite{physics4040092}.

\begin{figure}[htb!]
\centering
\includegraphics[width=0.5\textwidth]{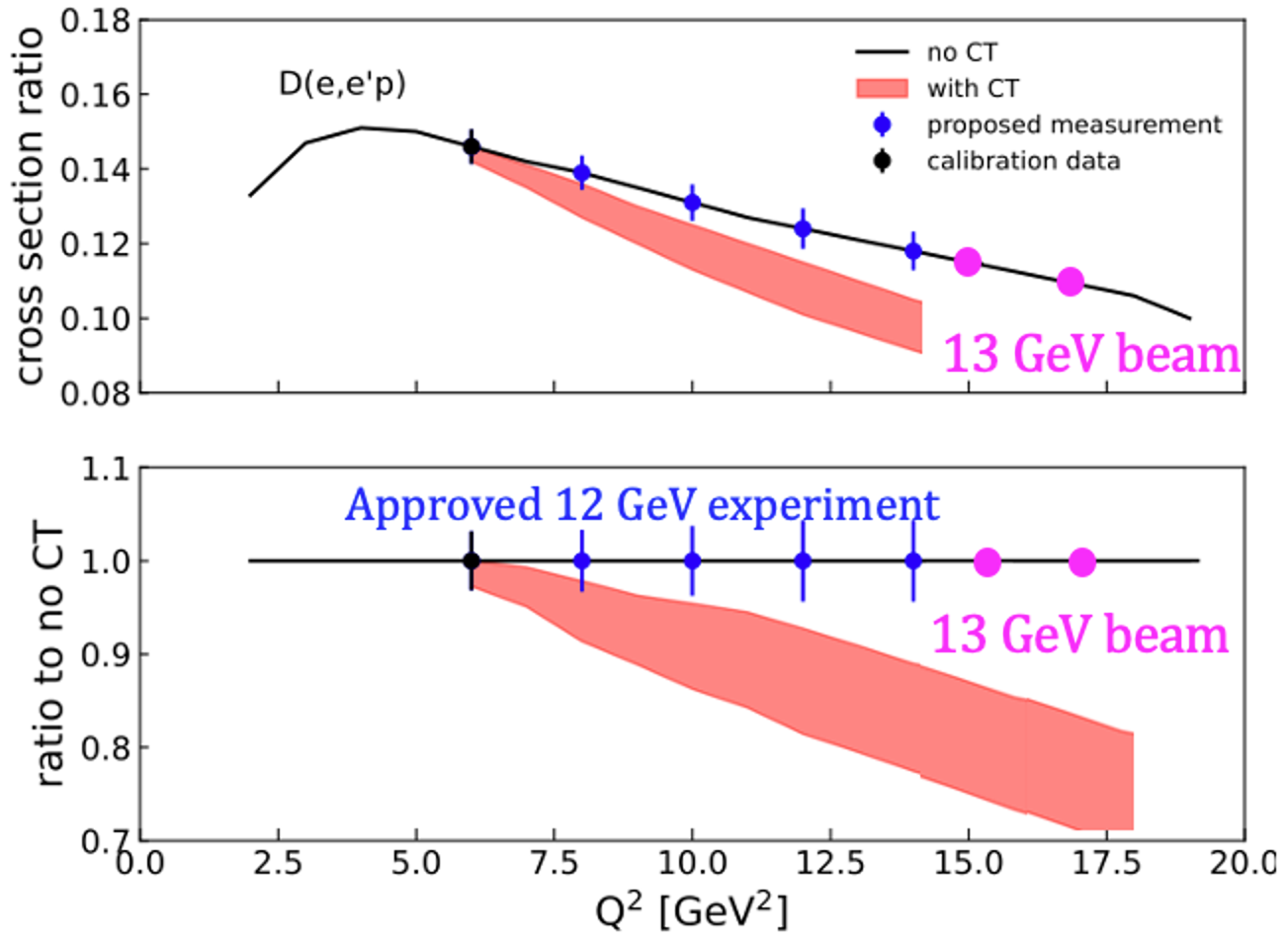}
\caption{Detected rescattered protons from deuterium are sensitive to shorter point-like configuration lifetimes.}
\label{fig:ct-pdct} 
\end{figure}

The approved experiment E12-23-010 will run in Hall C on a deuterium target and will explore up to $Q^2=14$~GeV$^2$. A 13~GeV electron beam would enable access to $Q^2$ as high as 17~GeV$^2$ as shown in 
Fig.~\ref{fig:ct-pdct}. Although achieving 3\% statistical precision in such an experiment would still require two and half months of beam time; this experiment would be highly motivated if the approved experiment E12-23-010 
observes a signal indicating the onset of CT. This experiment is intended to run before MOLLER runs in Hall A and would require approximately two years of analysis for results. If the experiment runs (optimistically) 
in late 2026, then results could be expected in 2029. 

Another possibility is that the observation of CT will be reaction-dependent. The recent Hall D experiment E12-19-003 studied photoproduction on nuclear targets of $^{12}$C, $^4$He, and deuterium. The first results of 
this experiment observed sub-threshold $J/\Psi$ photoproduction on nuclear targets~\cite{Pybus:2024ifi}. The analysis is currently in progress to look for color transparency effects in the reaction of $\gamma p \to \rho^0 p$ 
in the nuclear targets. 

\begin{figure*}[ht]
\begin{center}
\includegraphics[width=0.5\columnwidth]{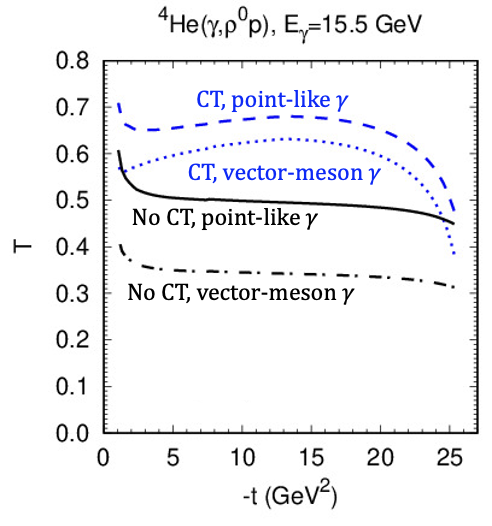}
\caption{Theoretical calculation for $^4$He$(\gamma p,\rho^0p)$ with a 15.5~GeV coherent photon beam.}
\label{fig:ct-halld}
\end{center}
\end{figure*}

While the particle identification in the GlueX detector and statistics limit the $|t|$ range to a maximum of about 4~GeV$^2$, the data will be sensitive enough to distinguish between CT and non-CT effects, as well as 
photon interactions as point-like and vector meson. Further simulation work is required to evaluate and optimize the photon peak for reactions at 22~GeV and to understand the particle identification limitations or 
potential for improvements. A modest assumption with the coherent photon peak at 15.5~GeV was used for theoretical calculations to show the reaction $^4$He$(\gamma p,\rho^0p)$ as shown in Fig.~\ref{fig:ct-halld} could 
be extended over a large range of $|t|$ for $|u|>1$~GeV$^2$. \\

\noindent\emph{\underline{The global landscape}}:
Additional reaction channels remain to be studied, and their results could further motivate studies at the 22~GeV upgrade. Furthermore, at 22~GeV, $J/\Psi$ is accessible in both photo- and electroproduction reactions, 
which could provide another channel to study in the mesonic sector, as well as a cross-check between different reaction mechanisms. This possibility will be studied in detail.
 
\subsection{Medium Modifications of Quark Propagation and Hadron Formation}

\begin{figure*}[ht]
\begin{center}
\includegraphics[width=0.5\columnwidth]{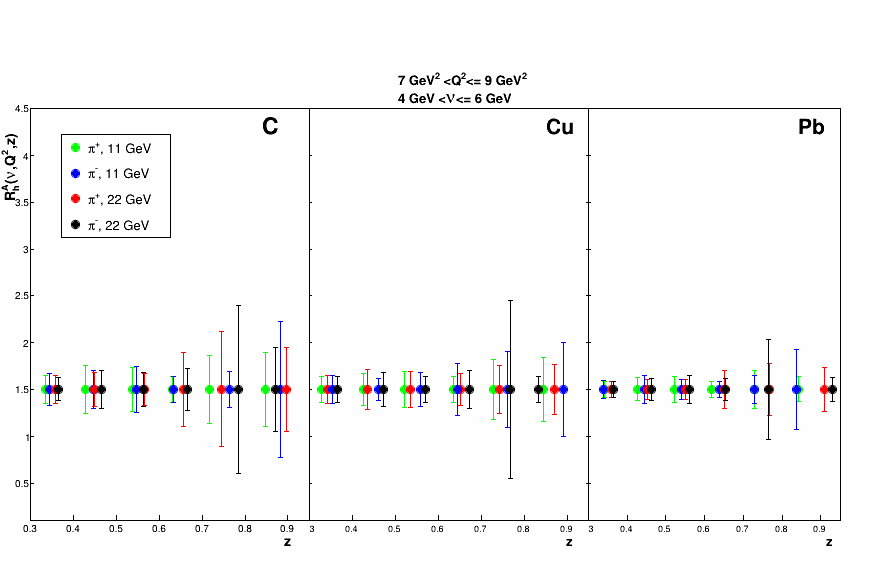}
\caption{A comparison of 11~GeV and 22~GeV multiplicity ratio projections for charged pions production off three nuclei illustrating the 22~GeV statistical precision gain, especially at high-$z$.}
\label{fig:MR22-11gev}
\end{center}
\end{figure*}

\noindent \emph{\underline{Recent developments since the White Paper}}: 
A set of Monte Carlo events was generated using the PyRad framework, which integrates Pythia for simulating SIDIS events in nuclei, utilizing nuclear parton distribution functions (LHAPDFs 6.5.4), and RADGEN to account 
for radiative corrections. The tuning of the Pythia parameters was performed using the CLAS12 RG-C 11 GeV dataset and subsequently applied to the 22~GeV data. After generating the events, we reconstructed them using the CLAS12 RG-E
implementation of the dual-target system along with the GEANT-based description of the detector. We compared the reconstructed 3D multiplicity ratios for positive pions across ($Q^2$, $\nu$, $z$) and ($Q^2$, $\nu$, 
$p_T^2$) bins for both 10.5~GeV and 22~GeV electron beams (see Fig.~\ref{fig:MR22-11gev}). By analyzing a luminosity of 10$^{35}$cm$^{-2}$s$^{-1}$ for 12 PAC days for each nucleus, the potential was identified for describing the kinematic 
dependencies of multiplicities in previously inaccessible kinematics, particularly those of high $\nu$ and high $Q^2$ regions of $p_T^2$ and $z$ dependencies. These regions offer a unique opportunity to explore and disentangle the complex
interplay between competing partonic and hadronic processes as they evolve across different energy regimes.\\

\noindent \emph{\underline{Future plans}}:
Increasing the energy to 22~GeV facilitates the study of rare and complex hadronic processes, offering valuable insights into mass, strangeness, and rank dependencies in hadron formation and color propagation. Based on 
event generation data (PyRad Monte Carlo), we confirm that the first observation of GeV-scale mesons, such as $D$ and $J/\psi$, as well as $\Omega$ baryons, becomes achievable as the center-of-mass energy reaches 
$\sqrt{s}$ = 6.4~GeV. The next step in these studies involves performing a complete reconstruction of the states using the current configuration of the CLAS12 detector, employing reconstruction packages that are currently undergoing improvements. This task also presents a challenge, as these events are rare; therefore, an exceptionally large set of simulations must be processed through the reconstruction pipeline. Additionally, the final 
states must be identified through invariant mass analysis to estimate the projected count rate.\\

\noindent\emph{\underline{The global landscape}}:
This energy upgrade provides a unique opportunity to investigate the dynamics of hadronization in the context of combinations of charm quarks and light quarks, examine strangeness in meson production, and contrast it with
baryon dynamics involving multiple strange quarks. These studies shed light on how varying combinations of charm and strangeness affect hadron formation and the propagation of color charge.

\subsection{Study of Tagged Processes with $^4$He and ALERT at 22~GeV}

\noindent \emph{\underline{Recent developments since the White Paper}}: 
The overarching goals of studies with the ALERT (a low energy recoil tracker) detector are essentially the same as described in the White Paper~\cite{Accardi:2023chb}. The recently completed 11~GeV CLAS12 ALERT experiment aims to study coherent exclusive DVCS
and DVMP to perform model-independent 3D tomography of the partonic structure of $^4$He and $^2$H. However, the experiment will also use spectator tagging as a pioneering process to probe medium modifications of the 
intrinsic structure of bound nucleons in nuclei.  

Furthermore, DVCS on $^4$He offers a straightforward method to access the GPDs of the nucleus. Given that $^4$He possesses only two chiral even ($\mathcal{H}$) and chiral odd ($\mathcal{H}_T$) GPDs, greatly simplifying the 
model-independent extraction of the coherent beam spin asymmetry (BSA). The latter is experimentally measured via scattering a longitudinally polarized lepton beam ($L$) on an unpolarized target ($U$) to obtain the DVCS
differential cross section whose ratio for positive ($+$) and negative ($-$) beam helicity states gets reduced to
\begin{equation}
    A_{LU} = \frac{1}{P_b}\frac{N^+ - N^-}{N^+ + N^-},
\end{equation}
where $P_b$ is the polarization of the incident electron beam and $N$ is the DVCS event yield. 

Since the White Paper, the simulation has been updated to include the latest CLAS12 reconstruction software, including all modifications of the ALERT detector insert structure and target length. The new study to extract
the 22~GeV BSA was carried out by generating about 20M events of nuclear DVCS reactions on $^4$He using the Monte Carlo event generator TOPEG (The Orsay Perugia Event Generator) assuming a luminosity of 
$2\times10^{35}$cm$^{-2}$s$^{-1}$ over 20 PAC days. These events were then passed to the CLAS12 GEANT4-based MC (GEMC) package to get a full estimation of the experimental apparatus acceptance and efficiency effects. 
As a result, the Bjorken $x_B$ and momentum transfer $-t$ bins have been modified to take into account the doubled CLAS12 nominal luminosity.\\

\begin{figure}[h]
\centering
\includegraphics[clip=true, trim= 0 0 0.25cm 0,width=0.505\textwidth]{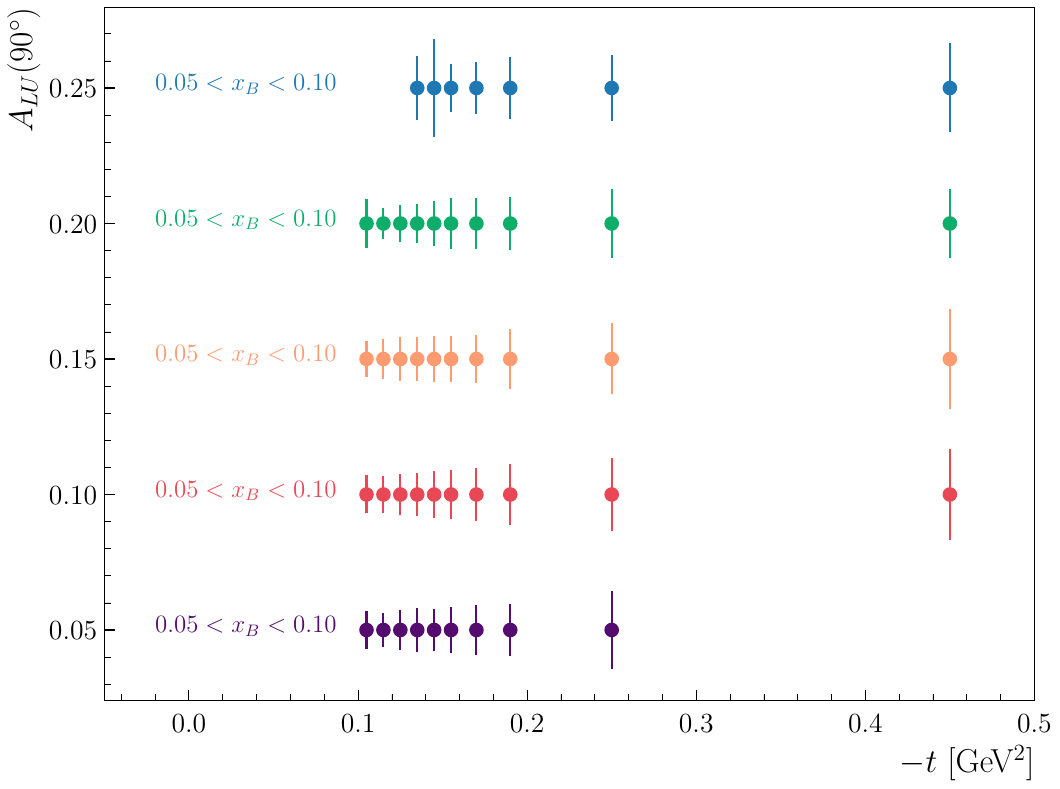}
\caption{The 22~GeV $t$-dependent BSA projections at $90^{\circ}$ for various $x_B$ bins for coherent exclusive DVCS from $^4$He.}
\label{fig:R1-8-MO-ALU-t}
\end{figure}

\noindent \emph{\underline{Future plans}}:
In the coherent BSA analysis, the low-momentum $^4$He nucleus recoils intact in ALERT, while the exclusivity is achieved by detecting the scattered electron and real photon in the forward CLAS12 detector and tagger. 
The BSA has been extracted for all kinematic bins by integrating over the full $Q^2$ range and assuming a beam polarization of $\approx$80\%; see the depicted $t$-dependent projection at 90$^\circ$ in 
Fig.~\ref{fig:R1-8-MO-ALU-t}. Further improvements of ALERT track finding (fitting) and particle identification based on Artificial Intelligence (Kalman filter) techniques should improve the obtained results. However, 
running more simulations to fix the issue of limited statistics in some kinematic bins, as well as adding background processes such as SIDIS and DV$\pi^0$P, is the envisioned next simulation step. \\

\noindent\emph{\underline{The global landscape}}:
Overall, the anticipated 22~GeV JLab upgrade is expected to significantly extend the $Q^2$ reach and enhance the statistical precision in the lower $x_B$ region compared to the current 11~GeV ALERT experiment. It will 
also bridge the gap between the JLab12 and the forthcoming EIC tagged DVCS measurements targeting much higher $Q^2$ and lower $x_B$ at the saturation region. 

\subsection{Study of Quark Distributions with $eA$ Scattering at 22 GeV}

\noindent \emph{\underline{Recent developments since the White Paper}}: 
A crucial strategy to address these open questions is to introduce new experimental observables, beyond inclusive DIS cross sections, to probe the partonic structure of a wide range of nuclei through high-energy 
electron scattering. In the White Paper~\cite{Accardi:2023chb}, we proposed utilizing 22 GeV electron beams incident on light to heavy nuclei to measure the production rates of various hadrons (e.g., $\pi^{\pm}$, $K^{\pm}$, 
$p^{\pm}$, $\Lambda^{\pm}$) in SIDIS processes.\\

Using a leading-order (LO) approximation, we demonstrated that the so-called super-ratio observables—defined as $R_1 = \frac{(\sigma_A^{h+}/\sigma_A^{h-})}{(\sigma_D^{h+}/\sigma_D^{h-})}$, $R_2 = \frac{(\sigma_A^{h+} \pm
\sigma_A^{h-})}{(\sigma_D^{h+} \pm \sigma_D^{h-})}$ —can potentially disentangle the medium modifications to the nPDFs from those to the nuclear fragmentation functions (nFFs), both of which affect SIDIS cross sections in
nuclei. Preliminary Monte Carlo simulation results further indicated that these super-ratio observables, particularly using pion and kaon production, are effective in isolating valence- and sea-quark PDFs.

We recently conducted a more sophisticated study based on a recent theoretical development~\cite{Ke:2023ixa}. In this work, the nPDF was obtained from the EPPS21-NLO global fit, while the nFF was modeled using the NNFF10lo
fit convoluted with a medium effect characterized by an Effective Glauber gluon density, $\rho_G$. To simulate the uncertainty in the medium effects on the nFF for the $^{56}\mathrm{Fe}$ nucleus, we varied $\rho_G$ between 
0.5~fm$^{-3}$ and 0.75~fm$^{-3}$, and examined how its variation affects the experimental observables.

\begin{figure}[ht]
\centering
\includegraphics[width=0.45\textwidth]{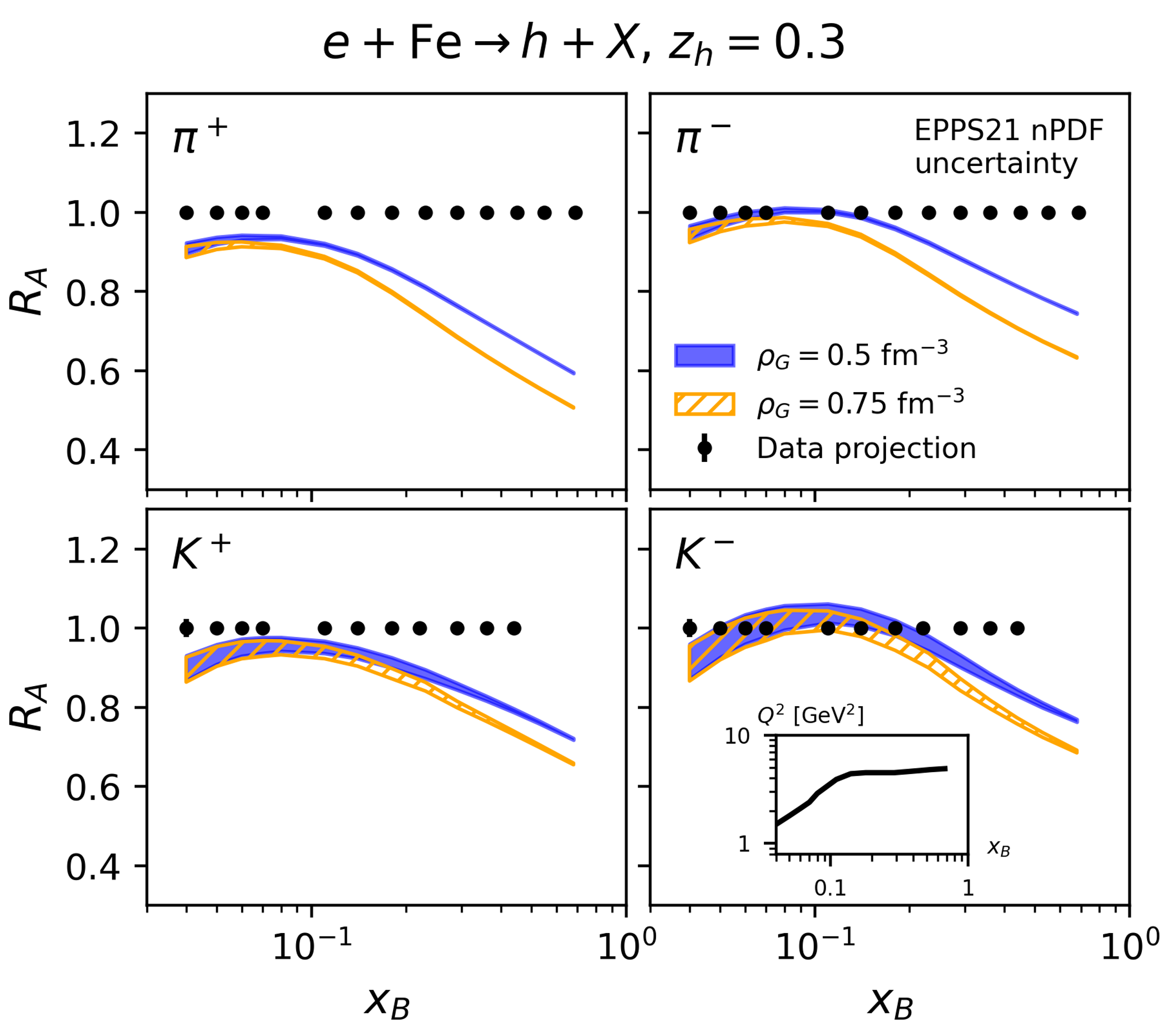}
\includegraphics[width=0.45\textwidth]{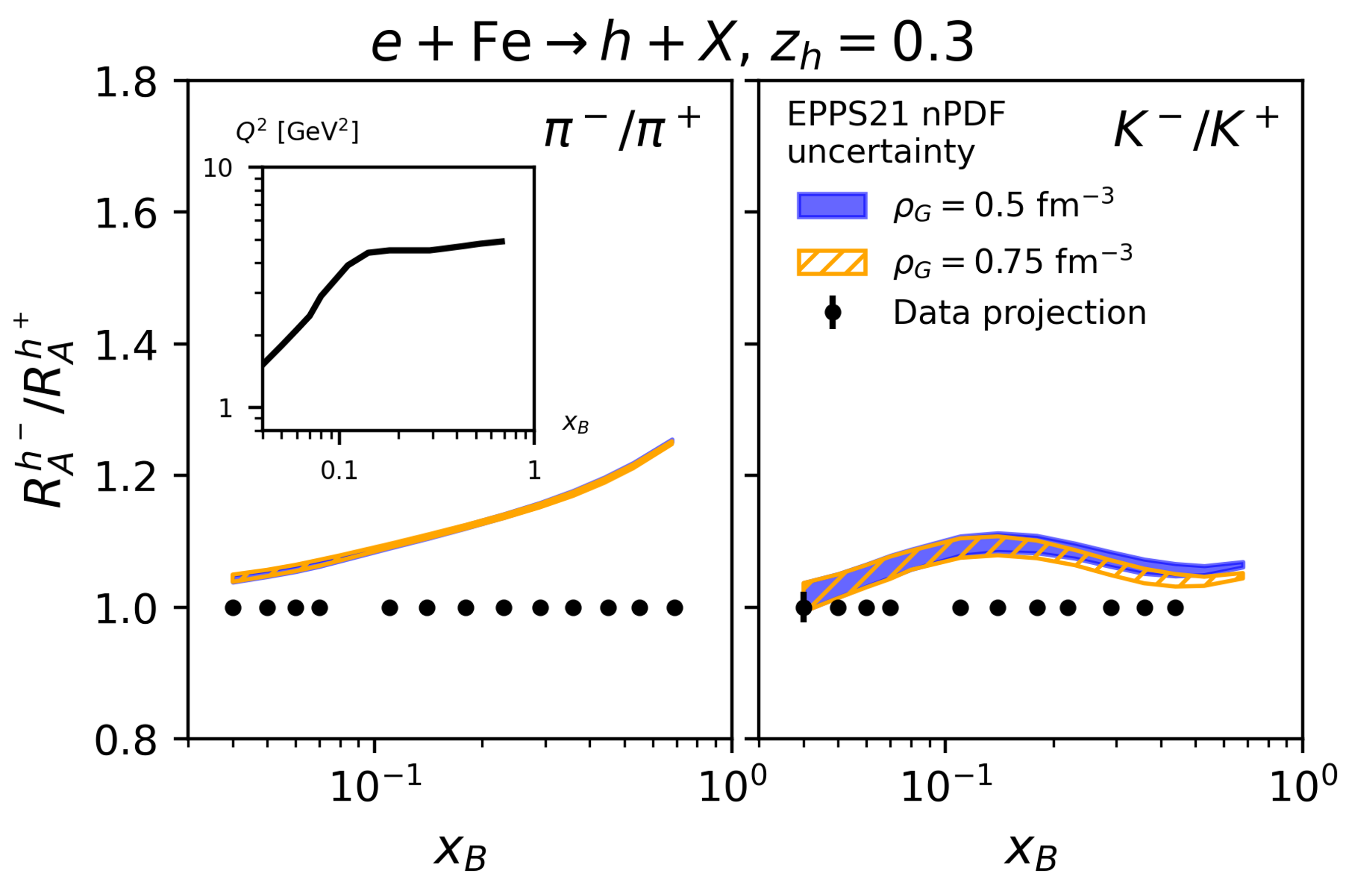}
\caption{The left four subplots are the multiplicity ratios, $R_A(x,z_h)=\frac{\sigma_A^{h+}(x,z_h)/\sigma_A^{DIS}(x)}{\sigma_D^{h+}(x,z_h)/\sigma_D^{DIS}(x)}$, as a function of $x$ at $Q^2$=3.9~GeV$^2$ and $z_h$=0.3, 
clearly illustrating the separation between two different nFF inputs. The black dots represent the projected SIDIS data points. The two subplots on the right display the double ratios, $R_A^{h-}/R_A^{h+}$, for pions and 
kaons, demonstrating that the medium effects from the nFFs largely cancel out in these ratios, leaving primarily the contributions from the nPDFs.}
\label{fig:fe_sidis_multi_ratio}
\end{figure}

As shown in Fig.~\ref{fig:fe_sidis_multi_ratio}, the multiplicity ratios, $R_A(x,z_h) = \frac{\sigma_A^{h+}(x,z_h)/\sigma_A^{DIS}(x)}{\sigma_D^{h+}(x,z_h)/\sigma_D^{DIS}(x)}$, exhibit a strong sensitivity to different 
nFF inputs. In contrast, the double ratios of the multiplicity ratios, $R_A^{h-}/R_A^{h+}$, show reduced sensitivity to the medium effects in the nFF. A combined global analysis of these two observables, extracted from 
SIDIS production on nuclei through $\pi^{\pm}$ and $K^{\pm}$ channels, has the potential to disentangle the contributions from the nPDF and nFF for valence and sea quarks in heavy nuclei.\\

\noindent \emph{\underline{Future plans}}:
In future work, we plan to perform more extensive simulations to study how to more effectively separate valence-quark and sea-quark nPDFs. This will involve simulating SIDIS processes with a broader range of hadron 
production (e.g., $p$, $\bar{p}$, $\Lambda^{\pm}$) and conducting a global analysis using pseudo-data under 22 GeV kinematics to examine how the nPDFs for different quark flavors can be extracted. Additionally, we will 
explore simulations of SIDIS with transverse momentum distributions, which will allow us to evaluate the extraction of nuclear TMDs~\cite{Alrashed:2021csd}.\\

\noindent\emph{\underline{The global landscape}}:
The partonic structure of a free nucleon has been extensively studied through DIS measurements, and the PDFs for various quark flavors and gluons have been successfully extracted via global analyses of the extensive 
world data. In contrast, our understanding of the partonic structure in nuclei remains at a relatively early stage. The scarcity of precise DIS data involving high-energy collisions with different nuclei has led to the
extraction of nuclear PDFs (nPDFs) with large uncertainties and under numerous theoretical assumptions. Although the EMC, anti-shadowing, and shadowing effects were discovered over four decades ago, many fundamental 
questions about their origins and mechanisms remain unresolved.

\clearpage

\section{Tests of Physics Beyond the Standard Model}

\underline{Recent developments since the White Paper}:
\begin{itemize}
\item In  Light Dark Matter theory, the key development is the recognition of the crucial role that electron motion plays in expanding the energy range accessible to positron-based experiments. In a series of papers—including the most recent work by Nardi, di Cortona, and Arias-Aragon~\cite{Arias-Aragon:2025xcc}, it was demonstrated that: 1) resonant production mechanisms, when properly accounting for electron momentum, can extend the mass reach for feebly interacting particles (FIPs) by at least a factor of three; 2) JLab can exploit this enhanced kinematic range to perform precise measurements of the $e^+ e^- \to$ hadrons cross section, providing critical input for addressing the current discrepancies between experimental and lattice QCD calculations of the hadronic vacuum polarization (HVP) contribution to the muon $(g-2)_\mu$. 

\item The integration of artificial intelligence/machine learning (AI/ML) techniques has become increasingly prominent in beyond the Standard Model (BSM) searches~(e.g. Refs.~\cite{Kriesten:2023uoi,Kriesten:2024are}), particularly for detecting rare or unexpected signals in large datasets. This shift allows for more efficient identification of potential BSM phenomena, such as axion-like particles (ALPs), by analyzing complex event topologies and reducing background noise. Advancements in lattice QCD calculations have improved the theoretical understanding of hadronic structures, enabling more precise predictions that can be tested against experimental data from the 22 GeV CEBAF upgrade. This synergy enhances the ability to probe fundamental symmetries and confinement in QCD.

\item  Recent exploratory studies, such as those conducted using the GlueX detector~\cite{GlueX:2021myx}, have demonstrated the feasibility of searching for ALPs through Primakoff production on nuclear targets. These studies have explored the mass region from 200–450~MeV and have set the stage for more sensitive future searches at higher energies. Parity-Violating Deep Inelastic Scattering (PVDIS): Simulations indicate that the 22~GeV upgrade will significantly enhance the kinematic coverage and precision of PVDIS measurements. This improvement will allow for more stringent tests of the Standard Model and increased sensitivity to potential BSM effects.

\item Compared to the White Paper, a detailed Monte Carlo simulation was performed to estimate the fluxes of secondary muon and neutrino beams produced by the interaction of the primary 22 GeV electron beam with the Hall A beam dump. The results of this study are reported in Ref.~\cite{Battaglieri:2023gvd}.

\item The application of ML algorithms, such as anomaly detection and pattern recognition, is being explored to identify rare BSM signals within large datasets. These techniques can uncover subtle deviations from the Standard Model that might otherwise go unnoticed.
\end{itemize}

\underline{Future plans}:
\begin{itemize}
\item On the theory side, improving the estimate of the electron momentum distribution in the deeply relativistic regime is essential. Refining the corresponding predictions for FIP production is a critical step toward transitioning from rough order-of-magnitude estimates to a fully quantitative framework for any ``electron-motion-motivated'' search strategy.

\item High-priority simulations are needed to quantify sensitivity to ALPs, dark scalars, and other light mediators using AI-enhanced classification. These include training supervised models, e.g. convolutional neural networks (CNNs) and Boosted Decision Trees (BDTs), to isolate Primakoff-produced resonances and unsupervised models (e.g. auto-encoders, normalizing flows) for anomaly detection in the $\gamma \gamma$, $e^+ e^-$, or $\pi^+ \pi^-$ channels. Simulations must incorporate end-to-end AI pipelines that combine calorimeter, tracking, and recoil signatures to enhance event-level inference for weakly coupled, displaced, or long-lived BSM decays.

\item To better assess the sensitivity of the Beam Dump Experiment (BDX) at 22 GeV, more realistic simulations of the full experimental setup are needed—particularly to evaluate the dominant beam-related backgrounds: muons and neutrinos. At 22 GeV, the muon flux is expected to be about an order of magnitude higher than at 12 GeV, with an energy spectrum extending up to 16~GeV. This necessitates the simulation of various shielding configurations and alternative strategies to effectively suppress the muon background at the detector. Similarly, the neutrino flux increases significantly at 22~GeV, introducing an irreducible background that requires detailed modeling. Currently, the BDX Collaboration is focused on optimizing the infrastructure and detector design for the 12 GeV configuration, with operations planned to begin in 2028~\cite{Celentano:2015rea}. The experience gained from this initial phase will be critical for evaluating the feasibility of a 22 GeV upgrade. To further explore the potential of high-intensity secondary beams, an international workshop, {\it Secondary Beams at Jefferson Lab - BDX \& Beyond}, took place in 2025~\cite{workshop-bdx}. The event brought together experts in neutrino, muon, and light dark matter physics from both theoretical and experimental communities. The workshop focused on maximizing the scientific output of secondary beams at JLab, and its conclusions will be summarized in a White Paper. These outcomes will help clarify the simulation and infrastructure requirements needed to pursue these advanced physics opportunities.

\item AI/ML methods must include probabilistic calibration to distinguish genuine new physics from statistical fluctuations. Bayesian neural networks and flow-based uncertainty modeling are crucial.

\item First measurements of parity-violating asymmetries from 12 GeV SoLID will serve as training targets for generative ML models that can simulate full detector response under BSM hypotheses. First public results are likely to come in 2025–2026, feeding into generative model training pipelines by 2027.

\item The 12 GeV BDX experiment will be crucial for optimizing and evaluating the feasibility of a future run at 22 GeV. Data collected at 12 GeV will provide essential insights into detector performance, shielding effectiveness, infrastructure needs, and background characterization. This experience will guide improvements and help identify the challenges of operating at higher energies, allowing for a more accurate assessment of the 22~GeV configuration’s potential and viability. BDX is scheduled to run parasitically alongside the MOLLER experiment in Hall~A, collecting data throughout its anticipated two-year duration. The first significant results from BDX are expected by around 2030.
\end{itemize}

\underline{The global landscape}:
\begin{itemize}
\item Regarding new physics searches, future updates on the X17 anomaly—particularly from the various experiments currently investigating it—will be crucial. A confirmation of the excess observed by the INFN PADME experiment~\cite{PADME:2025dla} would strongly motivate the development of detectors capable of reconstructing such a particle. At the same time, PADME has already demonstrated the feasibility of a light resonance search program using positron beams, a strategy that the 22 GeV positron program at JLab is uniquely positioned to extend to higher mass ranges. More broadly, it remains to be seen whether Belle-II (Japan) can surpass the current BaBar (SLAC) limits on dark photon-related new physics. Additionally, while on a longer timescale, the progress of the SHIP program at CERN should also be closely monitored~\cite{SHiP:2021nfo}.

\item MESA/P2 at Mainz will measure parity-violating asymmetries at very low $Q^2$ to extract $\sin^2 \theta_W$ precisely~\cite{Becker:2018ggl}, focusing on electroweak precision tests complementary to JLab. P2 commissioning is expected around 2027–2028. Belle-II will continue accumulating an unprecedented dataset (targeting 50~ab$^{-1}$ by $\sim$ 2035). It will dominate in searches for invisible decays of dark photons and ALPs in the 10–1000~MeV mass range through $e^+e^-$ collisions~\cite{Ferber:2022rsf,Acanfora:2024daq}.

\item In the coming years, BDX will operate at 12 GeV with a focus on searching for light dark matter. This effort complements a broader global program, with several other experiments—such as NA64 at CERN~\cite{NA64:2023wbi}—pursuing similar goals using different techniques and facilities. 

\item Primakoff Production on Electrons and Nuclei: The 22 GeV program provides unique, clean access to pseudoscalar $\gamma \gamma$ couplings ($\pi^0$, $\eta$, and $\eta'$) via real photon interactions with atomic electrons, a capability not available at Belle-II, LHC (CERN), or MESA. Sub-percent precision on decay widths and transition form factors is critical for $(g - 2)_\mu$ HLbL calculations and chiral anomaly tests. 

\item BSM Sensitivity in Electromagnetic Production Channels: Unlike collider experiments focused on $e^+e^-$ or $pp$ collisions, the 22 GeV CEBAF enables purely electromagnetic BSM production through the Primakoff effect  - allowing resonance searches for ALPs, scalars, and hidden sector particles with minimal QCD backgrounds. The combination of 22 GeV energy and extreme luminosity ($\sim$$10^{39}$~cm$^{-2}$s$^{-1}$) is unmatched, especially for PVDIS measurements with SoLID, offering competitive or superior constraints on effective BSM couplings at moderate $Q^2$.

\end{itemize}

\clearpage

\section{CEBAF Accelerator Energy Upgrade}

\subsection{Scientific Background}

Recently, the Cornell–Brookhaven National Laboratory (BNL)–ERL Test Accelerator Facility at Cornell~\cite{Bartnik:2020pos} has demonstrated eight-pass recirculation of an electron beam with energy recovery (four accelerating beam passes and four decelerating beam passes). All eight beams were recirculated by single arcs of FFA (Fixed Field Alternating Gradient) magnets. This exciting new technology would enable a cost-effective method to double the energy of CEBAF, enabling significant new scientific opportunities. Technical studies of the implementation of FFA technology at CEBAF are in progress~\cite{Bogacz:2024phg,Khan:2024uih,Bodenstein:2024tbc,Benesch:2023fug}. The community is planning several workshops designed to underscore the physics reach that such an upgrade could enable. Further resources are required to continue accelerator science and technology to study the possibility to increase the CEBAF beam energy in a cost-effective manner using an FFA configuration. 

The 22 GeV CEBAF is envisioned as a staged upgrade. Initially starting with 1.1~GeV per LINAC and adding 5 rather than 6 new FFA passes, the top energy of about 20~GeV will be reached. Then through cryomodule upgrades (replacing lower gradient cryomodules with upgraded C100 types) one would increase the energy per LINAC to 1.21~GeV, to bring the top energy to 22~GeV. This approach would assure adequate energy flexibility for the LINACs (as is currently the case with the 12-GeV machine), to slightly reduce energy, if, e.g., several cavities de-rated, or one of the cryomodules was bypassed. Furthermore, LINAC energy flexibility is essential to optimize polarization settings to multiple halls in combination with the Wien filter; with a fixed LINAC energy, only one hall can get polarized beam.

\subsection{Scope}

This project aims at extending the energy reach of CEBAF up to 22~GeV within the existing tunnel. The proposed energy upgrade envisions increasing the number of recirculations, while using the existing CEBAF cavity system. The energy gain per pass remains unchanged, while the number of passes through the accelerating cavities is nearly doubled. A proposal was formulated to replace the highest-energy arcs with FFA arcs. The new pair of arcs would support simultaneous transport of an additional six passes with energies spanning a factor of two, using the non-scaling FFA principle implemented with Halbach-derived permanent magnets - a novel magnet technology that significantly saves energy and lowers operating costs.

\subsection{Status}

A conceptual design scheme has been developed and the accelerator design process for its subsystems is underway. In particular, the following key components of the scheme are under study: 

\begin{itemize}
\item Recirculating FFA arcs configured with compact (3 m) FODO cells based on permanent multi-function Halbach magnets with dominant dipole and quadrupole fields. The optics has been designed for closely spaced orbits (4 cm spread) and low betas (a few meters) resulting from very strong focusing, reducing the horizontal dispersion function from meters in conventional separate function arcs down to a few cm in the FFA arc. The arc optics was optimized to facilitate individual adjustment of momentum compaction and the horizontal emittance dispersion, $H$ (to suppress adverse effects of the synchrotron radiation on beam quality).
\item The permanent magnet design features an open mid-plane geometry, in order for the synchrotron radiation to pass through the magnets, while minimizing radiation damage to the permanent magnet material. This novel magnet technology saves energy and lowers operating costs. 
\item The ``Splitter" beamlines, being essential for time-of-flight correction of each of the energies that pass through the FFA arcs, have been investigated. The current design of individual horizontal chicanes has been optimized for adjusting the momentum compaction and dispersion for each energy beam, while minimizing the horizontal emittance dispersion, $H$.
\item The multi-pass LINAC optics has been designed to accommodate twice the number of passes and to provide uniform focusing in a vast range of energies, using a fixed quad lattice. A strongly focusing triplet lattice, scaled up with increasing momentum along the LINAC has been optimized for 10 passes.
\item Replacement of the current 123 MeV injector with a higher energy, 650 MeV injector compatible with the new multi-pass LINACs. It is envisioned as a 3-pass recirculating injector based on the existing LERF infrastructure. The current injector design is synergistic with the electron source needed to produce positrons for Ce$^+$BAF~\cite{Grames:2023lxn}.
\item The ``transition" section between the FFA arcs and the LINACs has been designed as a two-step-system: the first step adiabatically matches the beam dispersion and orbit trajectories, while the second step aligns the Twiss parameters with those at the LINAC entrance. Given the tight spatial constraints and multiple matching requirements, a genetic algorithm is being explored to optimize beam optic matching.
\item A concept of a new extraction system and beam delivery to the experimental halls has been explored. Variations of the extraction system based on RF separators, capable of four-hall-operation has been studied based on the requirements of the experimental physics users. Individual hall beamlines are currently under investigation. Improvements to the magnetic septa are expected to be required.
\end{itemize}

To conclude, significant progress has been made in the design of the energy upgrade for CEBAF using FFA beam transport. Over the last three years, we have settled on a design concept, developed more detailed optics solutions for various machine sections, and optimized the overall design, as simulations were performed. While the full design is not yet completed, we are working toward that goal to be summarized in an upcoming Technical Design Report.

\subsection{Plans}

The accelerator design tasks planned for this project over the next 5 years (2026–2030):

\begin{itemize}
\item Complete optics design for a pair of FFA arcs, including merger transitions to the LINACs
\item Splitter design for the northeast and southwest arc ends, including momentum compaction compensation and beta matching
\item Extraction system design capable of delivering all FFA passes (1-5) to Halls A, B, C (same energy) and half a pass above to Hall D
\item Finalize multi-pass LINAC optics based on strongly focusing triplets
\item 3-pass, 650 MeV recirculating injector design (in the LERF vault)
\item Carry out start-to-end (S2E) tracking of the entire complex with field maps, including errors
\item Implement a robust correction system
\item Complete instrumentation requirements
\item Design and prototype special magnets (compact bends for the Splitter, extraction septa)
\item Study synchrotron radiation emittance growth (longitudinal and transverse) via tracking and implement mitigation
\item Study synchrotron radiation effects on polarization
\item Perform a beam transport demonstration for a prototype FFA cell with multi-GeV beams at CEBAF
\item Complete the Technical Design Report
\end{itemize}

\subsection{R\&D Activities}

The ongoing and proposed research and development activities for the near-term future (2024–2026) include:

\begin{itemize}
\item Ongoing: ``FFA@CEBAF Permanent Magnet Resiliency in Real Radiation Environment'' - Testing permanent magnet materials in a radiation environment similar to those in which they will operate. Samples of permanent magnet materials (neodymium iron boron and or samarium cobalt) will be installed around the CEBAF facility in a variety of different radiation environments, along with appropriate dosimetry.
\item Proposed: ``FFA Beam Transport Test at CEBAF – Measurements Scoping and Beamline Modifications" – Design and integration of an instrumented FFA magnet insert (full- or half-FFA cell) within the existing CEBAF infrastructure (beam switchyard dump line or Hall C line). The aim is to design a test demonstrating multi-energy beam transport through the FFA cell configured with permanent magnets, including orbit and optics functions measurements.
\end{itemize}

\clearpage

\begin{acknowledgments}

This work was supported in part by the the U.S. Department of Energy, Office of Science, Office of Nuclear Physics  under contract DE-AC05-06OR23177.

\noindent
A special thanks to Daniel Carman for his crucial role in compiling, drafting, and coherently editing this document.

\vskip 0.2cm

\noindent
{\bf Chairs}: David Dean, Cynthia Keppel, Patrizia Rossi, Matthew Shepherd\\
\vspace{-2.0mm}

\noindent
{\bf Local Organization Committee}: Marco Battaglieri, Marco Mirazita, Alessandro Pilloni, Patrizia Rossi \\
\vspace{-2.0mm}

\noindent
{\bf Program Committee}: Harut Avakian, Marco Battaglieri, Jian-Ping Chen, Lamiaa El Fassi, Latifa Elouadrhiri, Liping Gan, Dave Gaskell, Ralf Gothe, Garth Huber, Cynthia Keppel, Marco Mirazita, Ioana Niculescu,
Alessandro Pilloni, Patrizia Rossi, Misak Sargisian, Nobuo Sato, Matt Shepherd, Justin Stevens, Christian Weiss

\end{acknowledgments}


\bibliographystyle{unsrtnat}
\bibliography{main}

\clearpage

\begin{appendix}
    \section{Table of Contributions}

Contributions were organized into six different topical areas over the five-day workshop.
Parallel discussions were scheduled to synthesize the presented material in the context of
the goals of the workshop as outlined in this talk.  The final day included overview talks,
summaries from the working group conveners, and discussions about planning the next steps.
A list of all contributions appears in the table below.  Copies of the presentation materials
can be found by following the hyperlinked titles to the 
\href{https://agenda.infn.it/event/39742/timetable/#all.detailed}{INDICO site}.

\subsection*{Welcome}
\rowcolors{1}{white}{gray!10}
\begin{longtable}{p{10.4cm} >{\raggedright\arraybackslash}p{4.5cm}}
\toprule
\textbf{Title (links to contribution page)} & \textbf{Speaker} \\
\midrule
\href{https://agenda.infn.it/event/39742/contributions/248651/}{Welcome from the organizers} & Patrizia Rossi (Istituto Nazionale di Fisica Nucleare) \\
\href{https://agenda.infn.it/event/39742/contributions/248652/}{Welcome from the LNF director} & Paola Gianotti (LNF) \\
\bottomrule
\end{longtable}
\vspace{0.4cm}

\subsection*{Spectroscopy}
\rowcolors{1}{white}{gray!10}
\begin{longtable}{p{10.4cm} >{\raggedright\arraybackslash}p{4.5cm}}
\toprule
\textbf{Title (links to contribution page)} & \textbf{Speaker} \\
\midrule
\href{https://agenda.infn.it/event/39742/contributions/246854/}{Spectroscopy at $e^+e^-$ Machines in the JLab 22 GeV era} & Nils Hüsken (JGU Mainz) \\
\href{https://agenda.infn.it/event/39742/contributions/248653/}{Theoretical Overview of Heavy Meson Spectroscopy in Photoproduction} & Daniel Winney (Universität Bonn (HISKP)) \\
\href{https://agenda.infn.it/event/39742/contributions/248659/}{Spectroscopy with Quasi-Real Photoproduction} & Derek Ian Glazier \\
\href{https://agenda.infn.it/event/39742/contributions/248000/}{Experimental Expectations for XYZs at GlueX} & Malte Albrecht (Ruhr-Universitaet Bochum) \\
\href{https://agenda.infn.it/event/39742/contributions/246848/}{Exotic States at 22 GeV era Kaon Beams} & Mikhail Bashkanov (University of York) \\
\href{https://agenda.infn.it/event/39742/contributions/246849/}{Spectroscopy Experiment of Charmed and Multi-strange Baryons using Hadron Beam at the J-PARC Hadron Facility} & Dr Kotaro Shirotori (RCNP, Osaka University) \\
\bottomrule
\end{longtable}
\vspace{0.4cm}

\subsection*{Partonic Structure and Spin}
\rowcolors{1}{white}{gray!10}
\begin{longtable}{p{10.4cm} >{\raggedright\arraybackslash}p{4.5cm}}
\toprule
\textbf{Title (links to contribution page)} & \textbf{Speaker} \\
\midrule
\href{https://agenda.infn.it/event/39742/contributions/247844/}{Study Light Sea in Intermediate-x Region with SIDIS at JLab22 in SoLID and Hall C} & Jian-ping Chen (Jefferson Lab) \\
\href{https://agenda.infn.it/event/39742/contributions/247857/}{The SoLID-SIDIS Program at 22GeV} & Prof. Zhihong Ye (Tsinghua University) \\
\href{https://agenda.infn.it/event/39742/contributions/248661/}{Experiment: Parity Violating DIS for c-odd PDFs and BSM} & Michael Nycz (University of Virginia) \\
\href{https://agenda.infn.it/event/39742/contributions/247862/}{Determination of $\alpha_s$ with JLab at 22 GeV} & Alexandre Deur (Jefferson Lab) \\
\href{https://agenda.infn.it/event/39742/contributions/247841/}{Meson and Nucleon Form Factors} & Garth Huber (University of Regina) \\
\href{https://agenda.infn.it/event/39742/contributions/247852/}{Meson Parton Distributions at 22 GeV} & Patrick Barry (Argonne National Lab) \\
\href{https://agenda.infn.it/event/39742/contributions/248667/}{Discussion: Partonic Structure and Spin} & (session discussion) \\
\href{https://agenda.infn.it/event/39742/contributions/248662/}{Role of QED Effects} & Jianwei Qiu (Jefferson Lab) \\
\href{https://agenda.infn.it/event/39742/contributions/247847/}{Impact of JLab22 on Unpolarized PDFs at Large x} & Dr Matteo Cerutti (Christopher Newport U. and Jefferson Lab) \\
\bottomrule
\end{longtable}
\clearpage

\subsection*{Hadronization and Transverse Momentum}
\rowcolors{1}{white}{gray!10}
\begin{longtable}{p{10.4cm} >{\raggedright\arraybackslash}p{4.5cm}}
\toprule
\textbf{Title (links to contribution page)} & \textbf{Speaker} \\
\midrule
\href{https://agenda.infn.it/event/39742/contributions/248669/}{The Role of Multi-D Approach in TMD Studies: COMPASS Experience} & Bakur Parsamyan (INFN) \\
\href{https://agenda.infn.it/event/39742/contributions/248668/}{Vector Mesons} & Harut Avakian \\
\href{https://agenda.infn.it/event/39742/contributions/248671/}{SIDIS MC Including Polarization Effects} & Christopher Dilks (Jefferson Lab) \\
\href{https://agenda.infn.it/event/39742/contributions/248672/}{Polarized Collisions at LHC} & Pasquale Di Nezza (INFN) \\
\href{https://agenda.infn.it/event/39742/contributions/247697/}{Isolated Meson Electroproduction at High Transverse Momentum with 22~GeV Electrons} & Carl Carlson (William \& Mary) \\
\href{https://agenda.infn.it/event/39742/contributions/248673/}{Hadron Mass Corrections} & Alberto Accardi (Hampton U. and Jefferson Lab) \\
\href{https://agenda.infn.it/event/39742/contributions/247860/}{Tree Level Matching Relations for Next-to-leading Power TMDs with Mass Corrections} & Alessio Carmelo Alvaro (Università di Pavia) \\
\href{https://agenda.infn.it/event/39742/contributions/247700/}{Exploring SIDIS Kinematics Regions at Jefferson Lab 22 GeV} & Prof. Mariaelena Boglione (INFN) \\
\href{https://agenda.infn.it/event/39742/contributions/247693/}{Towards Pixel-Based Imaging of Transverse Momentum Distributions} & Marco Zaccheddu (Jefferson Lab) \\
\href{https://agenda.infn.it/event/39742/contributions/247696/}{Unveiling the Collins-Soper Kernel in Inclusive DIS at Threshold and Implications at JLab 22} & Andrea Simonelli \\
\href{https://agenda.infn.it/event/39742/contributions/248674/}{Back-to-back SIDIS} & Timothy Hayward \\
\href{https://agenda.infn.it/event/39742/contributions/248675/}{Kaon SIDIS} & Simone Vallarino (INFN) \\
\href{https://agenda.infn.it/event/39742/contributions/247692/}{MAPTMD24: First Global Flavor Dependent TMD Extractions} & Lorenzo Rossi (INFN) \\
\href{https://agenda.infn.it/event/39742/contributions/247695/}{Probing the Transverse Momentum of Longitudinally Polarized Quarks} & Dr Matteo Cerutti (Christopher Newport U. and Jefferson Lab) \\
\href{https://agenda.infn.it/event/39742/contributions/247694/}{Accessing Gluon Polarization with High-$P_T$ Hadrons in SIDIS} & Yiyu Zhou (University of Turin) \\
\bottomrule
\end{longtable}
\vspace{0.4cm}

\subsection*{Spatial Structure, Mechanical Properties, and Emergent Hadron Mass}
\rowcolors{1}{white}{gray!10}
\begin{longtable}{p{10.4cm} >{\raggedright\arraybackslash}p{4.5cm}}
\toprule
\textbf{Title (links to contribution page)} & \textbf{Speaker} \\
\midrule
\href{https://agenda.infn.it/event/39742/contributions/247861/}{J/$\psi$ Photoproduction Close to Threshold: Prospects at JLab22} & Lubomir Pentchev (Jefferson Lab) \\
\href{https://agenda.infn.it/event/39742/contributions/248678/}{EIC for Exclusive Processes in the Region of Large $x>0.05$} & Carlos Munoz Camacho \\
\href{https://agenda.infn.it/event/39742/contributions/248679/}{Meson Structure} & Rachel Montgomery (University of Glasgow) \\
\href{https://agenda.infn.it/event/39742/contributions/247851/}{Resonance Electroexcitations at High Momentum Transfers with Jefferson Lab at 22 GeV} & Patrick Achenbach (Jefferson Lab) \\
\href{https://agenda.infn.it/event/39742/contributions/247859/}{Nonperturbative Approach Towards Emergent Hadron Structure and Mass at JLab22} & Prof. Adnan Bashir (University of Michoacan and University of Huelva) \\
\href{https://agenda.infn.it/event/39742/contributions/248305/}{Refined Simulations of Double Pion Electroproduction for CLAS22} & Alexis Osmond (University of South Carolina) \\
\href{https://agenda.infn.it/event/39742/contributions/248304/}{Inclusive Electron Scattering in the Resonance Region at High $Q^2$} & Gabriel Niculescu (James Madison University) \\
\href{https://agenda.infn.it/event/39742/contributions/248703/}{Complementary Insights into the Pseudoscalar Meson and Baryon Structure from AMBER} & Oleg Denisov (Istituto Nazionale di Fisica Nucleare) \\
\href{https://agenda.infn.it/event/39742/contributions/247854/}{Future Hypernuclear Studies at J-PARC and JLab} & Prof. Satoshi N. Nakamura (The University of Tokyo) \\
\href{https://agenda.infn.it/event/39742/contributions/247853/}{Double Deeply Virtual Compton Scattering (DDVCS) at 22 GeV} & Rafayel Paremuzyan (Jefferson Lab) \\
\bottomrule
\end{longtable}

\clearpage

\subsection*{Nuclear Dynamics}
\rowcolors{1}{white}{gray!10}
\begin{longtable}{p{10.4cm} >{\raggedright\arraybackslash}p{4.5cm}}
\toprule
\textbf{Title (links to contribution page)} & \textbf{Speaker} \\
\midrule
\href{https://agenda.infn.it/event/39742/contributions/247843/}{Prospects of Medium Modification Studies at JLab22} & Ian Cloët (Argonne National Laboratory) \\
\href{https://agenda.infn.it/event/39742/contributions/247848/}{The EMC Effect of Light-Nuclei within the Light-Front Hamiltonian Dynamics} & Matteo Rinaldi (INFN Perugia) \\
\href{https://agenda.infn.it/event/39742/contributions/247849/}{Continuing the Search for 3N SRCs} & Nadia Fomin (University of Tennessee) \\
\href{https://agenda.infn.it/event/39742/contributions/247850/}{Studying the Tensor-Polarized Deuteron System in the 22 GeV era} & Nathaly Santiesteban (University of New Hampshire) \\
\href{https://agenda.infn.it/event/39742/contributions/247998/}{Probing the EMC Effect thru the Measurement of Super-fast Quarks in Nuclei} & John Arrington (Lawrence Berkeley National Laboratory) \\
\href{https://agenda.infn.it/event/39742/contributions/247842/}{Searching for Color Transparency Effects at 22 GeV} & Holly Szumila-Vance (Florida International University) \\
\href{https://agenda.infn.it/event/39742/contributions/247856/}{Medium Modifications of Quark Propagation and Hadron Formation Observables} & Taisiya Mineeva (CCTVaL, UTFSM) \\
\href{https://agenda.infn.it/event/39742/contributions/247891/}{Study of Tagged Processes with $^4$He and ALERT at 22 GeV} & Mathieu Ouillon (Mississippi State University) \\
\href{https://agenda.infn.it/event/39742/contributions/247858/}{Study Quark Structures with eA Scattering at 22 GeV} & Prof. Zhihong Ye (Tsinghua University) \\
\href{https://agenda.infn.it/event/39742/contributions/247899/}{Anti-Shadowing Exploration Opportunities with CEBAF at 22 GeV} & Narbe Kalantarians (Virginia Union University) \\
\bottomrule
\end{longtable}
\vspace{0.4cm}

\subsection*{QCD Confinement, Fundamental Symmetries and BSM}
\rowcolors{1}{white}{gray!10}
\begin{longtable}{p{10.4cm} >{\raggedright\arraybackslash}p{4.5cm}}
\toprule
\textbf{Title (links to contribution page)} & \textbf{Speaker} \\
\midrule
\href{https://agenda.infn.it/event/39742/contributions/247846/}{Pseudoscalar Mesons and Emergent Mass} & Khepani Raya (University of Huelva) \\
\href{https://agenda.infn.it/event/39742/contributions/248681/}{Probe QCD Confinement via $\pi^0$, $\eta$ and $\eta'$} & Karol Kampf (Charles University) \\
\href{https://agenda.infn.it/event/39742/contributions/248682/}{Overview of Sub-GeV Physics in the Dark Sector (theory side)} & Luc Darme (IP2I - Lyon 1 University) \\
\href{https://agenda.infn.it/event/39742/contributions/248683/}{Experimental Overview of Sub-GeV Physics in the Dark Sector} & Mariangela Bondi (Istituto Nazionale di Fisica Nucleare) \\
\href{https://agenda.infn.it/event/39742/contributions/248684/}{The BDX Experiment} & Marco Spreafico (Istituto Nazionale di Fisica Nucleare) \\
\href{https://agenda.infn.it/event/39742/contributions/248685/}{AI for BSM Physics Searches} & Patrick Moran (The College of William \& Mary) \\
\href{https://agenda.infn.it/event/39742/contributions/247898/}{Probing Light Dark Particles with $\eta$ and $\eta'$ Decays} & Sergi Gonzalez-Solis (University of Barcelona) \\
\bottomrule
\end{longtable}
\vspace{0.4cm}

\subsection*{Summaries and Path Forward}
\rowcolors{1}{white}{gray!10}
\begin{longtable}{p{10.4cm} >{\raggedright\arraybackslash}p{4.5cm}}
\toprule
\textbf{Title (links to contribution page)} & \textbf{Speaker} \\
\midrule
\href{https://agenda.infn.it/event/39742/contributions/248693/}{Gravitational Structure of the Nucleon} & Volker Burkert (Jefferson Lab) \\
\href{https://agenda.infn.it/event/39742/contributions/247861/}{Overview of the CEBAF Accelerator Upgrade} & Alex Bogacz \\
\href{https://agenda.infn.it/event/39742/contributions/248700/}{Positron Workshop Summary} & Marco Battaglieri (Istituto Nazionale di Fisica Nucleare) \\
\href{https://agenda.infn.it/event/39742/contributions/248701/}{Realizing JLab 22 GeV: Project Planning} & Allison Lung (Jefferson Lab) \\
\href{https://agenda.infn.it/event/39742/contributions/248692/}{Charmed Pentaquark Production at the EicC and JLab 12 to 22 GeV} & Feng-Kun Guo (Institute of Theoretical Physics, Chinese Academy of Sciences) \\
\href{https://agenda.infn.it/event/39742/contributions/248694/}{Spectroscopy: Summary} & Justin Stevens (William \& Mary) \\
\href{https://agenda.infn.it/event/39742/contributions/248695/}{Partonic Structure and Spin: Summary} & Nobuo Sato (Jefferson Lab) \\
\href{https://agenda.infn.it/event/39742/contributions/248696/}{Hadronization and Transverse Momentum: Summary} & Harut Avakian; Nobuo Sato (Jefferson Lab) \\
\href{https://agenda.infn.it/event/39742/contributions/248697/}{Spatial Structure, Mechanical Properties, and Emergent Hadron Mass: Summary} & Garth Huber (University of Regina) \\
\href{https://agenda.infn.it/event/39742/contributions/248698/}{QCD Confinement, Fundamental Symmetries and BSM: Summary} & Marco Battaglieri (Istituto Nazionale di Fisica Nucleare) \\
\href{https://agenda.infn.it/event/39742/contributions/248699/}{Nuclear Dynamics: Summary} & Lamiaa El Fassi; Misak Sargsian \\
\href{https://agenda.infn.it/event/39742/contributions/248702/}{Discussion and Path forward} & Cynthia Keppel (Thomas Jefferson National Accelerator Facility) \\
\href{https://agenda.infn.it/event/39742/contributions/251348/}{Closeout} & Patrizia Rossi (Istituto Nazionale di Fisica Nucleare) \\
\bottomrule
\end{longtable}

\end{appendix}

\end{document}